\documentclass[11pt]{article}
\usepackage{epsfig} 
\newcommand{\sect}[1]{ \section{#1} \setcounter{equation}{0} }

\newcommand{\Dslash}{D \! \! \! \! /}

\newcommand{\half}{\mbox{\small{$\frac{1}{2}$}}}

\newcommand{\MSbar}{\overline{\mbox{MS}}} 
 
\newcommand{\Nf}{N_{\!f}}
\newcommand{\NF}{N_{\!F}}
\newcommand{\NA}{N_{\!A}}

\begin{document}
\title{Loop calculations in the three dimensional Gribov-Zwanziger Lagrangian}
\author{J.A. Gracey, \\ Theoretical Physics Division, \\ 
Department of Mathematical Sciences, \\ University of Liverpool, \\ P.O. Box 
147, \\ Liverpool, \\ L69 3BX, \\ United Kingdom.} 
\date{} 
\maketitle 

\vspace{5cm} 
\noindent 
{\bf Abstract.} The three dimensional Gribov-Zwanziger Lagrangian is analysed
at one and two loops. Specifically, the two loop gap equation is evaluated and
the Gribov mass is expressed in terms of the coupling constant. The one loop
corrections to the propagators of all the fields are determined. It is shown 
that when the gap equation is satisfied the Faddeev-Popov ghost and both Bose 
and Grassmann localizing ghosts all enhance in the infrared limit at one loop. 
This verifies that the Kugo-Ojima confinement criterion holds to this order and
we also show that both Grassmann ghosts are enhanced at two loops. For the 
Bose ghost we determine the full form of the propagator in the zero momentum 
limit for both the transverse and longitudinal pieces and confirm Zwanziger's 
recent general analysis for the low energy behaviour. We provide an alternative
but equivalent version of the horizon condition expressing it as the vacuum 
expectation value of an operator involving only the localizing Bose ghost 
field. The one loop static potential is also determined. 

\vspace{-19.5cm}
\hspace{13.5cm}
{\bf LTH 887}

\newpage

\sect{Introduction.}

The Gribov-Zwanziger Lagrangian is a formulation of the Landau gauge fixed
Yang-Mills theories where the Gribov problem is incorporated in a localized
way, \cite{1,2,3,4}. This problem, \cite{5}, essentially relates to the 
difficulties in fixing a gauge {\em globally} for gauge theories with a 
non-abelian symmetry. In his seminal work, \cite{5}, Gribov demonstrated that 
globally different gauge configurations could satisfy the same gauge condition 
thereby introducing an ambiguity into the gauge fixing procedure. Such Gribov 
copies do not affect the local gauge fixing in Yang-Mills theories and hence 
the ultraviolet structure of such theories does not encounter gauge fixing 
difficulties. By contrast, the problem relates to global issues and hence the 
infrared r\'{e}gime of the theory. For non-abelian gauge theories, Gribov 
indicated that such gauge copies could be entangled with the problem of 
confinement, \cite{5}, which is sometimes referred to as infrared slavery. One
consequence of the analysis of \cite{5} is to overcome the copy problem in the 
main by restricting the path integral to a specific region of configuration 
space. This region, known as the Gribov region, denoted by $\Omega$ and 
containing the origin, is defined by the locus of points where the 
Faddeev-Popov operator is positive, \cite{5}. Geometrical aspects of the region
and their consequences have been explored in \cite{6,7,8,9,10,11}. As an aside
we note that such a restriction does not lead to unambiguous gauge 
configurations. Instead the Gribov region has a subregion called the 
Fundamental Modular Region, denoted by $\Lambda$, where the gauge {\em is} 
fixed uniquely globally. However, it has been argued in \cite{11} that Green's 
functions defined over $\Lambda$ and $\Omega$ are equivalent. To incorporate 
the path integral restriction to $\Omega$ Gribov modified the Yang-Mills action
to include a non-local operator which in effect cut off the domain of 
integration, \cite{5}. The presence of such an operator, referred to as the 
horizon or no pole condition, introduces an arbitrary mass scale, $\gamma$, 
which is known as the Gribov mass. However, it is not a new parameter of the 
theory but satisfies a gap equation defined by the defining horizon condition 
and is a function of the coupling constant, \cite{5}. The presence of this 
non-local operator and the Gribov mass alters the structure of the propagators 
of the theory. For instance, the gluon has a propagator which vanishes at zero 
momentum and depends on the Gribov mass. Though it has a non-fundamental form 
with the gluon being in effect massless but with a non-zero width, \cite{5}. 
This may appear to be contradictory but the gluon is not a fundamental field in
itself as it is confined. A second feature is that as a consequence of the gap 
equation for $\gamma$, the Faddeev-Popov ghost propagator has an enhanced or 
dipole behaviour in the zero momentum limit. The latter property was later 
encapsulated in the Kugo-Ojima confinement criterion, \cite{12,13}, for the 
Landau gauge. Indeed this condition has been re-examined in the 
Gribov-Zwanziger context in \cite{14}.

The relevance of the Gribov-Zwanziger Lagrangian to the Gribov path integral
restriction rests primarily in the reformulation of Gribov's Lagrangian in a 
localized way, \cite{1,2,3,4}. Gribov's analysis was in a semi-classical 
approximation but one cannot perform high order loop computations using a 
Lagrangian with a non-local operator. Instead Zwanziger managed to localize the
non-locality to produce a local renormalizable Lagrangian for the Landau gauge.
The renormalizability has been established in several articles, \cite{4,15,16}.
The localization introduces several extra fields which are known as ghosts. 
However, one set, $\{ \phi^{ab}_\mu, \bar{\phi}^{ab}_\mu \}$, have Bose 
statistics whilst their partners, $\{ \omega^{ab}_\mu, \bar{\omega}^{ab}_\mu 
\}$, are fermionic. The latter are crucial in maintaining the established
ultraviolet properties of the theory such as asymptotic freedom and ensure the
one and higher loop $\MSbar$ $\beta$-function, \cite{17,18,19,20,21}, is
unaltered. This is important since the Gribov operator in some sense relates to
the infrared structure of the theory and its presence therefore ought not to 
upset the ultraviolet structure where indeed there is no gauge fixing 
ambiguity. One consequence of the localization is that one can carry out 
explicit loop computations. In \cite{2,4} the one loop gap equation satisfied 
by $\gamma$ in the original Gribov action was reproduced. This has been 
extended to two loops in \cite{22} in the $\MSbar$ scheme and led to the check 
that the Kugo-Ojima confinement criterion holds to two loops explicitly. More 
recently the one loop static potential for heavy quarks was computed in the
Gribov-Zwanziger context in \cite{23} as well as the full one loop corrections 
to all the propagators of the fields. It transpires that the latter have been 
important for verifying a recent non-perturbative analysis of the 
Gribov-Zwanziger framework by Zwanziger, \cite{24}. 

Whilst the appearance of dipole propagators for the Faddeev-Popov and 
$\omega^{ab}_\mu$ ghost propagators is in keeping with the Kugo-Ojima
criterion, \cite{12,13,14}, such fields cannot play a role in the actual 
confinement of heavy quarks, for instance. This is purely due to their 
statistics and the lack of a (direct) coupling to quarks. Instead one would 
require an enhanced field with Bose statistics. Clearly the gluon cannot be 
that field due to its infrared suppression. However, using symmetry arguments 
which are valid to all orders, Zwanziger has argued that certain colour 
components of the imaginary part of the Bose ghost field, $\phi^{ab}_\mu$, are 
enhanced, \cite{24}. Indeed this structure has been confirmed at one loop in 
\cite{25}. There the explicit enhancement was shown for the transverse 
component. The longitudinal piece was not considered since it clearly could not 
play a role in the exchange particle for the static potential considered there.
However, it has been analysed subsequently (and referred to in passing in 
\cite{24}) and will be reported on in full in this article as part of a larger 
calculation. More specifically given the elegance of the Gribov-Zwanziger 
construction and its potential for being a working Lagrangian incorporating 
confinement, it is the main aim of this article to record in one place the full
analysis for the {\em three} dimensional theory. In the series of papers, 
\cite{22,23,25}, the two loop $\MSbar$ gap equation, static potential and all 
the propagator corrections were all given for four dimensions. However, if one 
is to have a full understanding of that case, it should also be the case that 
there are parallels in the lower dimensional version. Indeed the three 
dimensional theory has several interesting features deserving study in their 
own right. For instance, the Gribov mass plays the role of a natural infrared 
regulator in this superrenormalizable quantum field theory. Moreover, the 
ultraviolet finiteness means that the gap equation simply relates the Gribov 
mass to the (dimensionful) coupling constant. Therefore, we will provide all 
the quantities for the three dimensional Gribov-Zwanziger Lagrangian that have 
been computed in four dimensions to the same loop order.

It would be remiss not to discuss the relation of the current Gribov-Zwanziger
scenario with that found on the lattice for quantities such as the gluon and 
Faddeev-Popov ghost propagators. The present point of view is that the 
propagators are not respectively suppressed or enhanced at zero momentum. 
Instead the gluon propagator freezes to a non-zero value whilst there is 
clearly no dipole behaviour for the ghost. The evidence for this has been
provided over a number of years by several lattice collaborations,
\cite{26,27,28,29,30,31,32}. This decoupling scenario, \cite{33}, is in 
contrast to the conformal or scaling situation of gluon propagator suppression 
and ghost enhancement of the original Gribov set-up, \cite{5}. Whilst it is 
possible to model the decoupling situation by a condensate argument based on 
the Gribov-Zwanziger Lagrangian, \cite{34,35}, it is clear that the Kugo-Ojima 
confinement criterion cannot be satisfied. Moreover, there cannot be any 
enhanced propagators with either Bose or fermion statistics. However, one can 
argue that the positivity violating gluon propagator which the decoupling 
solution has, is sufficient to ensure a confining theory. Though a condensate 
explanation based on a perturbative vacuum would need to be extended to
incorporate non-perturbative aspects of the vacuum. Irrespective of this we 
believe that the debate has not been fully resolved and that to understand the 
theory one ought at the very least to have as much analysis available as is 
calculationally possible and in spacetime dimensions other than just four. 
 
The article is organised as follows. We review the main features of the
Gribov-Zwanziger Lagrangian in section two which are required for our three
dimensional analysis. The two loop correction to the gap equation satisfied
by $\gamma$ is given in the next section with an estimate for the non-zero one 
loop value of the renormalization group invariant strong coupling constant at 
zero momentum. Section four is devoted to the calculation of the one loop 
static potential of heavy quarks. Given that there is currently interest in the
behaviour of the Bose ghost at zero momentum we give the formal one loop
propagator corrections in section five. Whilst the transverse part was 
considered in \cite{25}, we concentrate on the longitudinal piece and extract
the Landau gauge behaviour in accord with Zwanziger's analysis of \cite{24}. 
The explicit one loop structure of the gluon and $\phi^{ab}_\mu$ propagator 
sector for both three and four dimensions and for arbitrary colour group are 
recorded in section six. We give our conclusions in section seven. There are 
three appendices. The first two record the explicit one loop corrections to all
the $2$-point functions for the transverse and longitudinal sectors 
respectively. The final appendix provides the complete structure of the one 
loop form factors appearing in the propagator of the real part of 
$\phi^{ab}_\mu$ given the most general possible $SU(N_c)$ colour structure of 
the corresponding $2$-point function.

\sect{Formalism.}

In this section we recall the basic formalism for the Gribov-Zwanziger 
Lagrangian, \cite{1,2,3,4}. First, the canonical QCD Lagrangian with a linear 
covariant gauge fixing term is given by  
\begin{equation} 
L^{\mbox{\footnotesize{QCD}}} ~=~ -~ \frac{1}{4} G_{\mu\nu}^a 
G^{a \, \mu\nu} ~-~ \frac{1}{2\alpha} (\partial^\mu A^a_\mu)^2 ~-~ 
\bar{c}^a \partial^\mu D_\mu c^a ~+~ i \bar{\psi}^{iI} \Dslash \psi^{iI} 
\label{lagqcd}
\end{equation} 
where $G^a_{\mu\nu}$ is the field strength for the gauge potential $A^a_\mu$, 
the covariant derivative, $D_\mu$, is defined by
\begin{eqnarray}
D_\mu c^a &=& \partial_\mu c^a ~-~ g f^{abc} A^b_\mu c^c \nonumber \\
D_\mu \psi^{iI} &=& \partial_\mu \psi^{iI} ~+~ i g T^a_{IJ} A^a_\mu \psi^{iJ} 
\end{eqnarray} 
$g$ is the coupling constant, $f^{abc}$ are the colour group structure
constants with generators $T^a$. The various indices are restricted to the
ranges $1$~$\leq$~$a$~$\leq$~$\NA$, $1$~$\leq$~$I$~$\leq$~$\NF$ and
$1$~$\leq$~$i$~$\leq$~$\Nf$ where $\NF$ and $\NA$ are the respective dimensions
of the fundamental and adjoint representations and $\Nf$ is the number of
massless quarks. Whilst the focus is primarily on the Yang-Mills Lagrangian we 
have incorporated massless quarks partly for completeness and also as an 
internal aid in checking some computations. The gauge parameter $\alpha$ is 
included in order to assist with determining the propagators of the theory. 
However, we will work in the Landau gauge throughout, where $\alpha$~$=$~$0$, 
which is assumed unless it is required at intermediate stages of computing one 
loop propagator corrections as will be the case in a later section. The
Lagrangian (\ref{lagqcd}) is the usual starting point for high energy 
computations and one does not need to be concerned with the fact that the gauge
is not fixed uniquely globally. In the ultraviolet r\'{e}gime Gribov copies do 
not alter physical predictions. However, to handle the ambiguity problem the
path integral restriction equates to modifying the action by the no pole
condition or equivalently the horizon condition. The boundaries of the Gribov 
regions are given by the zeroes of the Faddeev-Popov operator 
$\partial^\mu D^a_\mu$. So the interior of the first Gribov region is that set 
of points where the inverted Faddeev-Popov operator is finite. Whilst the 
original arguments of \cite{5} were based on a semi-classical approach this now
equates to extending (\ref{lagqcd}) to the Lagrangian 
\begin{equation}
L^{\mbox{\footnotesize{Grib}}} ~=~ L^{\mbox{\footnotesize{QCD}}} ~+~
\frac{C_A \gamma^4}{2} A^{a\,\mu} \frac{1}{\partial^\nu D_\nu} A^a_\mu ~-~
\frac{d \NA \gamma^4}{2g^2}
\label{laggrib}
\end{equation}
where $\gamma$ is the Gribov mass parameter. Originally the non-local operator
of (\ref{laggrib}) was only approximated by the Laplacian, \cite{5}, but this 
was later extended and made more concrete by Zwanziger in \cite{1}. The
parameter $\gamma$ is not an independent quantity in the Gribov theory. Instead
it is a function of the coupling constant and the relation between the two is 
defined by the horizon condition. For (\ref{laggrib}) this is, \cite{1,5}, 
\begin{equation}
\left\langle A^a_\mu(x) \frac{1}{\partial^\nu D_\nu} A^{a\,\mu}(x)
\right\rangle ~=~ \frac{d N_A}{C_A g^2}
\label{hordef}
\end{equation}
where $C_A$ is given by
\begin{equation}
f^{acd} f^{bcd} ~=~ C_A \delta^{ab} 
\end{equation}
and $d$ is the spacetime dimension. If one could handle the non-locality when
calculating the vacuum expectation value of (\ref{hordef}) then the gap 
equation satisfied by $\gamma$ would emerge. In four dimensions $\gamma$ is 
expressed as a non-analytic function of the coupling constant. It is important 
to stress that one cannot treat $\gamma$ as an independent parameter of the 
theory. The non-local theory cannot be regarded as a gauge theory, in the 
Landau gauge, unless $\gamma$ satisfies the gap equation, \cite{1,2,3,4,5}.

The key to resolving the calculational obstacle represented by the non-local
term of (\ref{laggrib}) was provided by Zwanziger in a series of interrelated 
articles, \cite{1,2,3,4,6,7,8,9}. By considering the properties of the Gribov 
region in the Landau gauge the non-local Lagrangian was transformed into a 
local Lagrangian which involved extra spin-$1$ fields. These localizing ghosts,
$\phi^{ab}_\mu$, $\bar{\phi}^{ab}_\mu$, $\omega^{ab}_\mu$ and 
$\bar{\omega}^{ab}_\mu$, where the first pair are bosonic and the latter 
Grassmannian, are additional to the gauge potential and the Faddeev-Popov 
ghosts. Their presence does not alter the ultraviolet properties of the theory
since, for instance, asymptotic freedom still holds in four dimensions. Instead
they become effective in the infrared limit as one approaches the Gribov 
boundary. More specifically, the localized version of the Gribov Lagrangian is 
\cite{1,2,3,4,8,9,36} 
\begin{eqnarray}
L^{\mbox{\footnotesize{GZ}}} &=& L^{\mbox{\footnotesize{QCD}}} ~+~ 
\frac{1}{2} \rho^{ab \, \mu} \partial^\nu \left( D_\nu \rho_\mu 
\right)^{ab} ~+~ \frac{i}{2} \rho^{ab \, \mu} \partial^\nu 
\left( D_\nu \xi_\mu \right)^{ab} ~-~ \frac{i}{2} \xi^{ab \, \mu} 
\partial^\nu \left( D_\nu \rho_\mu \right)^{ab} \nonumber \\
&& +~ \frac{1}{2} \xi^{ab \, \mu} \partial^\nu \left( D_\nu \xi_\mu 
\right)^{ab} ~-~ \bar{\omega}^{ab \, \mu} \partial^\nu \left( D_\nu \omega_\mu 
\right)^{ab} ~-~ \frac{1}{\sqrt{2}} g f^{abc} \partial^\nu 
\bar{\omega}^{ae}_\mu \left( D_\nu c \right)^b \rho^{ec \, \mu} \nonumber \\
&& -~ \frac{i}{\sqrt{2}} g f^{abc} \partial^\nu \bar{\omega}^{ae}_\mu 
\left( D_\nu c \right)^b \xi^{ec \, \mu} ~-~ i \gamma^2 f^{abc} A^{a \, \mu} 
\xi^{bc}_\mu ~-~ \frac{d \NA \gamma^4}{2g^2} 
\label{laggz}
\end{eqnarray} 
where there is a mixed $2$-point term involving the gluon. We have chosen to
follow the current convention and use the real and imaginary parts of the
Bose ghosts rather than the complex versions, \cite{24,36}. This is because in 
four dimensions the behaviour of the propagators of each component is 
significantly different and is difficult to extract cleanly in the original 
$\phi^{ab}_\mu$ and $\bar{\phi}^{ab}_\mu$ formulation. We take as the real and 
imaginary parts 
\begin{equation}
\phi^{ab}_\mu ~=~ \frac{1}{\sqrt{2}} \left( \rho^{ab}_\mu ~+~ i \xi^{ab}_\mu 
\right) ~~,~~ 
\bar{\phi}^{ab}_\mu ~=~ \frac{1}{\sqrt{2}} \left( \rho^{ab}_\mu ~-~ 
i \xi^{ab}_\mu \right) ~.
\end{equation} 
(For comparison $\rho^{ab}_\mu$ and $\xi^{ab}_\mu$ are respectively the
$U^{ab}_\mu$ and $V^{ab}_\mu$ fields of \cite{24,36}.) Although we have omitted
the $\alpha$ dependent term since (\ref{laggz}) corresponds to the Landau 
gauge, one requires that term to safely derive all the propagators. As we will 
be considering the infrared properties of the one loop corrections to the 
transverse and longitudinal parts of the propagators, we record for 
completeness the form of the intermediate propagators prior to taking the 
$\alpha$~$\rightarrow$~$0$ limit. These are 
\begin{eqnarray}
\langle A^a_\mu(p) A^b_\nu(-p) \rangle &=& -~ 
\frac{\delta^{ab}p^2}{[(p^2)^2+C_A\gamma^4]} P_{\mu\nu}(p) ~-~ 
\frac{\alpha\delta^{ab}p^2}{[(p^2)^2+\alpha C_A\gamma^4]} L_{\mu\nu}(p) 
\nonumber \\
\langle A^a_\mu(p) \xi^{bc}_\nu(-p) \rangle &=& 
\frac{i f^{abc}\gamma^2}{[(p^2)^2+C_A\gamma^4]} P_{\mu\nu}(p) ~+~ 
\frac{i \alpha f^{abc}\gamma^2}{[(p^2)^2+ \alpha C_A\gamma^4]} L_{\mu\nu}(p) 
\nonumber \\
\langle A^a_\mu(p) \rho^{bc}_\nu(-p) \rangle &=& 0 \nonumber \\ 
\langle \xi^{ab}_\mu(p) \xi^{cd}_\nu(-p) \rangle &=& -~ 
\frac{\delta^{ac}\delta^{bd}}{p^2}\eta_{\mu\nu} ~+~
\frac{f^{abe}f^{cde}\gamma^4}{p^2[(p^2)^2+C_A\gamma^4]} P_{\mu\nu}(p) ~+~
\frac{\alpha f^{abe}f^{cde}\gamma^4}{p^2[(p^2)^2+\alpha C_A\gamma^4]} 
L_{\mu\nu}(p) \nonumber \\ 
\langle \xi^{ab}_\mu(p) \rho^{cd}_\nu(-p) \rangle &=& 0 \nonumber \\ 
\langle \rho^{ab}_\mu(p) \rho^{cd}_\nu(-p) \rangle &=& 
\langle \omega^{ab}_\mu(p) \bar{\omega}^{cd}_\nu(-p) \rangle ~=~ -~ 
\frac{\delta^{ac}\delta^{bd}}{p^2} \eta_{\mu\nu} 
\label{propal}
\end{eqnarray} 
where 
\begin{equation}
P_{\mu\nu}(p) ~=~ \eta_{\mu\nu} ~-~ \frac{p_\mu p_\nu}{p^2} ~~~,~~~ 
L_{\mu\nu}(p) ~=~ \frac{p_\mu p_\nu}{p^2} 
\end{equation}
are the respective transverse and longitudinal projectors. Consequently, in our
Landau gauge calculations we will use 
\begin{eqnarray}
\langle A^a_\mu(p) A^b_\nu(-p) \rangle &=& -~ 
\frac{\delta^{ab}p^2}{[(p^2)^2+C_A\gamma^4]} P_{\mu\nu}(p) \nonumber \\
\langle A^a_\mu(p) \xi^{bc}_\nu(-p) \rangle &=& 
\frac{i f^{abc}\gamma^2}{[(p^2)^2+C_A\gamma^4]} P_{\mu\nu}(p) 
\nonumber \\
\langle A^a_\mu(p) \rho^{bc}_\nu(-p) \rangle &=& 0 \nonumber \\ 
\langle \xi^{ab}_\mu(p) \xi^{cd}_\nu(-p) \rangle &=& -~ 
\frac{\delta^{ac}\delta^{bd}}{p^2}\eta_{\mu\nu} ~+~
\frac{f^{abe}f^{cde}\gamma^4}{p^2[(p^2)^2+C_A\gamma^4]} P_{\mu\nu}(p) 
\nonumber \\ 
\langle \xi^{ab}_\mu(p) \rho^{cd}_\nu(-p) \rangle &=& 0 \nonumber \\ 
\langle \rho^{ab}_\mu(p) \rho^{cd}_\nu(-p) \rangle &=& 
\langle \omega^{ab}_\mu(p) \bar{\omega}^{cd}_\nu(-p) \rangle ~=~ -~ 
\frac{\delta^{ac}\delta^{bd}}{p^2} \eta_{\mu\nu} 
\label{propdef}
\end{eqnarray} 
as our propagators. The derivation of (\ref{propal}) and (\ref{propdef}) is
complicated by the mixed term of (\ref{laggz}) but was discussed at length in
\cite{25}. Though we note that for the real and imaginary Bose ghost there is a
clean split of the propagators with the real part propagator having a similar
form to that of the associated fermionic localizing ghost. When we examine the
propagator corrections in the infrared this feature will be preserved in
keeping with Zwanziger's recent general arguments, \cite{24}.

\sect{Gap equation.}

In this section we derive the two loop gap equation satisfied by the Gribov
mass in three dimensions. This computation is similar, for example, to that in 
four dimensions in terms of the Feynman diagrams to be computed. The method is
to evaluate the horizon condition (\ref{hordef}) but not in the non-local
version of the theory. Instead we consider the equivalent definition of the
condition in the localized theory, (\ref{laggz}). In terms of the real Bose
ghost fields this is 
\begin{equation}
f^{abc} \left\langle A^{a\,\mu} (x) \xi^{bc}_\mu(x) \right\rangle ~=~ 
\frac{i d\NA \gamma^2}{g^2} 
\label{gapdef}
\end{equation}
where the relation between the $A^a_\mu$ and $\xi^{ab}_\mu$ is established via
the equation of motion 
\begin{equation}
A^a_\mu ~=~ -~ \frac{i}{C_A \gamma^2} f^{abc} \left( \partial^\nu D_\nu \xi_\mu
\right)^{bc} ~. 
\label{xieom}
\end{equation}
Hence (\ref{gapdef}) clearly equates to (\ref{hordef}) using (\ref{xieom}).
Using this version of the gap equation one evaluates the Feynman diagrams of
the vacuum expectation value to two loops. At leading order there is one
Feynman graph and at two loops there are nineteen diagrams to evaluate. These 
are generated using the {\sc Qgraf} package, \cite{37}, and then converted into
{\sc Form} input notation where {\sc Form}, \cite{38}, is the symbolic 
manipulation language used to handle the associated algebra with the
computation. We follow the standard procedure of breaking the one and two loop
Feynman diagrams up into a sum of master vacuum bubble integrals by using 
tensor reduction and then substituting their explicit values. For three 
dimensions, all master massive vacuum bubble topologies to {\em three} loops 
have been computed in \cite{39} for all possible independent masses. It is 
relatively straightforward to extract the integrals required and include them 
in the {\sc Form} routines. However, one needs to be careful in the 
Gribov-Zwanziger case where the Gribov propagator is not the standard massive
propagator. One first has to apply partial fractions using, for example,
\begin{equation}
\frac{1}{[(p^2)^2+C_A\gamma^4]} ~=~ \frac{1}{2 i \sqrt{C_A} \gamma^2} 
\left[ \frac{1}{[p^2-i\sqrt{C_A}\gamma^2]} ~-~ 
\frac{1}{[p^2+i\sqrt{C_A}\gamma^2]} \right] 
\end{equation}
to obtain factors within the Feynman integrals of more standard form. However,
each involves an imaginary mass corresponding to massless unstable fields. The 
main issue, though, is in utilizing the master integrals of \cite{39} which we 
illustrate with the simple one loop integral. From  \cite{39}
\begin{equation}
\int \frac{d^dk}{(2\pi)^d} \, \frac{1}{[k^2+m^2]} ~=~ -~ \frac{m}{4\pi} \left[
1 ~+~ \left[ 2 + 2 \ln \left[ \frac{1}{2m} \right] \right] \epsilon ~+~
O(\epsilon^2) \right] 
\label{int1}
\end{equation}
where in dimensional regularization $d$~$=$~$3$~$-$~$2\epsilon$. The argument 
of the logarithm will be rendered dimensionless when the mass scale, $\mu$,
which is required to retain a dimensionless coupling constant in 
$d$-dimensions is included in the overall computation. In four dimensions the
overall factor of the evaluation would be dimension two but in three dimensions
the dimensionality reduces by one unit. Thus for the Gribov case one would
require the square root of the width. For each of $\pm i \sqrt{C_A} \gamma^2$
there are two possibilities resulting in four different underlying masses.
However, to exclude any potential ambiguity in the overall final expression for
the gap equation, which must be real and not complex, within our {\sc Form}
routines we have formally extended (\ref{int1}) to 
\begin{equation}
\int \frac{d^dk}{(2\pi)^d} \, \frac{1}{[k^2+i\sqrt{C_A}\gamma^2]} ~=~ -~ 
\frac{\sqrt{i\sqrt{C_A}\gamma^2}}{4\pi} \left[ 1 ~+~ \left[ 2 + 2 \ln \left[ 
\frac{1}{2\sqrt{i\sqrt{C_A}\gamma^2}} \right] \right] \epsilon ~+~
O(\epsilon^2) \right] 
\end{equation}
and its complex conjugate. Then in constructing our final overall gap equation
we merely use the unambiguous and elementary identifications
\begin{equation} 
\sqrt{i\sqrt{C_A}\gamma^2} \sqrt{i\sqrt{C_A}\gamma^2} ~=~ 
i\sqrt{C_A}\gamma^2 ~~~,~~~ 
\sqrt{i\sqrt{C_A}\gamma^2} \sqrt{-i\sqrt{C_A}\gamma^2} ~=~ 
\sqrt{C_A}\gamma^2 ~. 
\end{equation}
If these objects appear instead in a ratio then one first rationalizes before
using these trivial identities. For completeness we note that the basic two 
loop sunset topology in three dimensions with three distinct masses $m_1$, 
$m_2$ and $m_3$ is, \cite{39},
\begin{equation}
\int_{kl} \frac{1}{[k^2+m_1^2][l^2+m_2^2][(k-l)^2+m_3^2]} ~=~ 
\frac{1}{16\pi^2} \left[ \frac{1}{4\epsilon} + \frac{1}{2} + \ln \left[ 
\frac{1}{[m_1+m_2+m_3]} \right] + O(\epsilon) \right]  
\label{int2}
\end{equation}
which is the only non-trivial topology at two loops. The other main topology is
the product of two one loop vacuum bubbles.

The one loop integral, (\ref{int1}), illustrates one feature of the three
dimensional Gribov-Zwanziger Lagrangian which differs from the four dimensional
case and that is that to two loops the theory is ultraviolet finite. Though
master integrals, such as (\ref{int2}), can be divergent. Ordinarily in three 
dimensional Yang-Mills theory with a massless gluon one has to be aware that 
the theory is potentially infrared pathological. However, in (\ref{laggz}) the 
presence of the Gribov mass whilst not corresponding to a non-zero mass does 
act as an infrared regulator. This is relevant at two loops for the computation
of the gap equation since at one loop only (\ref{int1}) is required. At two 
loops within our routines we have been careful in noting the potentially 
infrared divergent master graphs and checked that they actually {\em cancel} 
among themselves in the overall sum of all the contributing integrals for a 
diagram. These arise, for instance, when there are propagators of the form 
$1/(k^2)^2$ which occur in the $\xi^{ab}_\mu$ propagator due to the transverse 
projector. Whilst it is clear that overall such a factor has a tensor structure
rendering the potential double pole as unproblematic, within the algebraic 
rearrangements to produce the master integrals one may have similar problematic
integrals which are only cancelled, for example, by using integration by parts.
Although the finiteness of (\ref{laggz}) may appear beneficial from a 
computational point of view it is actually a disadvantage. In a renormalizable 
ultraviolet divergent theory, such as the four dimensional version of 
(\ref{laggz}), the renormalization constants satisfy Slavnov-Taylor identities,
\cite{4,15,16}. These then serve as important checks on setting up the Feynman
diagrams and the actual explicit master integral evaluations. In the three
dimensional case we do not have these additional important internal checks.
However, the main parts of the {\sc Form} code are the same as the four 
dimensional work and we merely use these again as they have been checked. This
leaves us with minimizing the potential source of errors to be that of 
substituting for the master integrals correctly. Of course, for the gap 
equation we must have a real expression ultimately since $\gamma$ is a real
parameter. It is not an independent quantity since it is defined by the horizon
condition and will be a function of the coupling constant which is real. In 
three dimensions, of course, the coupling constant is dimensionful and hence 
the gap equation will relate quantities to ensure that overall there is only 
one independent dimensionful parameter in (\ref{laggz}).  
 
Given these considerations the two loop gap equation for $\gamma$ in 
(\ref{laggz}) is then  
\begin{eqnarray}
\frac{3}{4} &=& \frac{\sqrt{2} C_A^{3/4} g^2}{16\pi\gamma} ~+~ \left[ \left[ 
\frac{917\pi}{262144} + \frac{17}{98304} + \frac{545}{131072} \tan^{-1} 
\left[ \frac{3}{4} \right] \right] C_A ~-~ \frac{\pi}{256} T_F \Nf \right] 
\frac{C_A^{1/2} g^4}{\pi^2\gamma^2} \nonumber \\
&& +~ O(g^6)
\label{gap2}
\end{eqnarray}
for $\Nf$ massless quarks where $C_A$ is the usual adjoint Casimir. The
arctangent derives from the four complex masses defining the Gribov propagators
and the form of the finite part of (\ref{int2}) when, for instance, there are 
two propagators giving a mass $\sqrt{i\sqrt{C_A}\gamma^2}$ and one with
$\sqrt{-i\sqrt{C_A}\gamma^2}$. This produces a term $\ln (4+3i)$ and its
conjugate for the conjugate integral and within the overall computation it is
the imaginary part which is translated into the final gap equation. As a note
on our conventions each appearance of $\gamma$ is always with one factor of
$C_A^{1/4}$ so that the peculiar appearance of these factors and powers in
(\ref{gap2}) is actually consistent with the presence of $C_A$ which is what 
ordinarily appears from the group theory in the one loop term. As the three 
dimensional theory is ultraviolet finite then both the coupling constant and 
$\gamma$ do not run. Moreover, there are no logarithms involving $\gamma$ and 
the renormalization scale $\mu$ as there is in the four dimensional gap 
equation. In \cite{22} the four dimensional gap equation was solved in order to
write $\gamma$ as an explicit function of the coupling constant producing an 
explicitly non-perturbative function. For (\ref{gap2}) we can also relate these 
parameters. The way we have chosen to do this is to simply solve (\ref{gap2})
as a quadratic equation. As noted in \cite{25} there does not appear to be a
unique way of solving the gap equation. Choosing to solve as a quadratic here
is straightforward but if the explicit three loop or higher gap equation was 
known then it is not clear whether a numerical solution could be extracted in
those cases. Despite these caveats we have solved (\ref{gap2}) numerically for
both $SU(2)$ and $SU(3)$ for a variety of values of $\Nf$ and recorded the
results in Tables $1$ and $2$. More specifically we have introduced the
dimensionless variable $\lambda_n$ defined by
\begin{equation}
\lambda_n ~=~ \frac{g^2}{4\pi C_A^{1/4} \gamma} 
\end{equation}
where $n$ reflects the loop order. Each table provides the relation of $\gamma$
to the coupling constant and vice versa. Clearly from the tables using this 
method of solution the convergence does not appear to be very good. Although 
$SU(3)$ is better than for $SU(2)$. It would be interesting to see if the three
loop corrections improved the convergence but one glance at the explicit 
expression for the master three loop integral for the Benz vacuum bubble 
topology of \cite{39} would indicate how tediously complicated such a 
calculation would be.
\begin{table}[ht]
\begin{center}
\begin{tabular}{|c||c|c|c|c|}
\hline
$\Nf$ & $\lambda_1$ & $\lambda_2$ & $\lambda_1^{-1}$ & $\lambda_2^{-1}$ \\
\hline
$0$ & $1.60660$ & $0.60389$ & $0.62243$ & $1.65593$ \\
$2$ & $1.60660$ & $0.70962$ & $0.62243$ & $1.40920$ \\
$3$ & $1.60660$ & $0.79534$ & $0.62243$ & $1.25732$ \\
$4$ & $1.60660$ & $0.93636$ & $0.62243$ & $1.06797$ \\
\hline
\end{tabular}
\end{center}
\begin{center}
{Table 1. Numerical values for relation between $g$ and $\gamma$ for $SU(2)$.}
\end{center}
\end{table}
\begin{table}[ht]
\begin{center}
\begin{tabular}{|c||c|c|c|c|}
\hline
$\Nf$ & $\lambda_1$ & $\lambda_2$ & $\lambda_1^{-1}$ & $\lambda_2^{-1}$ \\
\hline
$0$ & $0.70711$ & $0.40260$ & $1.41421$ & $2.48385$ \\
$2$ & $0.70711$ & $0.44502$ & $1.41421$ & $2.24709$ \\
$3$ & $0.70711$ & $0.47308$ & $1.41421$ & $2.11381$ \\
$4$ & $0.70711$ & $0.50851$ & $1.41421$ & $1.96653$ \\
\hline
\end{tabular}
\end{center}
\begin{center}
{Table 2. Numerical values for relation between $g$ and $\gamma$ for $SU(3)$.}
\end{center}
\end{table}

Given we have obtained a relation between the mass parameter, $\gamma$, with
the coupling constant from a two loop gap equation in three dimensions, it is 
worth noting related work. One interest in three dimensional Yang-Mills theory
resides in the fact that it is relevant to the four dimensional finite 
temperature theory. In this situation it is believed that a non-zero magnetic
mass is generated dynamically non-perturbatively. Such a magnetic mass can be
accessed from gap equations. For instance, there are a variety of one loop
results available, \cite{40,41,42,43}, where \cite{41,42} used a non-local mass
operator but which was not of the Gribov form. These ideas were extended to two
loops in \cite{44}. There an estimate of the magnetic mass, $m_m$, was quoted 
as $m_m$~$\approx$~$0.34 g^2$ for $SU(2)$. In our case for the case with no 
quarks $C_A^{1/4} \gamma$~$=$~$0.132 g^2$ where the group factor is unevaluated
and included with $\gamma$ because of our conventions. The discrepancy in 
values here ought not to be taken seriously though as in the former case the 
method of attack is to have a canonical massive gluon propagator. By contrast 
we are considering a Gribov style of propagator where the mass is zero but the 
width is not. 

Next we  use the one loop gap equation to determine the leading order value of 
an effective coupling constant which has been shown to freeze to a finite 
value, \cite{24}. From the gauge potential and Faddeev-Popov ghost propagator 
form factors one can define a renormalization group invariant object which 
behaves as the coupling constant at high energy. This is primarily due to the 
fact that the gluon ghost vertex does not undergo any renormalization due to a 
Slavnov-Taylor identity, \cite{45}. The source of the zero momentum value being
non-zero derives from the momentum dependence of the form factors when the gap 
equation for $\gamma$ is set. More specifically, if we define the gluon 
propagator as 
\begin{equation}
\langle A^a_\mu(p) A^b_\nu(-p) \rangle ~=~ -~ \delta^{ab} \frac{D_A(p^2)}{p^2}
P_{\mu\nu}(p)
\end{equation}
where $D_A(p^2)$ is the form factor and 
\begin{equation}
\langle c^a(p) \bar{c}^b(-p) \rangle ~=~ \delta^{ab} \frac{D_c(p^2)}{p^2}
\end{equation}
for the Faddeev-Popov ghost then the renormalization group invariant effective 
coupling constant is defined by
\begin{equation}
\alpha^{\mbox{\footnotesize{eff}}}_s (p^2) ~=~ \alpha_s(\mu) D_A(p^2) \left(
D_c(p^2) \right)^2 ~. 
\end{equation}
We have computed the one loop corrections to both $D_A(p^2)$ and $D_c(p^2)$ for
(\ref{laggz}) in three dimensions and recorded the explicit functions for each 
in Appendices A and B. That for the Faddeev-Popov ghost is equivalent to
$Q_\xi$ due to the similarities between the real Bose ghost and 
$\omega^{ab}_\mu$ fields. This follows from the consequences of the underlying
Slavnov-Taylor identities for (\ref{laggz}) which have been discussed primarily
in the context of the four dimensional theory but which also are valid in the 
three dimensional case, \cite{4,15,16}. Therefore since the gluon propagator
vanishes as $O(p^2)$ as $p^2$~$\rightarrow$~$0$, meaning $D_A(p^2)$ is
$O((p^2)^2)$, and $D_c(p^2)$ is also $O(p^2)$ in the same limit one is left 
with a finite answer for the effective coupling constant at zero momentum. 
Specifically we have   
\begin{equation}
\alpha^{\mbox{\footnotesize{eff}}}_s (0) ~=~ \frac{3\sqrt{2}}{4} C_A^{1/4} 
\gamma ~.
\end{equation}
We recall that since we are in three dimensions the coupling constant carries a
dimension which is why $\gamma$ appears on the right hand side. However,
$\gamma$ is not independent and satisfies the gap equation being reexpressed
as a function of the dimensionful coupling constant. Although this is a leading
one loop calculation and therefore qualitative, the three dimensional set-up
may be more useful in exploring the Gribov-Zwanziger scenario further. For
instance, if one could obtain a non-zero estimate for a frozen effective 
coupling constant, then this would essentially fix the parameters of the theory
provided the gap equation was known sufficiently accurately. The latter would 
be essential if, for instance, one wanted to extract a reliable magnetic mass 
estimate.  

We close this section with an indication of an alternative way of computing the
gap equation. In (\ref{laggrib}) the definition of the horizon condition, 
(\ref{hordef}), involves the non-local operator which cannot be determined 
without localization. Reformulating (\ref{laggrib}) in terms of localized 
fields produces a local version of the horizon definition, (\ref{gapdef}), by 
virtue of (\ref{xieom}). Given this we can reformulate (\ref{gapdef}) again by 
eliminating $A^a_\mu$ within the vacuum expectation value to produce an 
expectation involving only $\xi^{ab}_\mu$ fields at leading order. In other 
words 
\begin{equation}
f^{abp} f^{cdp} \left\langle \xi^{ab\,\mu} (x) \left( \partial^\nu D_\nu 
\xi_\mu \right)^{cd} (x) \right\rangle ~=~ -~ \frac{d C_A \NA \gamma^4}{g^2}
\label{gapdefalt}
\end{equation}
should also be equivalent to the horizon definition and produce the same gap
equation as (\ref{gap2}). We emphasise that this gap equation is not the vacuum
expectation value of the $\xi^{ab}_\mu$ kinetic term due to the presence of the
structure constants. So there is no parallel definition for $\rho^{ab}_\mu$. 
Given this reasoning we have evaluated the one one loop and eighteen two loop 
vacuum bubble graphs contributing to (\ref{gapdefalt}). This uses the same 
basic one and two loop master integrals as that for (\ref{gapdef}) and it is 
satisfying to record that (\ref{gapdefalt}) does indeed reproduce (\ref{gap2}).
Moreover, we have also checked that the two loop $\MSbar$ gap equation of 
\cite{22} in the four dimensional theory is also recovered with the definition 
(\ref{gapdefalt}). This is more involved than the three dimensional case as one
has to correctly take account of the ultraviolet divergences. So we can
summarize the different definitions of the gap equation in the unifying
equivalences 
\begin{eqnarray}
\left\langle A^a_\mu(x) \frac{1}{\partial^\nu D_\nu} A^{a\,\mu}(x)
\right\rangle &=& -~ \frac{i}{C_A \gamma^2}  f^{abc} \left\langle 
A^{a\,\mu} (x) \xi^{bc}_\mu(x) \right\rangle \nonumber \\
&=& -~ \frac{f^{abp} f^{cdp}}{C_A^2 \gamma^4} \left\langle \xi^{ab\,\mu} (x) 
\left( \partial^\nu D_\nu \xi_\mu \right)^{cd} (x) \right\rangle ~=~
\frac{d N_A}{C_A g^2} ~.
\label{gapunif}
\end{eqnarray}
In the last definition, which involves two terms when the covariant derivative
is written explicitly, we have used (\ref{xieom}) to redefine the gluon field
within the vacuum expectation value. However, there is in principle no reason 
why one cannot repeat the substitution of (\ref{xieom}) in the covariant 
derivative of (\ref{gapdefalt}). This would produce a vacuum expectation value 
involving three terms and equate to a perturbative expansion where the final 
term will involve a gluon via the appearance of a new covariant derivative. 
Iterating this procedure one can replace the final horizon definition by an 
{\em infinite} series of terms involving only $\xi^{ab}_\mu$ fields in a 
perturbative expansion. For instance, the first few terms would be
\begin{eqnarray}
\frac{d C_A \NA \gamma^4}{g^2} &=& f_4^{abcd} \left\langle \partial^\nu 
\xi^{ab\,\mu} \left[ \partial_\nu \xi^{cd}_\mu ~-~ \frac{i g}{C_A \gamma^2} 
f_4^{cfrs} \left( \partial^\sigma \partial_\sigma \xi^{rs}_\nu \right) 
\xi^{fd}_\mu \right. \right. \nonumber \\
&& \left. \left. ~~~~~~~~~~~~~~~~~~~-~ \frac{g^2}{C_A^2 \gamma^4} f_4^{cfrs} 
f_4^{rqmn} \partial^\sigma \left[ \left( \partial^\rho \partial_\rho 
\xi^{mn}_\sigma \right) \xi^{qs}_\nu \right] \xi^{fd}_\mu \right. \right.
\nonumber \\
&& \left. \left. ~~~~~~~~~~~~~~~~~~~+~ O(g^3) \right] \right\rangle
\label{gapdefinf}
\end{eqnarray} 
where $f_4^{abcd}$~$\equiv$~$f^{abp} f^{cdp}$ to simplify notation and we have 
integrated by parts on the ordinary derivative. With this version of the
horizon definition we have evaluated the one one loop and eighteen two loop
graphs contributing to (\ref{gapdefinf}) in three {\em and} four dimensions
and found that the respective two loop expressions for the gap equation are
reproduced exactly. Within the context of the vacuum expectation value 
definition of the horizon condition one might regard the perturbative expansion
of (\ref{gapdefinf}) as an infinite series representation of the original 
non-local definition of Gribov, (\ref{hordef}). In other words in 
(\ref{gapunif}) the first and last vacuum expectation values are a field
theoretic type of geometric series with $\xi^{ab}_\mu$ regarded as a 
pseudo-dual field to the gluon.

Whilst these equivalences between the different formulations of the horizon 
equation are novel and suggest a type of duality between the $A^a_\mu$ and 
$\xi^{ab}_\mu$ fields due to (\ref{xieom}), one must be careful when it should 
be set. For instance, it is tempting to replace all appearances of $A^a_\mu$ in
the localized Lagrangian, (\ref{laggz}), in the hope of producing some sort of 
effective low energy field theory where the dominant field is $\xi^{ab}_\mu$. 
However, if one naively does this then the propagators of the theory have no 
relation to that of $\xi^{ab}_\mu$ in (\ref{propdef}). For instance, 
eliminating the gluon completely in this naive approach would produce the 
non-standard propagator, for non-zero $\alpha$,
\begin{eqnarray}
\langle \xi^{ab}_\mu(p) \xi^{cd}_\nu(-p) \rangle_{\mbox{\footnotesize{eff}}} 
&=& -~ \frac{\delta^{ac}\delta^{bd}}{p^2}\eta_{\mu\nu} ~+~
\frac{f^{abe}f^{cde}}{C_A p^2} \eta_{\mu\nu} \nonumber \\
&& +~ \frac{f^{abe} f^{cde}\gamma^4}{p^2[(p^2)^2+C_A\gamma^4]} 
P_{\mu\nu}(p) ~+~ 
\frac{f^{abe} f^{cde} \alpha\gamma^4}{p^2[(p^2)^2+C_A\alpha\gamma^4]} 
L_{\mu\nu}(p)
\end{eqnarray}
where we have introduced the subscript to avoid confusion with the set
(\ref{propdef}), with the propagator for $\rho^{ab}_\mu$ being unchanged. More 
specifically, in the Landau gauge the full set of propagators would be 
\begin{eqnarray}
\langle \xi^{ab}_\mu(p) \xi^{cd}_\nu(-p) \rangle_{\mbox{\footnotesize{eff}}}
&=& -~ \frac{\delta^{ac}\delta^{bd}}{p^2}\eta_{\mu\nu} ~+~
\frac{f^{abe}f^{cde}}{C_A p^2} \eta_{\mu\nu} ~+~ 
\frac{f^{abe} f^{cde}\gamma^4}{p^2[(p^2)^2+C_A\gamma^4]} 
P_{\mu\nu}(p) \nonumber \\ 
\langle \xi^{ab}_\mu(p) \rho^{cd}_\nu(-p) \rangle_{\mbox{\footnotesize{eff}}}
&=& 0 \nonumber \\ 
\langle \rho^{ab}_\mu(p) \rho^{cd}_\nu(-p) \rangle_{\mbox{\footnotesize{eff}}}
&=& \langle \omega^{ab}_\mu(p) \bar{\omega}^{cd}_\nu(-p) 
\rangle_{\mbox{\footnotesize{eff}}} ~=~ -~ 
\frac{\delta^{ac}\delta^{bd}}{p^2} \eta_{\mu\nu} ~. 
\end{eqnarray} 
Taking the colour adjoint projection of $\xi^{ab}_\mu$ gives
\begin{equation}
\langle f^{apq} \xi^{pq}_\mu(p) f^{brs} \xi^{rs}_\nu(-p) 
\rangle_{\mbox{\footnotesize{eff}}} ~=~ 
\frac{\delta^{ab} C_A^2 \gamma^4}{p^2[(p^2)^2+C_A\gamma^4]} P_{\mu\nu}(p) 
\label{xiproj}
\end{equation}
from which one could, in principle, recover the original Gribov propagator when
the leading order term of (\ref{xieom}) is applied to this. So one would have a
{\em hidden} gluon with the perturbative propagator emerging as usual as 
$\gamma$~$\rightarrow$~$0$. In exploring this naive elimination idea further in
general terms, and ignoring contributions from the path integral measure, it 
ought to be the case that in the infrared there is enhancement of various 
colour channels of the $\xi^{ab}_\mu$ propagator as has been observed recently,
\cite{24,25}. If this were the case then (\ref{xiproj}) might produce a hidden 
gluon propagator which freezes to a non-zero value as has been observed on the 
lattice by various authors, \cite{26,27,28,29,30,31,32,33,34,35}. However, in 
speculating about the potential of a notional effective Lagrangian involving 
only the fields of (\ref{laggz}) which are enhanced in the infrared we must be 
clear in stating that such a theory has not been constructed. Our naive 
elimination by equations of motion, despite the fact that here it is related to
the horizon constraint definition, is not the correct or accepted normal 
procedure to produce an effective quantum field theory. We offer it as a 
possible line of future investigation. In other words it might seem a natural 
way to proceed to study infrared properties by focusing on the actual fields 
which dominate at low energy. In toying with this idea it is certainly 
elementary to see that the Faddeev-Popov ghost, $\rho^{ab}_\mu$ and 
$\omega^{ab}_\mu$ still remain enhanced at one loop. This is primarily because 
their associated vertices are effectively unchanged at this order. The 
difficulty comes in the $\xi^{ab}_\mu$ sector where, although there is now an 
infinite number of interactions, which is not really a {\em calculational}
obstruction, it is not clear whether renormalizability is retained. Therefore, 
one cannot even begin to consider if various colour channels of the 
$\xi^{ab}_\mu$ propagator enhance. However, in models of the low momentum the 
latter property is not crucial since, for example, the L\"{u}scher term is 
derived from a non-renormalizable construction, \cite{46,47}. Though an 
effective low energy theory with an infinite number of couplings of the colour 
valued $\xi^{ab}_\mu$ fields to the quarks could be construed as a mimic of a 
flux tube model. Finally, irrespective of whether this naive use of an equation
of motion is correct or not in trying to construct a theory involving only the 
fields which enhance in the infrared, it seems clear that one has to retain the 
horizon condition in some form and hence the gap equation for $\gamma$. The 
equivalences of (\ref{gapunif}) would appear to be a useful observation in this
respect as we now have a vacuum expectation value definition of the Gribov 
horizon in terms of {\em local} function of only $\xi^{ab}_\mu$ albeit an 
infinite perturbative series one. Though there may be a non-perturbative 
definition and thence an alternative way of pursuing an effective theory of 
enhanced fields. 

\sect{Static potential.} 

In this section we concentrate on computing the static potential of heavy
quarks in the three dimensional Gribov-Zwanziger Lagrangian. First, the
calculational formalism was developed primarily for four dimensional QCD with 
massless quarks in the case of the canonical gluon propagator, \cite{48,49,50}.
It is based on the Wilson loop and a series of Feynman rules were constructed 
in coordinate space for the extreme case of a temporally long but spatially 
thin loop. With the one loop static potential emerging in \cite{48,49,50}, the 
two loop $\MSbar$ expression was produced later. This was derived first in the 
Feynman gauge, \cite{51,52}, and then repeated in an arbitrary linear covariant
gauge in \cite{53,54}, which verified the gauge independence of the potential. 
The latter computation is comprehensively detailed in \cite{53} to which we 
refer the interested reader for background to technical aspects which we assume
here. More recently, the three loop $\MSbar$ potential has been constructed in 
\cite{55,56,57,58,59,60} which represents the current state of the art. By 
contrast only the one loop potential has been determined in three dimensional 
QCD in \cite{53,61} for canonical gluons. More specifically the momentum space 
potential, $\tilde{V}(\mathbf{p})$, is, \cite{53,61}, 
\begin{equation}
\left. \frac{}{} \tilde{V}( \mathbf{p} )
\right|_{\mbox{\footnotesize{QCD}}} ~=~ -~ \frac{C_F g^2}{\mathbf{p}^2} 
\left[ 1 ~+~ \left[ \left[ \frac{7}{32} C_A - \frac{1}{8} T_F \Nf \right] 
\frac{\sqrt{\mathbf{p}^2}}{\mathbf{p}^2} \right] g^2 \right] ~+~ O(g^6) ~.
\label{pot30}
\end{equation}
Interestingly if one performs the inverse Fourier transform 
\begin{equation}
V(r) ~=~ \int \frac{d^2 \mathbf{k}}{(2\pi)^2} \, e^{i\mathbf{k}.\mathbf{r}} \,
\tilde{V}(\mathbf{k})
\label{invtra}
\end{equation}
then the first term of (\ref{pot30}) reproduces the usual two dimensional
Coulomb potential but the one loop correction rises linearly. However, as noted
in \cite{53,61} it is not clear what occurs for the two loop correction. Given
the dimensionality of the coupling constant it is not inconceivable that higher
loop corrections could lead to higher powers of the spatial separation $r$. It 
is worth noting that in four dimensions a linearly rising potential would 
correspond to a dipole term in the momentum space potential. By contrast, in 
three dimensions one requires a behaviour of $O(1/(\mathbf{p}^2)^{3/2})$ for a 
linearly rising potential, \cite{53}. Given that we are interested in examining
a theory, (\ref{laggz}), which is believed to be confining since it is
consistent with the Kugo-Ojima confinement criterion, \cite{12,13}, our aim 
here is to compute the $\gamma$ dependent extension of (\ref{pot30}) and 
examine its behaviour in the infrared limit. If a term of the form 
$O(1/(\mathbf{p}^2)^{3/2})$ emerged then that would correspond to the confining
potential being preserved in the presence of the Gribov horizon. Though for
comparison we note that in \cite{25} the analogous term, which would be a
dipole, did not occur in four dimensions. One feature which emerged for the 
leading order term there was that the presence of the width in the gluon 
propagator meant that the coordinate space potential crossed the axis. Although
this is present in the accepted form of the potential as computed say using 
lattice regularization, the potential actually crossed the $r$~$=$~$0$ axis at 
an infinity of places corresponding to a Friedel type of potential with a set 
of quasi-stable vacua. The key point was that the width was essential for this.
A model where the gluon solely has a mass, if one ignores briefly the 
contradiction with the non-abelian gauge principle, would lead to a Yukawa 
potential in coordinate space which is never positive in $r$~$>$~$0$.  

As we are extending the derivation of the static potential for the three
dimensional version of (\ref{laggz}) we briefly recall several of the key parts
of the formalism which were discussed in more detail in \cite{25}. First, the
potential is defined in terms of the Wilson loop which has a large time
separation in comparison with the radial distance $r$, \cite{48,49,50},
\begin{equation}
V(r) ~=~ -~ \lim_{T\rightarrow \infty} \frac{1}{iT} \ln \left\langle 0 \left|
\, \mbox{Tr} \, {\cal P} \, \exp \left( ig \oint dx^\mu \, A^a_\mu T^a \right) 
\right| 0 \right\rangle ~.
\label{potdef}
\end{equation}
As is known this is equivalent to the definition involving the path integral,
\cite{48,49,50},
\begin{equation}
V(r) ~=~ -~ \lim_{T\rightarrow\infty} \frac{1}{iT} 
\frac{\mbox{tr}Z[J]}{\mbox{tr}Z[0]}
\label{potdefz}
\end{equation}
where
\begin{equation}
Z[J] ~=~ \int {\cal D} A_\mu \, {\cal D} \psi \, {\cal D} \bar{\psi} \, 
{\cal D} c \, {\cal D} \bar{c} ~ \exp \left[ -\, \int d^3 x \, \left( 
L^{\mbox{\footnotesize{QCD}}} ~+~ J^{a\,\mu} A^a_\mu \right) \right]
\end{equation}
and the source term corresponds to placing heavy quarks according to
\begin{equation}
J^a_\mu(x) ~=~ g v_\mu T^a \left[ \delta^{(3)} \left( \mathbf{x} 
- \half \mathbf{r} \right) ~-~ \delta^{(3)} \left( \mathbf{x} - \half
\mathbf{r}^\prime \right) \right] ~.
\end{equation}
Here $v_\mu$~$=$~$\eta_{\mu 0}$ is a unit vector which projects out the time 
component of the gluon it couples to. We also define 
$r$~$=$~$|\mathbf{r}-\mathbf{r}^\prime|$. The presence of the sources 
introduces additional Feynman rules which are not dependent on the spacetime
dimension and are given in, for example, \cite{49,53}. Though we follow the 
more modern approach and perform our static potential computations in momentum 
space rather than directly in coordinate space. The former is connected via the 
inverse Fourier transform, (\ref{invtra}). The extension of the formalism to
the Gribov-Zwanziger case is to replace the Lagrangian (\ref{lagqcd}) in the 
path integral in (\ref{potdefz}) by the Lagrangian (\ref{laggz}) whence the 
measure is extended to include the localizing fields, 
\begin{equation}
Z[J] ~=~ \int {\cal D} A_\mu \, {\cal D} \psi \, {\cal D} \bar{\psi} \,
{\cal D} c \, {\cal D} \bar{c} \, {\cal D} \xi \, {\cal D} \rho \,
{\cal D} \omega \, {\cal D} \bar{\omega} ~ \exp \left[ -\, \int d^4 x \,
\left( L^{\mbox{\footnotesize{GZ}}} ~+~ J^{a\,\mu} A^a_\mu \right) \right] ~.
\end{equation}
As the localizing fields are completely internal they do not couple to the 
heavy quark sources. Hence at leading order the only field which is exchanged
is the gauge potential which means that its propagator essentially determines
the static potential at this order. The localizing Bose ghost plays no role
until loop corrections are included. So, for instance,  
\begin{equation}
\tilde{V}( \mathbf{p} )
~=~ -~ \frac{C_F \mathbf{p}^2 g^2}{[(\mathbf{p}^2)^2+C_A \gamma^4]} ~+~ O(g^4) 
\label{pottree}
\end{equation}
and performing the inverse Fourier transform gives 
\begin{eqnarray}
V(r) &=& \frac{C_F g^2}{16} \left[ 
Y_0 \left( \frac{(1+i)}{\sqrt{2}} C_A^{1/4} \gamma r \right) ~+~
Y_0 \left( \frac{(1-i)}{\sqrt{2}} C_A^{1/4} \gamma r \right) \right. 
\nonumber \\
&& \left. ~~~~~~~~~+~
Y_0 \left( \frac{-(1-i)}{\sqrt{2}} C_A^{1/4} \gamma r \right) ~+~
Y_0 \left( \frac{-(1+i)}{\sqrt{2}} C_A^{1/4} \gamma r \right) \right] ~+~ 
O(g^4)
\label{pot3}
\end{eqnarray}
where $Y_0(z)$ is the Bessel function of the second type or Neumann function 
which is an entire function. Interestingly the four roots of the algebraic 
equation
\begin{equation}
z^4 ~=~ -~ C_A \gamma^4
\end{equation}
emerge as the arguments of the functions. Taking the $\gamma$~$\rightarrow$~$0$
limit recovers the Coulomb behaviour noted in \cite{53,61} 
\begin{equation}
\lim_{\gamma \rightarrow 0} V(r) ~=~ \frac{C_F g^2}{2\pi} \ln \left( r 
\right) ~+~ O(g^4) ~.
\end{equation}
However, if one plots the functions of (\ref{pot3}) for non-zero $\gamma$ the 
coordinate space potential has a similar feature to the Friedel form of four 
dimensions. Although the Coulomb potential crosses the axis once when $\gamma$ 
is non-zero the static potential has an infinite number of crossing points
which would again lead to quasi-stable vacua. 

\begin{figure}[ht]
\hspace{2.6cm}
\epsfig{file=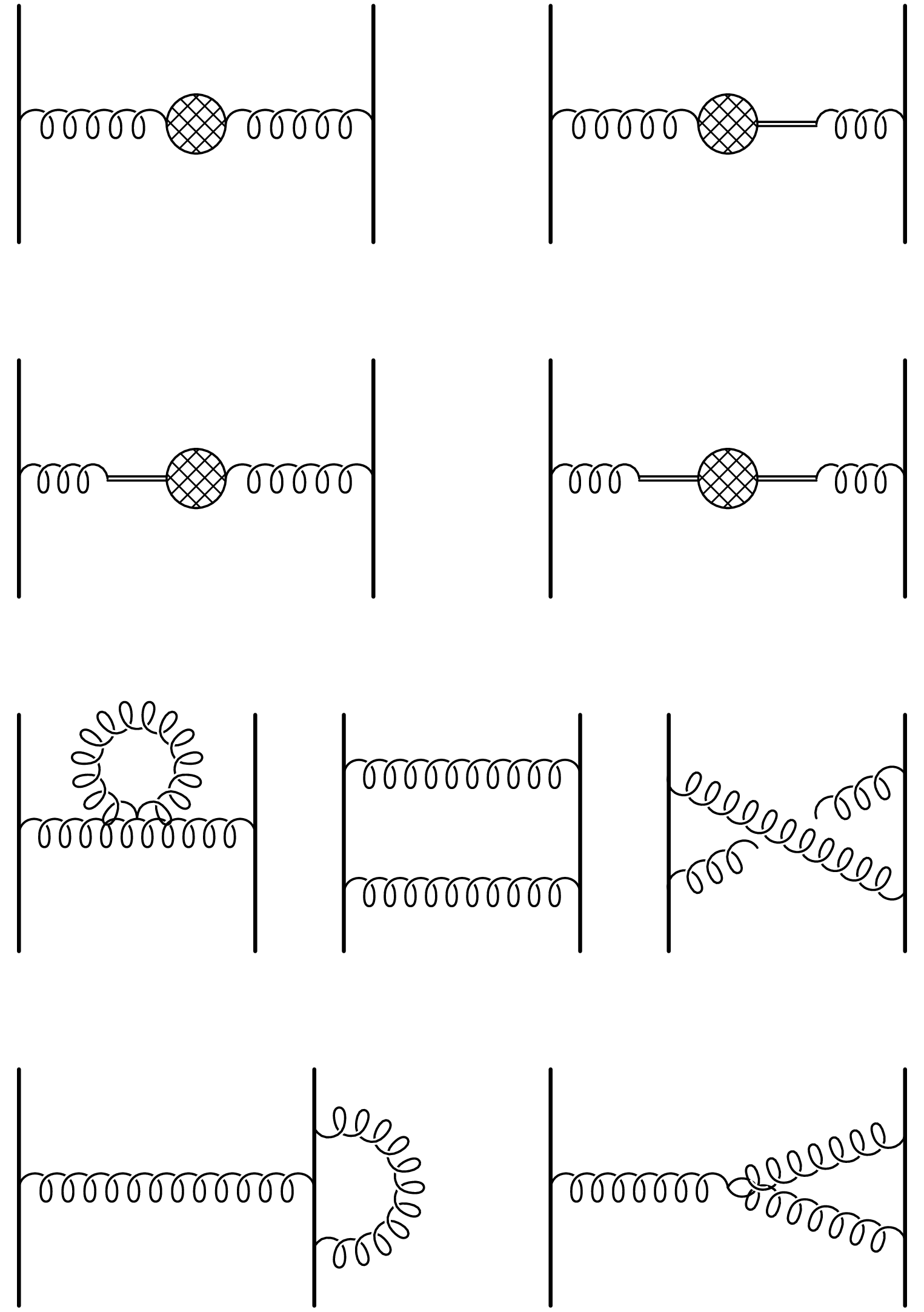,height=15cm}
\vspace{0.5cm}
\caption{One loop topologies contributing to the static potential.}
\end{figure}

We now turn to the one loop corrections of (\ref{pottree}). A representative
set of topologies for this calculation is given in Figure $1$ where we have 
displayed the full set of corrections to the $2$-point functions. In those
first four graphs the blob represents all the one loop corrections. However,
due to the mixed propagator the correction to the $\xi^{ab}_\mu$ $2$-point
function occurs even though there is no direct source $\xi^{ab}_\mu$ coupling.
The two box diagrams are important for the exponentiation implied in the
definition of the static potential, (\ref{potdefz}), and the asscociated
issues with have been discussed at length in \cite{48,49,50,62,63}. Therefore, 
it remains merely to compute the diagrams explicitly. Essential to this is the
automatic Feynman diagram package {\sc Qgraf}, \cite{37}, where the graphs are
generated electronically and then converted into {\sc Form} input notation. The
algorithm we use is to break up all the Feynman graphs into simple master
integrals and then identify the explicit functions for three dimensional
spacetime. A comprehensive analysis of such masters to three loops are given
in \cite{39}. However, for our situation there are only two main integrals but 
there is the complication of having to work with a non-standard propagator
which induces the canonical propagator to have a width after application of
simple partial fractions, such as
\begin{equation}
\frac{p^2}{[(p^2)^2+C_A\gamma^4]} ~=~
\frac{1}{2} \left[ \frac{1}{[p^2+i\sqrt{C_A}\gamma^2]} ~+~
\frac{1}{2} \frac{1}{[p^2-i\sqrt{C_A}\gamma^2]} \right] 
\end{equation} 
where now the momentum can involve the external momentum. In four dimensions 
the explicit expressions for Feynman integrals involved masses, $m$, and 
momenta, $p$, appearing in the form $m^2$ and $p^2$ respectively. However, in
our three dimensional case with the drop in one unit of the dimensionality of 
the integral measure, the dependence is purely in terms of $m$ {\em and}
$\sqrt{p^2}$. As discussed earlier for fields with a canonical mass term this 
is not a significant issue but when there is a width present one has to find 
the square root of the corresponding squared mass of the propagator. We follow
the procedure used previously but allowing for the presence of the external
momentum. Again ultimately one should obtain real and not complex expressions 
which is a check on our reasoning. So, for example, we have used the following 
intermediate expressions
\begin{eqnarray}
\int_k \frac{1}{k^2[(k-p)^2+i\sqrt{C_A}\gamma^2]} &=&
\frac{\sqrt{p^2}}{4\pi p^2} \tan^{-1} \left[ 
\frac{\sqrt{p^2}}{\sqrt{i\sqrt{C_A}\gamma^2}} \right] \nonumber \\ 
\int_k \frac{1}{[k^2+i\sqrt{C_A}\gamma^2][(k-p)^2+i\sqrt{C_A}\gamma^2]} &=&
\frac{\sqrt{p^2}}{4\pi p^2} \tan^{-1} \left[ 
\frac{\sqrt{p^2}}{2\sqrt{i\sqrt{C_A}\gamma^2}} \right] \nonumber \\ 
\int_k \frac{1}{[k^2+i\sqrt{C_A}\gamma^2][(k-p)^2-i\sqrt{C_A}\gamma^2]} &=&
\frac{\sqrt{p^2}}{4\pi p^2} \tan^{-1} \left[ 
\frac{\sqrt{p^2}}{[\sqrt{i\sqrt{C_A}\gamma^2}+\sqrt{-i\sqrt{C_A}\gamma^2}}
\right] 
\label{intex}
\end{eqnarray}
by adapting the results of \cite{39} in the same way as before, where 
$\int_k$~$=$~$d^dk/(2\pi)^d$. In (\ref{intex}) we will use our simplification 
identities which allow us to write the functions of a complex variable in terms
of a real and imaginary part. Although the three dimensional theory is finite, 
we still work in dimensional regularization with $d$~$=$~$3$~$-$~$2\epsilon$ as
some of the master diagrams have poles in $\epsilon$, \cite{39}. However, 
whilst the theory is ultraviolet finite we have been careful to check that 
there no (spurious) infrared infinities arise as a consequence of breaking the 
Feynman graphs up into scalar master integrals. For instance, such divergences 
could arise from the $1/p^2$ part of the propagators in the transverse 
projector or its powers but again we have checked that such potential terms 
cancel among themselves. This finiteness, at least to one loop, ensures that 
there is no source gluon renormalization constant as there is in the four 
dimensional arbitrary gauge calculation.

Given these considerations we are now in a position to record the one loop
correction to (\ref{pottree}) for (\ref{laggz}). We find  
\begin{eqnarray}
\tilde{V}( \mathbf{p} )
&=& -~ \frac{C_F \mathbf{p}^2 g^2}{[(\mathbf{p}^2)^2+C_A \gamma^4]}
\nonumber \\
&& +~ \left[ \left[ \frac{1}{2048} \eta_1(\mathbf{p}^2) 
- \frac{1}{4096} \eta_3(\mathbf{p}^2) \right] 
\frac{\sqrt{\mathbf{p}^2}}{\gamma^4}
- \frac{\sqrt{2} C_A^{5/4} \gamma}{256[(\mathbf{p}^2)^2+C_A \gamma^4]}
- \frac{\sqrt{2} C_A^{5/4} \gamma}{4[(\mathbf{p}^2)^2+16C_A \gamma^4]}
\right.
\nonumber \\
&& \left. ~~~~~+~ \frac{\sqrt{2} C_A^{5/4} \gamma} 
{8[(\mathbf{p}^2)^2-4C_A\gamma^4]} 
- \frac{\pi \sqrt{\mathbf{p}^2} T_F \Nf C_A \gamma^4}
{8[(\mathbf{p}^2)^2+C_A \gamma^4]^2} \right. \nonumber \\
&& \left. ~~~~~+~ \left[ \frac{545}{2048} \eta_1(\mathbf{p}^2) 
+ \frac{515}{4096} \eta_3(\mathbf{p}^2) \right] 
\frac{\sqrt{\mathbf{p}^2}C_A^2\gamma^4}{[(\mathbf{p}^2)^2+C_A\gamma^4]^2} 
+ \frac{49\pi \sqrt{\mathbf{p}^2} C_A^2 \gamma^4}
{1024[(\mathbf{p}^2)^2+C_A \gamma^4]^2}
\right. \nonumber \\
&& \left. ~~~~~-~ \frac{13 \sqrt{2} C_A^{9/4} \gamma^5}
{384[(\mathbf{p}^2)^2+C_A \gamma^4]^2}
+ \frac{\pi \sqrt{\mathbf{p}^2} T_F \Nf}
{8[(\mathbf{p}^2)^2+C_A \gamma^4]} \right. \nonumber \\
&& \left. ~~~~~+~ \left[ \frac{11}{256} \eta_2(\mathbf{p}^2) 
- \frac{313}{1024} \eta_1(\mathbf{p}^2) 
- \frac{121}{2048} \eta_3(\mathbf{p}^2) \right] 
\frac{\sqrt{\mathbf{p}^2}C_A}{[(\mathbf{p}^2)^2+C_A\gamma^4]} \right. 
\nonumber \\
&& \left. ~~~~~-~ \frac{49\pi \sqrt{\mathbf{p}^2} C_A}
{1024[(\mathbf{p}^2)^2+C_A \gamma^4]}
- \frac{\sqrt{\mathbf{p}^2} C_A}{512(\mathbf{p}^2)^2} \eta_2(\mathbf{p}^2)
\right. \nonumber \\
&& \left. ~~~~~-~ \left[ \frac{21}{1024} \eta_4(\mathbf{p}^2) 
+ \frac{3}{128} \eta_5(\mathbf{p}^2) \right] 
\frac{\sqrt{C_A\mathbf{p}^2}}{\mathbf{p}^2\gamma^2} 
+ \frac{\sqrt{2} C_A^{3/4}}{512\gamma\mathbf{p}^2} \right. \nonumber \\
&& \left. ~~~~~+~ \left[ \frac{11}{512} \eta_4(\mathbf{p}^2) 
+ \frac{47}{2048} \eta_5(\mathbf{p}^2) \right] 
\frac{\sqrt{C_A} (\mathbf{p}^2)^{3/2}}
{[(\mathbf{p}^2)^2+C_A\gamma^4]\gamma^2} 
- \frac{131 \sqrt{2} C_A^{3/4} \mathbf{p}^2}
{1024\gamma[(\mathbf{p}^2)^2+C_A \gamma^4]} \right. \nonumber \\
&& \left. ~~~~~+~ \frac{\sqrt{2} C_A^{3/4} \mathbf{p}^2}
{16\gamma[(\mathbf{p}^2)^2+16C_A \gamma^4]} 
+ \frac{\sqrt{2} C_A^{3/4} \mathbf{p}^2}
{16\gamma[(\mathbf{p}^2)^2-4C_A \gamma^4]} 
- \frac{265 C_A^{3/4} \gamma^2 (\mathbf{p}^2)^{3/2}}
{2048[(\mathbf{p}^2)^2+C_A \gamma^4]^2} \eta_5(\mathbf{p}^2)
\right. \nonumber \\
&& \left. ~~~~~+~ \frac{121 \sqrt{2} C_A^{7/4} \gamma^3 \mathbf{p}^2}
{3072[(\mathbf{p}^2)^2+C_A \gamma^4]^2} 
\right] \frac{C_Fg^4}{\pi} ~+~ O(g^6)
\label{potone}
\end{eqnarray}
where we have defined the intermediate functions $\eta_i(p^2)$ by 
\begin{eqnarray}
\eta_1(p^2) &=& \tan^{-1} \left[ \frac{\sqrt{2p^2}}{2 C_A^{1/4} \gamma}
\right] ~~~,~~~ 
\eta_2(p^2) ~=~ \tan^{-1} \left[ \frac{\sqrt{2 \sqrt{C_A} \gamma^2 p^2}}
{[\sqrt{C_A} \gamma^2 - p^2]} \right] \nonumber \\
\eta_3(p^2) &=& \tan^{-1} \left[ \frac{2\sqrt{2 \sqrt{C_A} \gamma^2 p^2}}
{[4\sqrt{C_A} \gamma^2 - p^2]} \right] ~~~,~~~ 
\eta_4(p^2) ~=~ \ln \left[ \frac{ p^2 + \sqrt{C_A} \gamma^2 
- \sqrt{2\sqrt{C_A} \gamma^2 p^2} }{ p^2 + \sqrt{C_A} \gamma^2 
+ \sqrt{2\sqrt{C_A} \gamma^2 p^2} } \right] \nonumber \\
\eta_5(p^2) &=& \ln \left[ \frac{ p^2 + 4 \sqrt{C_A} \gamma^2 
- 2 \sqrt{2\sqrt{C_A} \gamma^2 p^2} }{ p^2 + 4 \sqrt{C_A} \gamma^2
+ 2 \sqrt{2\sqrt{C_A} \gamma^2 p^2} } \right] ~.
\end{eqnarray}
Essentially these arise from taking the real and imaginary parts of expressions
such as those given in (\ref{intex}). 

As a check on (\ref{potone}) if we take the $\gamma$~$\rightarrow$~$0$ limit
we recover the one loop expression of \cite{53,61} given in (\ref{pot30}). We 
stress though that whilst (\ref{pot30}) was computed in an arbitrary linear 
covariant gauge we have been restricted to the Landau gauge as we have 
incorporated the Gribov problem within the Lagrangian. However, (\ref{potone}) 
represents the first non-trivial check on (\ref{pot30}). Given that 
(\ref{laggz}) is supposed to represent a confining theory we can now examine 
(\ref{potone}) to see if the dominant behaviour in the zero momentum limit 
could deliver a behaviour leading to a linearly rising potential. In three 
dimensions this would correspond to an $O(\sqrt{p^2}/(p^2)^2)$ type term and in 
(\ref{potone}) there is one term which appears with such a singularity. 
However, its numerator involves the function $\eta_2(p^2)$ which vanishes at
zero momentum and therefore, the appropriate singular behaviour does not seem
to emerge. More concretely as $p^2$~$\rightarrow$~$0$ we have 
\begin{equation}
\tilde{V} ( \mathbf{p} ) ~=~ -~ \frac{C_F \mathbf{p}^2 g^2}{C_A \gamma^4} ~-~
C_F \left[ \frac{\sqrt{2}C_A^{1/4}}{48\gamma^3} ~+~ 
\frac{113\sqrt{2}\mathbf{p}^2}{1920\pi C_A^{1/4} \gamma^5} \right] g^4 ~+~
O((\mathbf{p}^2)^2;g^6) ~.
\end{equation}
Therefore, similar to the four dimensional case, \cite{25}, the one loop 
correction freezes to a finite value, which is
\begin{equation}
\tilde{V} (0) ~=~ -~ \frac{\sqrt{2} C_F C_A^{1/4}g^4}{48 \pi \gamma^3} ~+~
O(g^6)
\end{equation}
and there is no net divergence whose presence would at least be necessary for a
rising potential. There is a degree of irony with this observation in that the 
$\gamma$~$=$~$0$ potential at one loop has a Fourier transform which produces a
linearly rising potential. Though in that case in is not clear what would 
transpire at two loops given the dimensionality of the coupling constant. 
However, as noted in \cite{24,25} the more appropriate route to proceed down 
would be to analyse the zero momentum behaviour of the propagators of the
localizing fields. Whilst the fermionic ghosts are both enhanced at zero 
momentum in (\ref{laggz}), it has recently been shown that the same is true for
certain colour components of the Bose ghost, \cite{24,25}. This property is 
independent of the spacetime dimension, \cite{25}, but it has not been fully 
determined what the implications are for the static potential. 

\sect{Formal propagator corrections.} 

We turn now to the enhancement of the Bose ghost fields. Recently the structure
of these fields was analysed non-perturbatively by Zwanziger in \cite{24} where
it was demonstrated that there was enhancement in certain colour channels. The 
result is based on the spontaneous breaking of the BRST symmetry but given the 
presence of the horizon condition, which equates to a constraint on the gluon 
and $\xi^{ab}_\mu$ fields, this requires a more careful analysis than usual. 
One key outcome is that the associated Goldstone bosons of this spontaneous 
breaking generate massless excitations non-perturbatively but crucially in the 
context of the Gribov-Zwanziger Lagrangian, these fields are enhanced in the 
infrared. One consequence is that this enhancement is present order by order in
perturbation theory and this was confirmed by explicit one loop calculations in
four dimensions in the $\MSbar$ scheme, \cite{25}. However, in \cite{25} the
main emphasis was on the transverse part of the propagators since the interest 
was in examining the implications for the static potential for heavy coloured 
objects. The longitudinal piece of the exchanged particle does not contribute 
in the particular configuration considered for the Wilson loop. In this section
we provide the formal construction of the full propagator for the gluon and 
$\xi^{ab}_\mu$ fields prior to considering the behaviour in the infrared limit 
when the gap equation is realised which is given specifically in the next 
section. Although we concentrate primarily on three dimensions in this article
we will also include the analysis for four dimensions in this section. This is
because the general reasoning of \cite{24} is dimension independent and we 
confirm this at one loop by considering both dimensions within our analysis 
here.

We begin by recalling how enhancement occurs for the simple situation of the
Faddeev-Popov ghost, $c^a$. As was demonstrated in Gribov's seminal
contribution, \cite{5}, one computes the ghost $2$-point function at one loop 
when the horizon condition is implemented in the path integral. The consequent 
presence of the Gribov mass in the gluon propagator produces an expression 
which differs from what one would obtain using a canonical propagator. 
Expanding the finite function of the momentum in a Taylor series around zero 
momentum it transpires that the leading term is equivalent to the one loop 
correction to the Gribov mass gap equation. As the theory has no meaning as a 
gauge theory unless $\gamma$ satisfies the gap equation, this implies that the 
leading term of the $2$-point function in this limit is not $O(p^2)$ but 
$O\left( (p^2)^2 \right)$. Hence, the low energy behaviour of Faddeev-Popov 
ghost propagator is not the usual perturbative form but the enhanced dipole 
form. Subsequently, this property of ghost enhancement has been translated into
the Kugo-Ojima confinement condition. This was originally established in
\cite{12,13} for the Landau gauge version of Yang-Mills involving only 
Faddeev-Popov ghosts. More recently the criterion has been extended to the 
Gribov-Zwanziger context where there are additional fermionic ghosts, 
$\omega^{ab}_\mu$, which are clearly absent in the original Kugo-Ojima BRST 
analysis, as well as the Bose ghosts, $\rho^{ab}_\mu$ and $\xi^{ab}_\mu$,
\cite{14}. In the context of (\ref{laggz}) it has now been established that 
$\omega^{ab}_\mu$ also enhances. It has been demonstrated in the full analysis 
of \cite{24,36} and in explicit two loop $\MSbar$ computations in four 
dimensions, \cite{23}. We note at this point that we have repeated the latter 
calculations for $\omega^{ab}_\mu$ in the three dimensional version of 
(\ref{laggz}) and have verified that when $\gamma$ satisfies the two loop gap 
equation, (\ref{gap2}), the fermionic ghost enhances.

In focusing on the Faddeev-Popov enhancement derivation the key ingredient is
the zero momentum limit of the $2$-point functions. As we have evaluated the
$2$-point function corrections exactly at one loop for all the fields we can
now consider the situation for the Bose ghost fields. For ease we consider the
$\rho^{ab}_\mu$ field first. In the zero momentum limit we have,
\begin{equation}
\langle \rho^{ab}_\mu(p) \rho^{cd}_\nu(-p) \rangle^{-1} ~=~ -~ \delta^{ac} 
\delta^{bd} \eta_{\mu\nu} \left[ 1 ~-~ 
\frac{\sqrt{2} C_A^{3/4} g^2}{12\pi\gamma} ~+~ 
\frac{\sqrt{2} C_A^{1/4} g^2}{60\pi\gamma^3} p^2 ~+~ O \left( (p^2)^2 \right)
\right] p^2 ~+~ O(g^4) 
\label{rho2pt}
\end{equation}
whence it is elementary to observe that the leading one loop correction is 
precisely that which appears in the gap equation, (\ref{gap2}). Hence, when 
$\gamma$ fulfils that condition the leading term of the $2$-point function is 
the $O\left( (p^2)^2 \right)$ part of the {\em one loop} term. Consequently
when one inverts the coefficient of the $\eta_{\mu\nu}$ tensor in this zero 
momentum limit then the $\rho^{ab}_\mu$ field enhances. One could, of course, 
consider the transverse and longitudinal parts of (\ref{rho2pt}) separately
which is how the $A^a_\mu$ and $\xi^{ab}_\mu$ system was considered in 
\cite{25}. However, the upshot is that the $\rho^{ab}_\mu$ propagator at low
momentum is 
\begin{equation}
\langle \rho^{ab}_\mu(p) \rho^{cd}_\nu(-p) \rangle ~ \sim ~ -~ 
\frac{30 \sqrt{2} \pi \gamma^3}{C_A^{1/4} (p^2)^2 g^2} \delta^{ac} \delta^{bd}
\eta_{\mu\nu} ~.
\label{xirhoenh}
\end{equation}
Although the explicit expressions for the two loop corrections to the $2$-point
functions are not available, it is possible to determine their zero momentum
behaviour using the vacuum bubble expansion. A similar procedure was followed
in the four dimensional case and we note that at two loops the leading momentum
part of the $\rho^{ab}_\mu$ $2$-point function is precisely the two loop gap
equation (\ref{gap2}). Therefore, the enhancement of $\rho^{ab}_\mu$ is 
present at next order in exact agreement with the general BRST arguments of 
\cite{24}.
 
Whilst this demonstrates that a part of the Bose ghost can enhance we need to
complete the analysis by considering the imaginary component which is more
involved due to it being entwined with $A^a_\mu$. In three dimensions the zero
momentum limit of the $2$-point function is 
\begin{eqnarray}
\langle \xi^{ab}_\mu(p) \xi^{cd}_\nu(-p) \rangle^{-1} &=& -~ \left[
\delta^{ac} \delta^{bd} \left[ 1 ~-~ \frac{\sqrt{2} C_A^{3/4} g^2}{12\pi\gamma}
\right] p^2 ~+~ \frac{7\sqrt{2}g^2}{2880\pi C_A^{1/4} \gamma} f^{ace} f^{bde} 
p^2 \right. \nonumber \\
&& \left. ~~~~+~ \frac{\sqrt{2}g^2}{720\pi C_A^{1/4} \gamma} f^{abe} f^{cde} 
p^2 ~+~ \frac{7\sqrt{2}g^2}{480\pi C_A^{5/4} \gamma} d_A^{abcd} p^2 ~+~ O(g^4) 
\right] P_{\mu\nu}(p) \nonumber \\
&& -~ \left[ \delta^{ac} \delta^{bd} \left[ 1 ~-~ 
\frac{\sqrt{2} C_A^{3/4} g^2}{12\pi\gamma} \right] p^2 ~+~ 
\frac{\sqrt{2} g^2}{180\pi C_A^{1/4} \gamma} f^{ace} f^{bde} p^2 \right. 
\nonumber \\
&& \left. ~~~~-~ \frac{\sqrt{2}g^2}{360\pi C_A^{1/4} \gamma} f^{abe} f^{cde} 
p^2 ~+~ \frac{\sqrt{2}g^2}{30\pi C_A^{5/4} \gamma} 
d_A^{abcd} p^2 ~+~ O(g^4) \right] L_{\mu\nu}(p) \nonumber \\
&& +~ O\left((p^2)^2\right) ~.
\label{xi2pt}
\end{eqnarray}
Clearly the piece analogous to (\ref{rho2pt}) would equate to enhancement if
one could perform the zero momentum inversion of (\ref{xi2pt}) in the absence 
of the additional colour channels. However, not only would this not be correct
it would ignore the fact that the construction of the propagators actually
requires $A^a_\mu$ to be included due to the mixed quadratic term of
(\ref{laggz}). In order to do this we consider the problem of the inversion in
a general context. First, we recall the procedure for the transverse sector,
\cite{25}, and define the matrix of colour amplitudes of the $2$-point 
functions of the set of fields $\{ A^a_\mu, \xi^{ab}_\mu, \rho^{ab}_\mu \}$ by
\begin{equation}
\Lambda^{\{ab|cd\}} ~=~ \left(
\begin{array}{ccc}
{\cal X} \delta^{ac} & {\cal U} f^{acd} & 0 \\
{\cal U} f^{cab} & {\cal Q}^{abcd}_\xi & 0 \\
0 & 0 & {\cal Q}^{abcd}_\rho \\
\end{array}
\right) 
\label{tradec}
\end{equation} 
where
\begin{eqnarray}
{\cal Q}^{abcd}_\xi &=& {\cal Q}_\xi \delta^{ac} \delta^{bd} ~+~ {\cal W}_\xi
f^{ace} f^{bde} ~+~ {\cal R}_\xi f^{abe} f^{cde} ~+~ 
{\cal S}_\xi d_A^{abcd} ~+~ {\cal P}_\xi \delta^{ab} \delta^{cd} ~+~ 
{\cal T}_\xi \delta^{ad} \delta^{bc} \nonumber \\ 
{\cal Q}^{abcd}_\rho &=& {\cal Q}_\rho \delta^{ac} \delta^{bd} ~+~ 
{\cal W}_\rho f^{ace} f^{bde} ~+~ {\cal R}_\rho f^{abe} f^{cde} ~+~ 
{\cal S}_\rho d_A^{abcd} ~+~ {\cal P}_\rho \delta^{ab} \delta^{cd} ~+~ 
{\cal T}_\rho \delta^{ad} \delta^{bc} \nonumber \\ 
\end{eqnarray}
and, \cite{64}, 
\begin{equation}
d_A^{abcd} ~=~ \frac{1}{6} \mbox{Tr} \left( T_A^a T_A^{(b} T_A^c T_A^{d)}
\right)
\end{equation}
is totally symmetric. The subscript on the group generator $T_A^a$ indicates 
that it is in the adjoint representation. We have omitted the common transverse
projector from (\ref{tradec}). Including $\rho^{ab}_\mu$ in the basis here may 
not appear to be necessary since (\ref{tradec}) is block diagonal, and we have 
treated it already, but it is relevant when we examine the longitudinal sector 
since not all the corresponding zero entries of (\ref{tradec}) remain zero. In 
introducing general amplitudes for the colour decomposition we note that each 
represents the leading term of the $2$-point function as well as the loop 
corrections. In each of the rank four colour decompositions we have included 
structures which do not arise in the explicit computations at one loop. Aside 
from ensuring we work with a complete basis such structures may occur at a 
higher loop order but they will also give us an insight into the effect such 
pieces have on the form of the inverse of the matrix which is the propagators. 
For this we formally define the inverse in a similar way with  
\begin{equation}
\Pi^{\{cd|pq\}} ~=~ \left(
\begin{array}{ccc}
{\cal A} \delta^{cp} & {\cal B} f^{cpq} & 0 \\
{\cal B} f^{pcd} & {\cal D}^{cdpq}_\xi & 0 \\
0 & 0 & {\cal D}^{cdpq}_\rho \\
\end{array}
\right) 
\end{equation} 
where
\begin{eqnarray}
{\cal D}^{cdpq}_\xi &=& {\cal D}_\xi \delta^{cp} \delta^{dq} ~+~ {\cal J}_\xi 
f^{cpe} f^{dqe} ~+~ {\cal K}_\xi f^{cde} f^{pqe} ~+~ 
{\cal L}_\xi d_A^{cdpq} ~+~ {\cal M}_\xi \delta^{cd} \delta^{pq} ~+~ 
{\cal N}_\xi \delta^{cq} \delta^{dp} \nonumber \\ 
{\cal D}^{cdpq}_\rho &=& {\cal D}_\rho \delta^{cp} \delta^{dq} ~+~ 
{\cal J}_\rho f^{cpe} f^{dqe} ~+~ {\cal K}_\rho f^{cde} f^{pqe} ~+~ 
{\cal L}_\rho d_A^{cdpq} ~+~ {\cal M}_\rho \delta^{cd} \delta^{pq} ~+~ 
{\cal N}_\rho \delta^{cq} \delta^{dp} \nonumber \\ 
\end{eqnarray} 
and the transverse projector is again omitted. As $\Pi^{\{ab|cd\}}$ is the 
inverse colour matrix, it satisfies 
\begin{equation}
\Lambda^{\{ab|cd\}} \Pi^{\{cd|pq\}} ~=~ \left(
\begin{array}{ccc}
\delta^{cp} & 0 & 0 \\
0 & \delta^{cp} \delta^{dq} & 0 \\
0 & 0 & \delta^{cp} \delta^{dq} \\
\end{array}
\right) 
\end{equation} 
where the right hand side is the unit matrix in the colour vector space of the
basis of fields we use. For the inversion relating to the Lorentz structure, we
recall the trivial identity
\begin{equation}
\eta_{\mu\nu} ~=~ P_{\mu\nu}(p) ~+~ L_{\mu\nu}(p)
\end{equation}
and note that the first term on the right side is where our current focus is.

The method to find the formal inverse is to multiply out the matrices and solve
the resulting relations between the amplitudes algebraically. In order to do
this we note that the products involving $d_A^{abcd}$ can be simplified with 
the relations, \cite{25}, 
\begin{eqnarray}
d_A^{abpq} d_A^{cdpq} &=& a_1 \delta^{ab} \delta^{cd} ~+~
a_2 \left( \delta^{ac} \delta^{bd} ~+~ \delta^{ad} \delta^{bc} \right) ~+~
a_3 \left( f^{ace} f^{bde} ~+~ f^{ade} f^{bce} \right) ~+~
a_4 d_A^{abcd} \nonumber \\
f^{ape} f^{bqe} d_A^{cdpq} &=& b_1 \delta^{ab} \delta^{cd} ~+~
b_2 \left( \delta^{ac} \delta^{bd} ~+~ \delta^{ad} \delta^{bc} \right)
\nonumber \\
&& +~ b_3 \left( f^{ace} f^{bde} ~+~ f^{ade} f^{bce} \right) ~+~
b_4 d_A^{abcd} 
\end{eqnarray}
where the coefficients are defined by, \cite{25}, 
\begin{eqnarray}
a_1 &=& -~ \left[ 540 C_A^2 \NA (\NA-3) d_A^{abcd} d_A^{cdpq} d_A^{abpq}
+ 144 (2\NA+19) \left( d_A^{abcd} d_A^{abcd} \right)^2 \right. \nonumber \\
&& \left. ~~~~-~ 150 C_A^4 \NA (3\NA+11) d_A^{abcd} d_A^{abcd}
+ 625 C_A^8 \NA^2 \right] \nonumber \\
&& ~~~ \times ~
\frac{1}{54\NA(\NA-3) [ 12 (\NA + 2) d_A^{efgh} d_A^{efgh} - 25 C_A^4 \NA]}
\nonumber \\
a_2 &=& \left[ 144 (11\NA - 8) \left( d_A^{abcd} d_A^{abcd} \right)^2 \right.
- 1080 C_A^2 \NA (\NA - 3) d_A^{abcd} d_A^{cdpq} d_A^{abpq} \nonumber \\
&& \left. ~+~ 625 C_A^8 \NA^2 - 3000 C_A^4 \NA d_A^{abcd} d_A^{abcd} \right]
\nonumber \\
&& \times ~
\frac{1}{108\NA(\NA-3) [ 12 (\NA + 2) d_A^{efgh} d_A^{efgh} - 25 C_A^4 \NA]}
\nonumber \\
a_3 &=& \frac{[ 12 (\NA + 2) d_A^{abcd} d_A^{abcd} - 25 C_A^4 \NA]}
{54 C_A \NA (\NA-3)} \nonumber \\
a_4 &=& \frac{[ 216 (\NA + 2) d_A^{abcd} d_A^{cdpq} d_A^{abpq} - 125 C_A^6 \NA
- 360 C_A^2 d_A^{abcd} d_A^{abcd}]}{18[ 12 (\NA + 2) d_A^{efgh} d_A^{efgh}
- 25 C_A^4 \NA]}
\end{eqnarray}
and 
\begin{equation}
b_1 ~=~ -~ 2 b_2 ~=~ \frac{[5C_A^4 \NA - 12 d_A^{abcd} d_A^{abcd}]}
{9 C_A \NA (\NA - 3)} ~~,~~
b_3 ~=~ \frac{[6 (\NA-1) d_A^{abcd} d_A^{abcd} - 5 C_A^4 \NA]}
{9 C_A^2 \NA (\NA - 3)} ~~,~~ b_4 ~=~ \frac{C_A}{3} ~. 
\end{equation}
Multiplying out the matrices results in the relations 
\begin{eqnarray}
1 &=& {\cal AX} ~+~ C_A {\cal UB} ~~~,~~~ 0 ~=~ {\cal XB} ~+~ 
\left( {\cal D}_\xi - {\cal N}_\xi + C_A {\cal K}_\xi 
+ \frac{1}{2} C_A {\cal J}_\xi \right) {\cal U} \nonumber \\
0 &=& {\cal A} {\cal U} ~+~ \left( {\cal Q}_\xi + C_A {\cal R}_\xi 
+ \frac{1}{2} C_A {\cal W}_\xi - {\cal T}_\xi \right) {\cal B} \nonumber \\
1 &=& {\cal Q}_\xi {\cal D}_\xi ~+~ b_2 {\cal L}_\xi {\cal W}_\xi ~+~ 
b_2 {\cal S}_\xi {\cal J}_\xi ~+~ a_2 {\cal S}_\xi {\cal L}_\xi ~+~
{\cal T}_\xi {\cal N}_\xi \nonumber \\ 
0 &=& \left( {\cal Q}_\xi +  C_A {\cal W}_\xi + \frac{5}{6} C_A^2 {\cal S}_\xi 
+ \NA {\cal P}_\xi + {\cal T}_\xi \right) {\cal M}_\xi ~+~ 
\left( C_A {\cal J}_\xi + \frac{5}{6} C_A^2 {\cal L}_\xi + {\cal D}_\xi 
+ {\cal N}_\xi \right) {\cal P}_\xi \nonumber \\
&& +~ b_1 {\cal W}_\xi {\cal L}_\xi ~+~ b_1 {\cal S}_\xi {\cal J}_\xi ~+~ 
a_1 {\cal S}_\xi {\cal L}_\xi \nonumber \\ 
0 &=& b_2 {\cal L}_\xi {\cal W}_\xi ~+~ b_2 {\cal S}_\xi {\cal J}_\xi ~+~
a_2 {\cal S}_\xi {\cal L}_\xi ~+~ {\cal Q}_\xi {\cal N}_\xi ~+~ 
{\cal T}_\xi {\cal D}_\xi \nonumber \\ 
0 &=& {\cal Q}_\xi {\cal L}_\xi ~+~ {\cal W}_\xi {\cal J}_\xi ~+~ 
{\cal S}_\xi {\cal D}_\xi ~+~ b_4 {\cal W}_\xi {\cal L}_\xi ~+~ 
b_4 {\cal S}_\xi {\cal J}_\xi ~+~ a_4 {\cal S}_\xi {\cal L}_\xi ~+~ 
{\cal S}_\xi {\cal N}_\xi ~+~ {\cal T}_\xi {\cal L}_\xi \nonumber \\
0 &=& {\cal W}_\xi {\cal D}_\xi ~+~ {\cal Q}_\xi {\cal J}_\xi ~+~ 
\frac{1}{6} C_A {\cal W}_\xi {\cal J}_\xi ~+~ 
2 b_3 {\cal S}_\xi {\cal J}_\xi ~+~ 2 a_3 {\cal S}_\xi {\cal L}_\xi ~+~ 
2 b_3 {\cal W}_\xi {\cal L}_\xi ~+~ {\cal W}_\xi {\cal N}_\xi ~+~ 
{\cal T}_\xi {\cal J}_\xi \nonumber \\
0 &=& {\cal UB} ~+~ {\cal Q}_\xi {\cal K}_\xi ~+~ \frac{1}{6} C_A {\cal W}_\xi 
{\cal J}_\xi ~+~ \frac{1}{2} C_A {\cal W}_\xi {\cal K}_\xi ~+~ 
\frac{1}{2} C_A {\cal R}_\xi {\cal J}_\xi ~+~ 
C_A {\cal R}_\xi {\cal K}_\xi ~-~ b_3 {\cal W}_\xi {\cal L}_\xi \nonumber \\
&& -~ {\cal W}_\xi {\cal N}_\xi ~+~ {\cal R}_\xi {\cal D}_\xi ~-~ 
{\cal R}_\xi {\cal N}_\xi ~-~ b_3 {\cal S}_\xi {\cal J}_\xi ~-~ 
a_3 {\cal S}_\xi {\cal L}_\xi ~-~ {\cal T}_\xi {\cal J}_\xi ~-~ 
{\cal T}_\xi {\cal K}_\xi  
\end{eqnarray} 
for the $A^a_\mu$ and $\xi^{ab}_\mu$ sector. The full set for the 
$\rho^{ab}_\mu$ sector is 
\begin{eqnarray}
1 &=& {\cal Q}_\rho {\cal D}_\rho ~+~ b_2 {\cal L}_\rho {\cal W}_\rho ~+~ 
b_2 {\cal S}_\rho {\cal J}_\rho ~+~ a_2 {\cal S}_\rho {\cal L}_\rho ~+~
{\cal T}_\rho {\cal N}_\rho \nonumber \\ 
0 &=& \left( {\cal Q}_\rho +  C_A {\cal W}_\rho 
+ \frac{5}{6} C_A^2 {\cal S}_\rho + \NA {\cal P}_\rho 
+ {\cal T}_\rho \right) {\cal M}_\rho ~+~ \left( C_A {\cal J}_\rho 
+ \frac{5}{6} C_A^2 {\cal L}_\rho + {\cal D}_\rho 
+ {\cal N}_\rho \right) {\cal P}_\rho \nonumber \\
&& +~ b_1 {\cal W}_\rho {\cal L}_\rho ~+~ b_1 {\cal S}_\rho {\cal J}_\rho ~+~ 
a_1 {\cal S}_\rho {\cal L}_\rho \nonumber \\ 
0 &=& b_2 {\cal L}_\rho {\cal W}_\rho ~+~ b_2 {\cal S}_\rho {\cal J}_\rho ~+~
a_2 {\cal S}_\rho {\cal L}_\rho ~+~ {\cal Q}_\rho {\cal N}_\rho ~+~ 
{\cal T}_\rho {\cal D}_\rho \nonumber \\ 
0 &=& {\cal Q}_\rho {\cal L}_\rho ~+~ {\cal W}_\rho {\cal J}_\rho ~+~ 
{\cal S}_\rho {\cal D}_\rho ~+~ b_4 {\cal W}_\rho {\cal L}_\rho ~+~ 
b_4 {\cal S}_\rho {\cal J}_\rho ~+~ a_4 {\cal S}_\rho {\cal L}_\rho ~+~ 
{\cal S}_\rho {\cal N}_\rho ~+~ {\cal T}_\rho {\cal L}_\rho \nonumber \\
0 &=& {\cal W}_\rho {\cal D}_\rho ~+~ {\cal Q}_\rho {\cal J}_\rho ~+~ 
\frac{1}{6} C_A {\cal W}_\rho {\cal J}_\rho ~+~ 
2 b_3 {\cal S}_\rho {\cal J}_\rho \nonumber \\
&& +~ 2 a_3 {\cal S}_\rho {\cal L}_\rho ~+~ 2 b_3 {\cal W}_\rho 
{\cal L}_\rho ~+~ {\cal W}_\rho {\cal N}_\rho ~+~ {\cal T}_\rho {\cal J}_\rho 
\nonumber \\
0 &=& {\cal Q}_\rho {\cal K}_\rho ~+~ \frac{1}{6} C_A {\cal W}_\rho 
{\cal J}_\rho ~+~ \frac{1}{2} C_A {\cal W}_\rho {\cal K}_\rho ~+~ \frac{1}{2} 
C_A {\cal R}_\rho {\cal J}_\rho ~+~ C_A {\cal R}_\rho {\cal K}_\rho ~-~ 
b_3 {\cal W}_\rho {\cal L}_\rho \nonumber \\
&& -~ {\cal W}_\rho {\cal N}_\rho ~+~ {\cal R}_\rho {\cal D}_\rho ~-~ 
{\cal R}_\rho {\cal N}_\rho ~-~ b_3 {\cal S}_\rho {\cal J}_\rho ~-~ 
a_3 {\cal S}_\rho {\cal L}_\rho ~-~ {\cal T}_\rho {\cal J}_\rho ~-~ 
{\cal T}_\rho {\cal K}_\rho ~. 
\label{rhoprop}
\end{eqnarray} 
Whilst this set appears formally similar to the corresponding equations of the
$A^a_\mu$ and $\xi^{ab}_\mu$ subset, the final equation has one fewer term.

For an arbitrary colour group the full solution to both sets of equations are
cumbersome. For instance, by way of illustration for the slightly simpler case 
of the $\rho^{ab}_\mu$ sector we have recorded the full expressions for 
$SU(N_c)$ in Appendix C. For the top sector the gluon and mixed propagators are
simple for an arbitrary colour group giving 
\begin{eqnarray}
{\cal A} &=& \frac{[{\cal Q}_\xi + C_A {\cal R}_\xi + \half C_A {\cal W}_\xi]}
{[ ({\cal Q}_\xi + C_A {\cal R}_\xi + \half C_A {\cal W}_\xi) {\cal X} 
- C_A {\cal U}^2]} \nonumber \\
{\cal B} &=& -~ \frac{{\cal U}}
{[ ({\cal Q}_\xi + C_A {\cal R}_\xi + \half C_A {\cal W}_\xi) {\cal X} 
- C_A {\cal U}^2]} ~. 
\end{eqnarray}
However, for the $\xi^{ab}_\mu$ propagator we only present the expressions for
$SU(3)$ which are 
\begin{eqnarray}
{\cal D}_\xi &=& \frac{1}{2{\cal Q}_\xi} \left[ 3 ( 3 {\cal S}_\xi 
- 2 {\cal W}_\xi ) ( {\cal S}_\xi + 2 {\cal W}_\xi ) ( {\cal S}_\xi 
+ {\cal W}_\xi ) + 8 ( 7 {\cal S}_\xi + 3 {\cal W}_\xi ) {\cal Q}_\xi^2 \right.
\nonumber \\
&& \left. ~~~~~+~ 16 {\cal Q}_\xi^3 + 2 ( 27 {\cal S}_\xi^3 + 20 {\cal S}_\xi 
{\cal W}_\xi - 8 {\cal W}_\xi^2 ) {\cal Q}_\xi \right] \nonumber \\
&& ~ \times \left[ 2 {\cal Q}_\xi + 3 {\cal S}_\xi + 3 {\cal W}_\xi 
\right]^{-1} \left[ 2 {\cal Q}_\xi + 3 {\cal S}_\xi - 2 {\cal W}_\xi 
\right]^{-1} \left[ 2 {\cal Q}_\xi + {\cal S}_\xi + 2 {\cal W}_\xi \right]^{-1}
\nonumber \\ 
{\cal J}_\xi &=& -~ \frac{4 {\cal W}_\xi}{[ 2 {\cal Q}_\xi + 3 {\cal S}_\xi
+ 3 {\cal W}_\xi ] [ 2 {\cal Q}_\xi + 3 {\cal S}_\xi - 2 {\cal W}_\xi ]} 
\nonumber \\ 
{\cal K}_\xi &=& \frac{1}{{\cal Q}_\xi} \left[ ( 4 ( 3 {\cal S}_\xi 
- {\cal W}_\xi ) {\cal Q}_\xi + 3 ( 3 {\cal S}_\xi - 2 {\cal W}_\xi ) 
( {\cal S}_\xi + {\cal W}_\xi ) ) ( 2 {\cal U}^2 - {\cal W}_\xi {\cal X} 
- 2 {\cal R}_\xi {\cal X} ) \right. \nonumber \\
&& \left. ~~~~-~ 8 ( {\cal R}_\xi {\cal X} - {\cal U} )^2 {\cal Q}_\xi^2 
\right] \nonumber \\
&& ~ \times \left[ 2 {\cal Q}_\xi {\cal X} + 6 {\cal R}_\xi {\cal X} 
- 6 {\cal U}^2 + 3 {\cal W}_\xi {\cal X} \right]^{-1} \left[ 2 {\cal Q}_\xi 
+ 3 {\cal S}_\xi + 3 {\cal W}_\xi \right]^{-1} \left[ 2 {\cal Q}_\xi 
+ 3 {\cal S}_\xi - 2 {\cal W}_\xi \right]^{-1} \nonumber \\ 
{\cal L}_\xi &=& -~ 4 \left[ 2 {\cal Q}_\xi {\cal S}_\xi + ( 3 {\cal S}_\xi 
+ 2 {\cal W}_\xi ) ( {\cal S}_\xi - {\cal W}_\xi ) \right] \nonumber \\
&& ~~ \times \left[ 2 {\cal Q}_\xi + 3 {\cal S}_\xi + 3 {\cal W}_\xi 
\right]^{-1} \left[ 2 {\cal Q}_\xi + 3 {\cal S}_\xi - 2 {\cal W}_\xi 
\right]^{-1} \left[ 2 {\cal Q}_\xi + {\cal S}_\xi + 2 {\cal W}_\xi 
\right]^{-1} \nonumber \\
{\cal M}_\xi &=& 6 \left[ 21 {\cal S}_\xi^3 + {\cal S}_\xi^2 {\cal W}_\xi 
- 12 {\cal S}_\xi {\cal W}_\xi^2 - 4 {\cal W}_\xi^3 + 2 ( 7 {\cal S}_\xi 
+ 4 {\cal W}_\xi ) {\cal Q}_\xi {\cal S}_\xi \right] 
\left[ 2 {\cal Q}_\xi + 15 {\cal S}_\xi + 6 {\cal W}_\xi \right]^{-1} 
\nonumber \\
&& \times \left[ 2 {\cal Q}_\xi + 3 {\cal S}_\xi + 3 {\cal W}_\xi \right]^{-1} 
\left[ 2 {\cal Q}_\xi + 3 {\cal S}_\xi - 2 {\cal W}_\xi \right]^{-1} 
\left[ 2 {\cal Q}_\xi + {\cal S}_\xi + 2 {\cal W}_\xi \right]^{-1} \nonumber \\
{\cal N}_\xi &=& -~ \frac{3}{2{\cal Q}_\xi} \left[ ( 3 {\cal S}_\xi 
- 2 {\cal W}_\xi ) ( {\cal S}_\xi + 2 {\cal W}_\xi ) ( {\cal S}_\xi 
+ {\cal W}_\xi ) + 2 ( {\cal S}_\xi + 4 {\cal W}_\xi ) {\cal Q}_\xi 
{\cal S}_\xi \right] \nonumber \\
&& ~~~ \times \left[ 2 {\cal Q}_\xi + 3 {\cal S}_\xi + 3 {\cal W}_\xi 
\right]^{-1} \left[ 2 {\cal Q}_\xi + 3 {\cal S}_\xi - 2 {\cal W}_\xi 
\right]^{-1} \left[ 2 {\cal Q}_\xi + {\cal S}_\xi + 2 {\cal W}_\xi 
\right]^{-1} ~. 
\end{eqnarray}

Next we formally repeat the procedure for the longitudinal part of the matrix
of $2$-point functions and denote the corresponding quantities with a 
superscript ${}^L$. For instance, the matrix of $2$-point functions is now 
\begin{equation}
\Lambda^{L\,\{ab|cd\}} ~=~ \left(
\begin{array}{ccc}
{\cal X}^L \delta^{ac} & {\cal U}^L f^{acd} & {\cal V}^L f^{acd} \\
{\cal U}^L f^{cab} & {\cal Q}^{L\, abcd}_\xi & 0 \\
{\cal V}^L f^{cab} & 0 & {\cal Q}^{L\, abcd}_\rho \\
\end{array}
\right) 
\end{equation} 
where 
\begin{eqnarray}
{\cal Q}^{L\, abcd}_\xi &=& {\cal Q}^L_\xi \delta^{ac} \delta^{bd} ~+~ 
{\cal W}^L_\xi f^{ace} f^{bde} ~+~ {\cal R}^L_\xi f^{abe} f^{cde} ~+~ 
{\cal S}^L_\xi d_A^{abcd} ~+~ {\cal P}^L_\xi \delta^{ab} \delta^{cd} ~+~ 
{\cal T}^L_\xi \delta^{ad} \delta^{bc} \nonumber \\ 
{\cal Q}^{L\, abcd}_\rho &=& {\cal Q}^L_\rho \delta^{ac} \delta^{bd} ~+~ 
{\cal W}^L_\rho f^{ace} f^{bde} ~+~ {\cal R}^L_\rho f^{abe} f^{cde} ~+~ 
{\cal S}^L_\rho d_A^{abcd} ~+~ {\cal P}^L_\rho \delta^{ab} \delta^{cd} ~+~ 
{\cal T}^L_\rho \delta^{ad} \delta^{bc} ~. \nonumber \\
\end{eqnarray}
In writing the formal longitudinal part we are making no assumptions at the
outset concerning the form of the $2$-point functions. For instance, from
(\ref{rho2pt}) and (\ref{xi2pt}) it is clear that 
\begin{equation}
{\cal Q}_\xi ~=~ {\cal Q}^L_\xi ~+~ O(a^2) ~=~ {\cal Q}_\rho ~+~ O(a^2) ~=~ 
{\cal Q}^L_\rho ~+~ O(a^2) 
\end{equation}
to one loop but we do not impose that condition initially. There is a non-zero 
entry in the $A^a_\mu$-$\rho^{ab}_\mu$ slot due to a non-zero one loop 
contribution to the longitudinal part of this $2$-point function. Thus the 
inverse has to be more general than that for the transverse sector and we take 
\begin{equation}
\Pi^{L\,\{cd|pq\}} ~=~ \left(
\begin{array}{ccc}
{\cal A}^L \delta^{cp} & {\cal B}^L f^{cpq} & {\cal C}^L f^{cpq} \\
{\cal B}^L f^{pcd} & {\cal D}^{L\, cdpq}_\xi & {\cal E}^{L\, cdpq}_\xi \\
{\cal C}^L f^{pcd} & {\cal E}^{L\, cdpq}_\xi & {\cal E}^{L\, cdpq}_\rho \\
\end{array}
\right) 
\end{equation} 
where  
\begin{eqnarray}
{\cal D}^{L\, cdpq}_\xi &=& {\cal D}^L_\xi \delta^{cp} \delta^{dq} ~+~ 
{\cal J}^L_\xi f^{cpe} f^{dqe} ~+~ {\cal K}^L_\xi f^{cde} f^{pqe} ~+~ 
{\cal L}^L_\xi d_A^{cdpq} ~+~ {\cal M}^L_\xi \delta^{cd} \delta^{pq} ~+~ 
{\cal N}^L_\xi \delta^{cq} \delta^{dp} \nonumber \\ 
{\cal E}^{L\, cdpq}_\xi &=& {\cal E}^L_\xi \delta^{cp} \delta^{dq} ~+~ 
{\cal F}^L_\xi f^{cpe} f^{dqe} ~+~ {\cal G}^L_\xi f^{cde} f^{pqe} ~+~ 
{\cal H}^L_\xi d_A^{cdpq} ~+~ {\cal Y}^L_\xi \delta^{cd} \delta^{pq} ~+~ 
{\cal Z}^L_\xi \delta^{cq} \delta^{dp} \nonumber \\ 
{\cal E}^{L\, cdpq}_\rho &=& {\cal E}^L_\rho \delta^{cp} \delta^{dq} ~+~ 
{\cal F}^L_\rho f^{cpe} f^{dqe} ~+~ {\cal G}^L_\rho f^{cde} f^{pqe} ~+~ 
{\cal H}^L_\rho d_A^{cdpq} ~+~ {\cal Y}^L_\rho \delta^{cd} \delta^{pq} ~+~ 
{\cal Z}^L_\rho \delta^{cq} \delta^{dp} ~. \nonumber \\ 
\end{eqnarray} 
The inverse containing the longitudinal sector of the propagators satisfies a
similar equation to that for the transverse sector, 
\begin{equation}
\Lambda^{L\,\{ab|cd\}} \Pi^{L\,\{cd|pq\}} ~=~ \left(
\begin{array}{ccc}
\delta^{cp} & 0 & 0 \\
0 & \delta^{cp} \delta^{dq} & 0 \\
0 & 0 & \delta^{cp} \delta^{dq} \\
\end{array}
\right) ~.
\end{equation} 
However, in order to reduce the size of the algebraic equations we will have to
solve eventually, for the longitudinal sector we set 
\begin{equation}
{\cal W}^L_\rho ~=~ {\cal R}^L_\rho ~=~ {\cal S}^L_\rho ~=~ {\cal P}^L_\rho ~=~ 
{\cal T}^L_\rho ~=~ 0
\end{equation}
at the outset since it is evident from the explicit computations at one loop
that these relations are valid. If at higher loop order it turns out that any
of these is non-zero then one would have a different set of equations to solve
for the propagators. We find 
\begin{eqnarray}
1 &=& {\cal A}^L {\cal X}^L ~+~ C_A {\cal U}^L {\cal B}^L ~+~ C_A {\cal V}^L
{\cal C}^L \nonumber \\
0 &=& {\cal X}^L {\cal B}^L ~+~ \left( {\cal D}^L_\xi - {\cal N}^L_\xi 
+ C_A {\cal K}^L_\xi + \frac{1}{2} C_A {\cal J}^L_\xi \right) {\cal U}^L ~+~
\left( {\cal E}^L_\xi - {\cal Z}^L_\xi + C_A {\cal G}^L_\xi 
+ \frac{1}{2} C_A {\cal F}^L_\xi \right) {\cal V}^L \nonumber \\
0 &=& {\cal X}^L {\cal C}^L ~+~ \left( {\cal E}^L_\xi - {\cal Z}^L_\xi 
+ C_A {\cal G}^L_\xi + \frac{1}{2} C_A {\cal F}^L_\xi \right) {\cal U}^L ~+~
\left( {\cal E}^L_\rho - {\cal Z}^L_\rho + C_A {\cal G}^L_\rho
+ \frac{1}{2} C_A {\cal F}^L_\rho \right) {\cal V}^L \nonumber \\
0 &=& {\cal A}^L {\cal U}^L ~+~ \left( {\cal Q}^L_\xi + C_A {\cal R}^L_\xi 
+ \frac{1}{2} C_A {\cal W}^L_\xi - {\cal T}^L_\xi \right) {\cal B}^L 
\nonumber \\
1 &=& {\cal Q}^L_\xi {\cal D}^L_\xi ~+~ b_2 {\cal L}^L_\xi {\cal W}^L_\xi ~+~ 
b_2 {\cal S}^L_\xi {\cal J}^L_\xi ~+~ a_2 {\cal S}^L_\xi {\cal L}^L_\xi ~+~
{\cal T}^L_\xi {\cal N}^L_\xi \nonumber \\ 
0 &=& \left( {\cal Q}^L_\xi +  C_A {\cal W}^L_\xi 
+ \frac{5}{6} C_A^2 {\cal S}^L_\xi + \NA {\cal P}^L_\xi + {\cal T}^L_\xi 
\right) {\cal M}^L_\xi ~+~ \left( C_A {\cal J}^L_\xi 
+ \frac{5}{6} C_A^2 {\cal L}^L_\xi + {\cal D}^L_\xi 
+ {\cal N}^L_\xi \right) {\cal P}^L_\xi \nonumber \\
&& +~ b_1 {\cal W}^L_\xi {\cal L}^L_\xi ~+~ b_1 {\cal S}^L_\xi 
{\cal J}^L_\xi ~+~ a_1 {\cal S}^L_\xi {\cal L}^L_\xi \nonumber \\ 
0 &=& b_2 {\cal L}^L_\xi {\cal W}^L_\xi ~+~ b_2 {\cal S}^L_\xi 
{\cal J}^L_\xi ~+~ a_2 {\cal S}^L_\xi {\cal L}^L_\xi ~+~ 
{\cal Q}^L_\xi {\cal N}^L_\xi ~+~ {\cal T}^L_\xi {\cal D}^L_\xi \nonumber \\ 
0 &=& {\cal Q}^L_\xi {\cal L}^L_\xi ~+~ {\cal W}^L_\xi {\cal J}^L_\xi ~+~ 
{\cal S}^L_\xi {\cal D}^L_\xi ~+~ b_4 {\cal W}^L_\xi {\cal L}^L_\xi ~+~ 
b_4 {\cal S}^L_\xi {\cal J}^L_\xi ~+~ a_4 {\cal S}^L_\xi {\cal L}^L_\xi ~+~ 
{\cal S}^L_\xi {\cal N}^L_\xi ~+~ {\cal T}^L_\xi {\cal L}^L_\xi \nonumber \\
0 &=& {\cal W}^L_\xi {\cal D}^L_\xi ~+~ {\cal Q}^L_\xi {\cal J}^L_\xi ~+~ 
\frac{1}{6} C_A {\cal W}^L_\xi {\cal J}^L_\xi ~+~ 
2 b_3 {\cal S}^L_\xi {\cal J}^L_\xi \nonumber \\
&& +~ 2 a_3 {\cal S}^L_\xi {\cal L}^L_\xi ~+~ 2 b_3 {\cal W}^L_\xi 
{\cal L}^L_\xi ~+~ {\cal W}^L_\xi {\cal N}^L_\xi ~+~ {\cal T}^L_\xi 
{\cal J}^L_\xi \nonumber \\
0 &=& {\cal U}^L {\cal B}^L ~+~ {\cal Q}^L_\xi {\cal K}^L_\xi ~+~ 
\frac{1}{6} C_A {\cal W}^L_\xi {\cal J}^L_\xi ~+~ \frac{1}{2} C_A 
{\cal W}^L_\xi {\cal K}^L_\xi ~+~ \frac{1}{2} C_A {\cal R}^L_\xi 
{\cal J}^L_\xi \nonumber \\
&& +~ C_A {\cal R}^L_\xi {\cal K}^L_\xi ~-~ b_3 {\cal W}^L_\xi 
{\cal L}^L_\xi ~-~ {\cal W}^L_\xi {\cal N}^L_\xi ~+~ 
{\cal R}^L_\xi {\cal D}^L_\xi ~-~ {\cal R}^L_\xi {\cal N}^L_\xi \nonumber \\
&& -~ b_3 {\cal S}^L_\xi {\cal J}^L_\xi ~-~ a_3 {\cal S}^L_\xi 
{\cal L}^L_\xi ~-~ {\cal T}^L_\xi {\cal J}^L_\xi ~-~ {\cal T}^L_\xi 
{\cal K}^L_\xi \nonumber \\ 
0 &=& {\cal Q}^L_\xi {\cal E}^L_\xi ~+~ b_2 {\cal H}^L_\xi {\cal W}^L_\xi ~+~ 
b_2 {\cal S}^L_\xi {\cal F}^L_\xi ~+~ a_2 {\cal S}^L_\xi {\cal H}^L_\xi ~+~
{\cal T}^L_\xi {\cal Z}^L_\xi \nonumber \\ 
0 &=& \left( {\cal Q}^L_\xi +  C_A {\cal W}^L_\xi 
+ \frac{5}{6} C_A^2 {\cal S}^L_\xi + \NA {\cal P}^L_\xi + {\cal T}^L_\xi 
\right) {\cal Y}^L_\xi ~+~ \left( C_A {\cal F}^L_\xi 
+ \frac{5}{6} C_A^2 {\cal H}^L_\xi + {\cal E}^L_\xi 
+ {\cal Z}^L_\xi \right) {\cal P}^L_\xi \nonumber \\
&& +~ b_1 {\cal W}^L_\xi {\cal H}^L_\xi ~+~ b_1 {\cal S}^L_\xi 
{\cal F}^L_\xi ~+~ a_1 {\cal H}^L_\xi {\cal L}^L_\xi \nonumber \\ 
0 &=& b_2 {\cal H}^L_\xi {\cal W}^L_\xi ~+~ b_2 {\cal S}^L_\xi 
{\cal F}^L_\xi ~+~ a_2 {\cal S}^L_\xi {\cal H}^L_\xi ~+~ 
{\cal Q}^L_\xi {\cal Z}^L_\xi ~+~ {\cal T}^L_\xi {\cal E}^L_\xi \nonumber \\ 
0 &=& {\cal Q}^L_\xi {\cal L}^L_\xi ~+~ {\cal W}^L_\xi {\cal F}^L_\xi ~+~ 
{\cal S}^L_\xi {\cal E}^L_\xi ~+~ b_4 {\cal W}^L_\xi {\cal H}^L_\xi ~+~ 
b_4 {\cal S}^L_\xi {\cal F}^L_\xi ~+~ a_4 {\cal S}^L_\xi {\cal H}^L_\xi ~+~ 
{\cal S}^L_\xi {\cal Z}^L_\xi ~+~ {\cal T}^L_\xi {\cal H}^L_\xi \nonumber \\
0 &=& {\cal W}^L_\xi {\cal E}^L_\xi ~+~ {\cal Q}^L_\xi {\cal F}^L_\xi ~+~ 
\frac{1}{6} C_A {\cal W}^L_\xi {\cal F}^L_\xi ~+~ 
2 b_3 {\cal S}^L_\xi {\cal F}^L_\xi \nonumber \\
&& +~ 2 a_3 {\cal S}^L_\xi {\cal H}^L_\xi ~+~ 2 b_3 {\cal W}^L_\xi 
{\cal H}^L_\xi ~+~ {\cal W}^L_\xi {\cal Z}^L_\xi ~+~ {\cal T}^L_\xi 
{\cal F}^L_\xi \nonumber \\
0 &=& {\cal U}^L {\cal C}^L ~+~ {\cal Q}^L_\xi {\cal G}^L_\xi ~+~ 
\frac{1}{6} C_A {\cal W}^L_\xi {\cal F}^L_\xi ~+~ \frac{1}{2} C_A 
{\cal W}^L_\xi {\cal G}^L_\xi ~+~ \frac{1}{2} C_A {\cal R}^L_\xi 
{\cal F}^L_\xi \nonumber \\
&& +~ C_A {\cal R}^L_\xi {\cal G}^L_\xi ~-~ b_3 {\cal W}^L_\xi 
{\cal H}^L_\xi ~-~ {\cal W}^L_\xi {\cal Z}^L_\xi ~+~ 
{\cal R}^L_\xi {\cal E}^L_\xi ~-~ {\cal R}^L_\xi {\cal Z}^L_\xi \nonumber \\
&& -~ b_3 {\cal S}^L_\xi {\cal F}^L_\xi ~-~ a_3 {\cal S}^L_\xi 
{\cal H}^L_\xi ~-~ {\cal T}^L_\xi {\cal F}^L_\xi ~-~ {\cal T}^L_\xi 
{\cal G}^L_\xi \nonumber \\ 
0 &=& {\cal V}^L {\cal A}^L ~+~ {\cal Q}^L_\rho {\cal C}^L ~~~,~~~ 
0 ~=~ {\cal Q}^L_\rho {\cal E}^L ~~~,~~~ 
0 ~=~ {\cal Q}^L_\rho {\cal Y}^L_\xi ~~~,~~~ 
0 ~=~ {\cal Q}^L_\rho {\cal Z}^L_\xi ~~~,~~~ 
0 ~=~ {\cal Q}^L_\rho {\cal H}^L_\xi \nonumber \\
0 &=& {\cal Q}^L_\rho {\cal F}^L_\xi ~~~,~~~ 
0 ~=~ {\cal V}^L {\cal B}^L ~+~ {\cal Q}^L_\rho {\cal G}^L_\xi ~~~,~~~ 
1 ~=~ {\cal Q}^L_\rho {\cal E}^L_\rho ~~~,~~~ 
0 ~=~ {\cal Q}^L_\rho {\cal Y}^L_\rho ~~~,~~~ 
0 ~=~ {\cal Q}^L_\rho {\cal Z}^L_\rho \nonumber \\
0 &=& {\cal Q}^L_\rho {\cal H}^L_\rho ~~~,~~~ 
0 ~=~ {\cal Q}^L_\rho {\cal F}^L_\rho ~~~,~~~
0 ~=~ {\cal V}^L {\cal C}^L ~+~ {\cal Q}^L_\rho {\cal G}^L_\rho ~. 
\end{eqnarray} 
These are clearly more involved than the transverse sector and lead to more
complicated forms for the explicit amplitudes. In addition to our earlier
nullifications, with ${\cal T}^L_\xi$~$=$~$0$ at the outset we find
\begin{eqnarray}
{\cal A}^L &=& \frac{[{\cal Q}^L_\xi + C_A {\cal R}^L_\xi 
+ \half C_A {\cal W}^L_\xi] {\cal Q}^L_\rho}
{\left[ {\cal Q }^L_\rho [ ({\cal Q}^L_\xi + C_A {\cal R}^L_\xi 
+ \half C_A {\cal W}^L_\xi) {\cal X}^L - C_A ({\cal U}^L)^2] - C_A [
{\cal Q}^L_\xi + C_A {\cal R}^L_\xi + \half C_A {\cal W}^L_\xi ] ({\cal V}^L)^2
\right]} \nonumber \\ 
{\cal B}^L &=& -~ \frac{{\cal U}^L {\cal Q}^L_\rho}
{\left[ {\cal Q }^L_\rho [ ({\cal Q}^L_\xi + C_A {\cal R}^L_\xi 
+ \half C_A {\cal W}^L_\xi) {\cal X}^L - C_A ({\cal U}^L)^2] - C_A [
{\cal Q}^L_\xi + C_A {\cal R}^L_\xi + \half C_A {\cal W}^L_\xi ] ({\cal V}^L)^2
\right]} \nonumber \\ 
{\cal C}^L &=& -~ \frac{[{\cal Q}^L_\xi + C_A {\cal R}^L_\xi 
+ \half C_A {\cal W}^L_\xi] {\cal V}^L}
{\left[ {\cal Q }^L_\rho [ ({\cal Q}^L_\xi + C_A {\cal R}^L_\xi 
+ \half C_A {\cal W}^L_\xi) {\cal X}^L - C_A ({\cal U}^L)^2] - C_A [
{\cal Q}^L_\xi + C_A {\cal R}^L_\xi + \half C_A {\cal W}^L_\xi ] ({\cal V}^L)^2
\right]} \nonumber \\  
\end{eqnarray}
for an arbitrary colour group. For the remaining amplitudes restricting to
$SU(3)$ produces  
\begin{eqnarray}
{\cal D}^L_\xi &=& \frac{1}{2{\cal Q}^L_\xi} \left[ 3 ( 3 {\cal S}^L_\xi 
- 2 {\cal W}^L_\xi ) ( {\cal S}^L_\xi + 2 {\cal W}^L_\xi ) ( {\cal S}^L_\xi 
+ {\cal W}^L_\xi ) + 8 ( 7 {\cal S}^L_\xi + 3 {\cal W}^L_\xi ) 
({\cal Q}^L_\xi)^2 \right. \nonumber \\
&& \left. ~~~~~+~ 16 ({\cal Q}^L_\xi)^3 + 2 ( 27 ({\cal S}^L_\xi)^3 
+ 20 {\cal S}^L_\xi {\cal W}^L_\xi - 8 ({\cal W}^L_\xi)^2 ) {\cal Q}^L_\xi 
\right] \nonumber \\
&& ~ \times \left[ 2 {\cal Q}^L_\xi + 3 {\cal S}^L_\xi + 3 {\cal W}^L_\xi 
\right]^{-1} \left[ 2 {\cal Q}^L_\xi + 3 {\cal S}^L_\xi - 2 {\cal W}^L_\xi 
\right]^{-1} \left[ 2 {\cal Q}^L_\xi + {\cal S}^L_\xi + 2 {\cal W}^L_\xi 
\right]^{-1} \nonumber \\ 
{\cal J}^L_\xi &=& -~ \frac{4 {\cal W}^L_\xi}{[ 2 {\cal Q}^L_\xi 
+ 3 {\cal S}^L_\xi + 3 {\cal W}^L_\xi ] [ 2 {\cal Q}^L_\xi + 3 {\cal S}^L_\xi 
- 2 {\cal W}^L_\xi ]} \nonumber \\ 
{\cal K}^L_\xi &=& -~ \frac{1}{{\cal Q}^L_\xi} \left[ 
8 ({\cal Q}^L_\xi)^2 {\cal Q}^L_\rho {\cal R}^L_\xi {\cal X}^L 
- 8 ({\cal Q}^L_\xi)^2 {\cal Q}^L_\rho ({\cal U}^L)^2 
- 24 ({\cal Q}^L_\xi)^2 {\cal R}^L_\xi ({\cal V}^L)^2 
+ 24 {\cal Q}^L_\xi {\cal Q}^L_\rho {\cal R}^L_\xi {\cal S}^L_\xi {\cal X}^L 
\right. \nonumber \\
&& \left. ~~~~~~~~~
-~ 8 {\cal Q}^L_\xi {\cal Q}^L_\rho {\cal R}^L_\xi {\cal W}^L_\xi {\cal X}^L 
- 24 {\cal Q}^L_\xi {\cal Q}^L_\rho {\cal S}^L_\xi ({\cal U}^L)^2 
+ 12 {\cal Q}^L_\xi {\cal Q}^L_\rho {\cal S}^L_\xi {\cal W}^L_\xi {\cal X}^L 
\right. \nonumber \\
&& \left. ~~~~~~~~~
+~ 8 {\cal Q}^L_\xi {\cal Q}^L_\rho {\cal W}^L_\xi ({\cal U}^L)^2 
- 4 {\cal Q}^L_\xi {\cal Q}^L_\rho {\cal X}^L ({\cal W}^L_\xi)^2 
- 72 {\cal Q}^L_\xi {\cal R}^L_\xi {\cal S}^L_\xi ({\cal V}^L)^2 
\right. \nonumber \\
&& \left. ~~~~~~~~~
+~ 24 {\cal Q}^L_\xi {\cal R}^L_\xi {\cal W}^L_\xi ({\cal V}^L)^2 
- 36 {\cal Q}^L_\xi {\cal S}^L_\xi {\cal W}^L_\xi ({\cal V}^L)^2 
+ 12 {\cal Q}^L_\xi ({\cal V}^L)^2 ({\cal W}^L_\xi)^2 
\right. \nonumber \\
&& \left. ~~~~~~~~~
+~ 18 {\cal Q}^L_\rho {\cal R}^L_\xi ({\cal S}^L_\xi)^2 {\cal X}^L 
+ 6 {\cal Q}^L_\rho {\cal R}^L_\xi {\cal S}^L_\xi {\cal W}^L_\xi {\cal X}^L 
- 12 {\cal Q}^L_\rho {\cal R}^L_\xi ({\cal W}^L_\xi)^2 {\cal X}^L 
\right. \nonumber \\
&& \left. ~~~~~~~~~
-~ 18 {\cal Q}^L_\rho ({\cal S}^L_\xi)^2 ({\cal U}^L)^2 
+ 9 {\cal Q}^L_\rho {\cal W}^L_\xi ({\cal S}^L_\xi)^2 {\cal X}^L 
- 6 {\cal Q}^L_\rho {\cal W}^L_\xi ({\cal U}^L)^2 {\cal S}^L_\xi 
\right. \nonumber \\
&& \left. ~~~~~~~~~
+~ 3 {\cal Q}^L_\rho {\cal S}^L_\xi ({\cal W}^L_\xi)^2 {\cal X}^L
+ 12 {\cal Q}^L_\rho ({\cal W}^L_\xi)^2 ({\cal U}^L)^2 
- 6 {\cal Q}^L_\rho ({\cal W}^L_\xi)^3 {\cal X}^L
\right. \nonumber \\
&& \left. ~~~~~~~~~
-~ 54 {\cal R}^L_\xi ({\cal S}^L_\xi)^2 ({\cal V}^L)^2 
- 18 {\cal R}^L_\xi {\cal S}^L_\xi ({\cal V}^L)^2 {\cal W}^L_\xi
+ 36 {\cal R}^L_\xi ({\cal V}^L)^2 ({\cal W}^L_\xi)^2
\right. \nonumber \\
&& \left. ~~~~~~~~~
-~ 27 {\cal W}^L_\xi ({\cal V}^L)^2 ({\cal S}^L_\xi)^2
- 9 {\cal S}^L_\xi ({\cal V}^L)^2 ({\cal W}^L_\xi)^2
+ 18 ({\cal V}^L)^2 ({\cal W}^L_\xi)^3
\right] \nonumber \\
&& ~~ \times \left[ 2 {\cal Q}^L_\xi {\cal Q}^L_\rho {\cal X}^L
- 6 {\cal Q}^L_\rho ({\cal V}^L)^2 + 6 {\cal Q}^L_\rho {\cal R}^L_\xi
{\cal X}^L - 6 {\cal Q}^L ({\cal U}^L)^2 + 3 {\cal Q}^L_\rho {\cal W}^L_\xi
{\cal X}^L \right. \nonumber \\
&& \left. ~~~~~~- 18 {\cal R}^L_\xi ({\cal V}^L)^2 - 9 ({\cal V}^L)^2 
{\cal W}^L_\xi \right]^{-1} \left[ 2 {\cal Q}^L_\xi 
+ 3 {\cal S}^L_\xi + 3 {\cal W}^L_\xi \right]^{-1} \left[ 2 {\cal Q}^L_\xi 
+ 3 {\cal S}^L_\xi - 2 {\cal W}^L_\xi \right]^{-1} \nonumber \\ 
{\cal L}^L_\xi &=& -~ 4 \left[ 2 {\cal Q}^L_\xi {\cal S}^L_\xi 
+ ( 3 {\cal S}^L_\xi + 2 {\cal W}^L_\xi ) ( {\cal S}^L_\xi - {\cal W}^L_\xi ) 
\right] \nonumber \\
&& ~~ \times \left[ 2 {\cal Q}^L_\xi + 3 {\cal S}^L_\xi + 3 {\cal W}^L_\xi 
\right]^{-1} \left[ 2 {\cal Q}^L_\xi + 3 {\cal S}^L_\xi - 2 {\cal W}^L_\xi 
\right]^{-1} \left[ 2 {\cal Q}^L_\xi + {\cal S}^L_\xi + 2 {\cal W}^L_\xi 
\right]^{-1} \nonumber \\
{\cal M}^L_\xi &=& 6 \left[ 21 ({\cal S}^L_\xi)^3 + ({\cal S}^L_\xi)^2 
{\cal W}^L_\xi - 12 {\cal S}^L_\xi ({\cal W}^L_\xi)^2 - 4 ({\cal W}^L_\xi)^3 
+ 2 ( 7 {\cal S}^L_\xi + 4 {\cal W}^L_\xi ) {\cal Q}^L_\xi {\cal S}^L_\xi 
\right] \nonumber \\
&& \times \left[ 2 {\cal Q}^L_\xi + 15 {\cal S}^L_\xi + 6 {\cal W}^L_\xi 
\right]^{-1} \left[ 2 {\cal Q}^L_\xi + 3 {\cal S}^L_\xi + 3 {\cal W}^L_\xi 
\right]^{-1} \nonumber \\
&& \times \left[ 2 {\cal Q}^L_\xi + 3 {\cal S}^L_\xi - 2 {\cal W}^L_\xi 
\right]^{-1} \left[ 2 {\cal Q}^L_\xi + {\cal S}^L_\xi + 2 {\cal W}_\xi 
\right]^{-1} \nonumber \\
{\cal N}^L_\xi &=& -~ \frac{3}{2{\cal Q}^L_\xi} \left[ ( 3 {\cal S}^L_\xi 
- 2 {\cal W}^L_\xi ) ( {\cal S}^L_\xi + 2 {\cal W}^L_\xi ) ( {\cal S}^L_\xi 
+ {\cal W}^L_\xi ) + 2 ( {\cal S}^L_\xi + 4 {\cal W}^L_\xi ) {\cal Q}^L_\xi 
{\cal S}^L_\xi \right] \nonumber \\
&& ~~~ \times \left[ 2 {\cal Q}^L_\xi + 3 {\cal S}^L_\xi + 3 {\cal W}^L_\xi 
\right]^{-1} \left[ 2 {\cal Q}^L_\xi + 3 {\cal S}^L_\xi - 2 {\cal W}^L_\xi 
\right]^{-1} \left[ 2 {\cal Q}_\xi + {\cal S}_\xi + 2 {\cal W}_\xi 
\right]^{-1} \nonumber \\
{\cal E}^L_\xi &=& {\cal F}^L_\xi ~=~ {\cal H}^L_\xi ~=~ {\cal Y}^L_\xi ~=~ 
{\cal Z}^L_\xi ~=~ 0 \nonumber \\
{\cal G}^L_\xi &=& 2 {\cal U}^L {\cal V}^L \left[
2 [ {\cal Q}^L_\rho {\cal X}^L - 3 ({\cal V}^L)^2 ] {\cal Q}^L_\xi
- 9 [ 2 {\cal R}^L_\xi + {\cal W}^L_\xi ] \right. \nonumber \\
&& \left. ~~~~~~~~~~-~ 3 [ 2 ({\cal U}^L)^2 - {\cal W}^L_\xi {\cal X}^L
- 2 {\cal R}^L_\xi {\cal X}^L ] {\cal Q}^L_\rho \right]^{-1} 
({\cal Q}^L_\rho)^{-1} \nonumber \\ 
{\cal E}^L_\rho &=& \frac{1}{{\cal Q}^L_\rho} ~~~,~~~  
{\cal F}^L_\rho ~=~ {\cal H}^L_\rho ~=~ {\cal Y}^L_\rho ~=~ 
{\cal Z}^L_\rho ~=~ 0 \nonumber \\
{\cal G}^L_\rho &=& \left[ 3 [ 2 {\cal R}^L_\xi + {\cal W}^L_\xi ]
+ 2 {\cal Q}^L_\xi \right] ({\cal V}^L)^2 ({\cal Q}^L_\rho)^{-1}  \nonumber \\
&& \times \left[ 2 [ {\cal Q}^L_\rho {\cal X}^L - 3 ({\cal V}^L)^2 ] 
{\cal Q}^L_\xi - 9 [ 2 {\cal R}^L_\xi + {\cal W}^L_\xi ] 
- 3 [ 2 ({\cal U}^L)^2 - {\cal W}^L_\xi {\cal X}^L
- 2 {\cal R}^L_\xi {\cal X}^L ] {\cal Q}^L_\rho \right]^{-1} ~. \nonumber \\
\end{eqnarray}
As a check on these solutions we have verified that the actual propagators,
(\ref{propdef}), are correctly reproduced when the $a$ independent values of 
the $2$-point function are inserted.

\sect{$\xi^{ab}_\mu$ and $\rho^{ab}_\mu$ enhancement.} 

Equipped with these solutions for both sectors we can now examine the specific
problem of enhancement. For the transverse sector it is a straightforward 
exercise to substitute the explicit zero momentum behaviour from the $2$-point 
functions (\ref{rho2pt}) and (\ref{xi2pt}). As was noted in \cite{25} this 
produces an enhanced $\xi^{ab}_\mu$ propagator in the transverse sector as 
expected given our parallel reasoning for the Faddeev-Popov ghost propagator. 
Moreover, the enhancement in the three dimensional case is similar to that of 
the four dimensional analysis, \cite{25}. For the longitudinal sector the 
derivation of the zero momentum behaviour for $\xi^{ab}_\mu$ requires care. 
Whilst the explicit expressions for the colour channel amplitudes in the 
longitudinal sector are general they mask the fact that ultimately we are in 
the Landau gauge. As is well known in order to construct the Landau gauge 
propagators in the non-Gribov scenario the gauge fixing term includes a 
parameter, $\alpha$. When this vanishes one is in the Landau gauge. However, 
such a term is required in order to prevent a non-singular matrix inversion 
such as that needed for deriving (\ref{propal}). As $\alpha$ occurs in the 
longitudinal term in that case, we cannot omit it in the inversion for the full
set of longitudinal $2$-point functions. Specifically, the leading term of 
${\cal X}^L$ is the only place where $\alpha$ appears. However, in order to 
extract the correct zero momentum behaviour of the propagators one must be 
careful in taking the limit to the Landau gauge. It transpires that this must 
be taken first in all our expressions for the propagator amplitudes and then 
the zero momentum limit taken. The order of the limits is not commutative. 
Though to examine the problem of enhancement we must set the gap equation 
initially. Following this procedure we find the zero momentum behaviour of the 
propagators is  
\begin{eqnarray}
\langle \xi^{ab}_\mu(p) \xi^{cd}_\nu(-p) \rangle & \sim & 
\frac{15\sqrt{2} \pi \gamma^5}{C_A^{1/4} (p^2)^2 g^2} \left[ 
\delta^{ad} \delta^{bc} - \delta^{ac} \delta^{bd} \right] \eta_{\mu\nu} ~+~ 
\frac{30\sqrt{2} \pi \gamma^5}{C_A^{5/4} (p^2)^2 g^2} f^{abe} f^{cde} 
P_{\mu\nu}(p) \nonumber \\ 
\langle \rho^{ab}_\mu(p) \rho^{cd}_\nu(-p) \rangle & \sim & -~ 
\frac{30\sqrt{2} \pi \gamma^5}{C_A^{1/4} (p^2)^2 g^2} \delta^{ac} \delta^{bd} 
\eta_{\mu\nu} ~. 
\end{eqnarray} 
Clearly the colour structures of the transverse and longitudinal parts of the
$\xi^{ab}_\mu$ propagator are different. However, if one contracts either field
of either propagator with a structure function then the enhancement disappears.
This loss of enhancement for this colour projection is completely in accord 
with Zwanziger's all orders observations from the BRST symmetry considerations 
in \cite{24}. Though it should be noted that there are $O(1/p^2)$ pieces which 
remain. More specifically, retaining the next term of the expansion as 
$p^2$~$\rightarrow$~$0$, we have 
\begin{eqnarray}
\langle A^a_\mu(p) A^b_\nu(-p) \rangle & \sim & -~ 
\frac{\sqrt{2} p^2 g^2}{384 \pi C_A^{1/4} \gamma^5} \delta^{ab} P_{\mu\nu}(p) 
\nonumber \\
\langle A^a_\mu(p) \xi^{bc}_\nu(-p) \rangle & \sim & 
\frac{i}{C_A \gamma^2} \left[ 1 + \frac{\sqrt{2} C_A^{1/4} g^2}{12 \pi \gamma}
\right] f^{abc} P_{\mu\nu}(p) \nonumber \\
\langle A^a_\mu(p) \rho^{bc}_\nu(-p) \rangle & = & 0 \nonumber \\ 
\langle \xi^{ab}_\mu(p) \xi^{cd}_\nu(-p) \rangle & \sim & -~ \left[ 
\frac{15\sqrt{2}\pi\gamma^3}{C_A^{1/4} (p^2)^2 g^2} \right. \nonumber \\
&& \left. ~~~~~-~ \frac{5\sqrt{2}\pi\gamma}{21 C_A^{3/4} p^2 g^2} 
\left[ 6 C_A^4 a_4 - 12 C_A^4 b_3 + C_A^5 b_4 - 432 C_A^2 a_3 b_4 
\right. \right. \nonumber \\
&& \left. \left. ~~~~~~~~~~~~~~~~~~~~~~~~~~-~ 72 C_A^3 a_3 - 108 a_2 b_3 
+ 9 C_A a_2 b_4 \right. \right. \nonumber \\
&& \left. \left. ~~~~~~~~~~~~~~~~~~~~~~~~~~-~ 9 C_A a_4 b_2 + 432 C_A^2 a_4 b_3 
+ 108 a_3 b_2 \right] \right. \nonumber \\
&& \left. ~~~~~~~~~\times \left[ 12 a_2 b_3 - C_A a_2 b_4 - 12 a_3 b_2 
+ C_A a_4 b_2 \right]^{-1} \frac{}{} \right] \delta^{ac}\delta^{bd} 
P_{\mu\nu}(p) \nonumber \\
&& -~ \left[ \frac{15\sqrt{2}\pi\gamma^3}{C_A^{1/4} (p^2)^2 g^2} \right. 
\nonumber \\
&& \left. ~~~~~-~ \frac{5\sqrt{2}\pi\gamma}{336 C_A^{3/4} p^2 g^2} 
\left[ 42 C_A^4 a_4 - 84 C_A^4 b_3 + 7 C_A^5 b_4 - 3024 C_A^2 a_3 b_4 
\right. \right. \nonumber \\
&& \left. \left. ~~~~~~~~~~~~~~~~~~~~~~~~~~-~ 504 C_A^3 a_3 - 1728 a_2 b_3 
+ 144 C_A a_2 b_4 \right. \right. \nonumber \\
&& \left. \left. ~~~~~~~~~~~~~~~~~~~~~~~~~~-~ 144 C_A a_4 b_2 
+ 3024 C_A^2 a_4 b_3 + 1728 a_3 b_2 \right] \right. \nonumber \\
&& \left. ~~~~~~~~~\times \left[ 12 a_2 b_3 - C_A a_2 b_4 - 12 a_3 b_2 
+ C_A a_4 b_2 \right]^{-1} \frac{}{} \right] \delta^{ac}\delta^{bd} 
L_{\mu\nu}(p) \nonumber \\
&& +~ \frac{20 \sqrt{2} \pi C_A^{5/4} \gamma [ 72 a_3 - 6 C_A a_4 + 12 C_A b_3 
- C_A^2 b_4 ]}{7[ 12 a_2 b_3 - C_A a_2 b_4 - 12 a_3 b_2 
+ C_A a_4 b_2 ] p^2 g^2} f^{ace} f^{bde} P_{\mu\nu}(p) \nonumber \\
&& +~ \frac{5 \sqrt{2} \pi C_A^{5/4} \gamma [ 72 a_3 - 6 C_A a_4 + 12 C_A b_3 
- C_A^2 b_4 ]}{2[ 12 a_2 b_3 - C_A a_2 b_4 - 12 a_3 b_2 
+ C_A a_4 b_2 ] p^2 g^2} f^{ace} f^{bde} L_{\mu\nu}(p) \nonumber \\
&& +~ \left[ \frac{30\sqrt{2}\pi\gamma^3}{C_A^{5/4} (p^2)^2 g^2} \right. 
\nonumber \\
&& \left. ~~~~~-~ \frac{10\sqrt{2}\pi\gamma}{7 C_A^{7/4} p^2 g^2}
\left[ 6 C_A^4 a_4 - 12 C_A^4 b_3 + C_A^5 b_4 - 3 C_A a_2 b_4 \right. \right. 
\nonumber \\
&& \left. \left. ~~~~~~~~~~~~~~~~~~~~~~~~+~ 36 a_2 b_3 - 72 C_A^3 a_3 
- 36 a_3 b_2 + 3 C_A a_4 b_2 \right] \right. \nonumber \\
&& \left. ~~~~~~~~~\times
\left[ 12 a_2 b_3 - C_A a_2 b_4 - 12 a_3 b_2 + C_A a_4 b_2 \frac{}{} 
\right]^{-1} \right] f^{abe} f^{cde} P_{\mu\nu}(p) \nonumber \\
&& -~ \frac{5 \sqrt{2} \pi C_A^{5/4} \gamma [ 72 a_3 - 6 C_A a_4 + 12 C_A b_3 
- C_A^2 b_4 ]}{2[ 12 a_2 b_3 - C_A a_2 b_4 - 12 a_3 b_2 
+ C_A a_4 b_2 ] p^2 g^2} f^{abe} f^{cde} L_{\mu\nu}(p) \nonumber \\
&& +~ \frac{120\sqrt{2}\pi C_A^{5/4} \gamma [ C_A b_4 - 12 C_A b_3 ]}
{7[ 12 a_2 b_3 - C_A a_2 b_4 - 12 a_3 b_2 + C_A a_4 b_2 ] p^2 g^2} 
d_A^{abcd} P_{\mu\nu}(p) \nonumber \\
&& +~ \frac{15\sqrt{2}\pi C_A^{5/4} \gamma [ C_A b_4 - 12 C_A b_3 ]}
{[ 12 a_2 b_3 - C_A a_2 b_4 - 12 a_3 b_2 + C_A a_4 b_2 ] p^2 g^2} 
d_A^{abcd} L_{\mu\nu}(p) \nonumber \\
&& +~ \frac{120 \sqrt{2} \pi \gamma [ 12 a_1 b_3 - C_A a_1 b_4 - 12 a_3 b_1 
+ C_A a_4 b_1 ]}
{7 C_A^{3/4} [ 12 a_2 b_3 - C_A a_2 b_4 - 12 a_3 b_2 + C_A a_4 b_2 ] p^2 g^2}
\delta^{ab} \delta^{cd} P_{\mu\nu}(p) \nonumber \\
&& +~ \frac{15 \sqrt{2} \pi \gamma [ 12 a_1 b_3 - C_A a_1 b_4 - 12 a_3 b_1 
+ C_A a_4 b_1 ]}
{2 C_A^{3/4} [ 12 a_2 b_3 - C_A a_2 b_4 - 12 a_3 b_2 + C_A a_4 b_2 ] p^2 g^2}
\delta^{ab} \delta^{cd} L_{\mu\nu}(p) \nonumber \\
&& +~ \left[ \frac{15 \sqrt{2} \pi \gamma^3}{C_A^{1/4} (p^2)^2 g^2} \right. 
\nonumber \\
&& \left. ~~~~~+~ \frac{5\sqrt{2}\pi\gamma}{21 C_A^{3/4} p^2 g^2} 
\left[ 6 C_A^4 a_4 - 12 C_A^4 b_3 + C_A^5 b_4 - 432 C_A^2 a_3 b_4 
\right. \right. \nonumber \\
&& \left. \left. ~~~~~~~~~~~~~~~~~~~~~~~~~~-~ 72 C_A^3 a_3 + 108 a_2 b_3 
- 9 C_A a_2 b_4 \right. \right. \nonumber \\
&& \left. \left. ~~~~~~~~~~~~~~~~~~~~~~~~~~+~ 9 C_A a_4 b_2 + 432 C_A^2 a_4 b_3 
- 108 a_3 b_2 \right] \right. \nonumber \\
&& \left. ~~~~~~~~~\times \left[ 12 a_2 b_3 - C_A a_2 b_4 - 12 a_3 b_2 
+ C_A a_4 b_2 \right]^{-1} \frac{}{} \right] \delta^{ad}\delta^{bc} 
P_{\mu\nu}(p) \nonumber \\
&& +~ \left[ \frac{15 \sqrt{2} \pi \gamma^3}{C_A^{1/4} (p^2)^2 g^2} \right. 
\nonumber \\
&& \left. ~~~~~+~ \frac{5\sqrt{2}\pi\gamma}{336 C_A^{3/4} p^2 g^2} 
\left[ 42 C_A^4 a_4 - 84 C_A^4 b_3 + 7 C_A^5 b_4 - 3024 C_A^2 a_3 b_4 
\right. \right. \nonumber \\
&& \left. \left. ~~~~~~~~~~~~~~~~~~~~~~~~~~~-~ 504 C_A^3 a_3 + 1728 a_2 b_3 
- 144 C_A a_2 b_4 \right. \right. \nonumber \\
&& \left. \left. ~~~~~~~~~~~~~~~~~~~~~~~~~~~+~ 144 C_A a_4 b_2 
+ 3024 C_A^2 a_4 b_3 - 1728 a_3 b_2 \right] \right. \nonumber \\
&& \left. ~~~~~~~~~\times \left[ 12 a_2 b_3 - C_A a_2 b_4 - 12 a_3 b_2 
+ C_A a_4 b_2 \right]^{-1} \frac{}{} \right] \delta^{ad}\delta^{bc} 
L_{\mu\nu}(p) \nonumber \\
\langle \xi^{ab}_\mu(p) \rho^{cd}_\nu(-p) \rangle & = & 0 \nonumber \\ 
\langle \rho^{ab}_\mu(p) \rho^{cd}_\nu(-p) \rangle & \sim & -~ 
\left[ \frac{30 \sqrt{2} \pi \gamma^3}{C_A^{1/4} (p^2)^2 g^2} ~+~ 
\frac{30\sqrt{2}\pi\gamma}{7C_A^{3/4} p^2 g^2} \right] 
\delta^{ac} \delta^{bd} \eta_{\mu\nu} \nonumber \\ 
\langle \omega^{ab}_\mu(p) \bar{\omega}^{cd}_\nu(-p) \rangle & \sim & -~ 
\left[ \frac{30 \sqrt{2} \pi \gamma^3}{C_A^{1/4} (p^2)^2 g^2} ~+~ 
\frac{30\sqrt{2}\pi\gamma}{7C_A^{3/4} p^2 g^2} \right] 
\delta^{ac} \delta^{bd} \eta_{\mu\nu} ~. 
\label{propsth}
\end{eqnarray} 
In order to compare with the colour adjoint projection of \cite{24} the leading
behaviour of the bosonic ghost in the zero momentum limit for an arbitrary 
colour group is 
\begin{eqnarray}
f^{apq} f^{brs} \langle \xi^{pq}_\mu(p) \xi^{rs}_\nu(-p) \rangle & \sim & 
-~ \delta^{ab} \left[ \frac{p^2}{\gamma^4} ~+~ 
\frac{53 \sqrt{2} C_A^{5/4} g^2}{384 \gamma^3} ~-~
\frac{[ 21 C_A - 64 T_F \Nf ] \sqrt{p^2} g^2}{512 \gamma^4} \right] 
P_{\mu\nu}(p) \nonumber \\ 
&& -~ \delta^{ab} \left[ 
\frac{30 \sqrt{2} \pi C_A^{3/4} \gamma^3}{(p^2)^2 g^2} ~+~ 
\frac{30\sqrt{2} C_A^{1/4} \pi\gamma}{7 p^2 g^2} \right. \nonumber \\
&& \left. ~~~~~~~~~+~ 
\frac{100\sqrt{2}\pi}{49C_A^{1/4} \gamma g^2} \right] L_{\mu\nu}(p) ~. 
\label{adjprojth}
\end{eqnarray} 
Not only does the enhancement disappear but the simple pole is also absent
leaving a finite non-zero value at zero momentum. Although the tree term is 
absent at zero momentum the loop corrections lead to a finite value in this
limit. Whilst it is tempting to assert that the freezing of the transverse part
of this correlation of a spin-$1$ field carrying one adjoint colour label is 
what is observed on the lattice and regarded as the frozen gluon propagator of 
the decoupling solution, the absence of a transverse propagator would exclude 
this as an alternative explanation. As far as we are aware numerical work 
observes a {\em tranverse} infrared frozen gluon propagator with no 
longitudinal part which is enhanced or otherwise. Moreover, the enhancement of
the Faddeev-Popov ghost still remains contrary to what the lattice observes,
\cite{26,27,28,29,30,31,32}. Just for completeness if we perform the same 
colour contraction on the original propagator, (\ref{propdef}), we have
\begin{equation}  
f^{apq} f^{brs} \langle \xi^{pq}_\mu(p) \xi^{rs}_\nu(-p) \rangle ~=~ 
-~ \frac{C_A p^2}{[(p^2)^2+C_A\gamma^4]} \delta^{ab} P_{\mu\nu}(p) ~-~
\frac{C_A}{p^2} \delta^{ab} L_{\mu\nu}(p) ~.
\label{adprxi}
\end{equation}
So that the effect of implementing the gap equation in deriving the zero
momentum behaviour of the propagator, appears to reduce the momentum structure
of both the transverse and longitudinal components by one power of momentum
for this specific limit. Though writing the original propagator with the
adjoint projection in this way demonstrates the existence of a massless pole
only in the longitudinal sector. So given that massless poles seem to lead to
enhancement in other situations, (\ref{adjprojth}) seems to be consistent with 
this observation. As an aside we draw attention to the contrasting structures 
of (\ref{adprxi}) and our earlier naive effective propagator, (\ref{xiproj}).

Whilst we have concentrated on the zero momentum behaviour for the propagators
when the gap equation has been implemented, it is worth noting some general
features of the full one loop corrections to the gluon propagator itself. In
(\ref{propsth}) we gave the leading order behaviour of the gluon. Unlike the
localizing fields there is no divergence in the zero momentum limit. Moreover, 
the propagator vanishes at one loop similar to the original propagator derived 
from the Lagrangian. This is in keeping with \cite{24} which showed that the
gluon form factor vanishes and hence is the key to showing positivity violation
for (\ref{laggz}). That our calculations reproduced this at one loop is in fact
a consistency check on \cite{24}. However, given the form of the equations for 
the longitudinal sector of the previous sector, it also turns out that there is
no longitudinal component for the gluon propagator which therefore remains
transverse at one loop similar to the original propagator. In fact this is due
to the longitudinal correction being proportional to the gauge parameter which
vanishes for the Landau gauge. Hence, these remarks are independent of the
dimension and the same property is present for the four dimensional gluon
propagator in (\ref{laggz}). Equally the mixed $A^a_\mu$-$\xi^{bc}_\nu$ 
propagator remains transverse at one loop in the Feynman gauge in either
spacetime dimension.

As \cite{25} only considered the enhancement of the transverse propagator, 
because the focus was on the implications for the static potential, we also 
record the situation with the enhanced propagators in {\em four} dimensions. At
leading order in the zero momentum limit we have 
\begin{eqnarray}
\langle \xi^{ab}_\mu(p) \xi^{cd}_\nu(-p) \rangle & \sim & 
\frac{4 \gamma^2}{\pi \sqrt{C_A} (p^2)^2 a} \left[ \delta^{ad} \delta^{bc}
- \delta^{ac} \delta^{bd} \right] \eta_{\mu\nu} ~+~ 
\frac{8 \gamma^2}{\pi C_A^{3/2} (p^2)^2 a} f^{abe} f^{cde} 
P_{\mu\nu}(p) \nonumber \\ 
\langle \rho^{ab}_\mu(p) \rho^{cd}_\nu(-p) \rangle & \sim & -~ 
\frac{8 \gamma^2}{\pi \sqrt{C_A} (p^2)^2 a} \delta^{ac} \delta^{bd} 
\eta_{\mu\nu} ~. 
\end{eqnarray} 
Thus the propagators have the same colour structure as the three dimensional 
case and so the colour adjoint projected fields are clearly not enhanced in 
this situation either. Including the subsequent term in the series we have, for
a general colour group, 
\begin{eqnarray}
\langle A^a_\mu(p) A^b_\nu(-p) \rangle & \sim & -~ 
\frac{p^2 a}{16 \gamma^4} \delta^{ab} P_{\mu\nu}(p) \nonumber \\
\langle A^a_\mu(p) \xi^{bc}_\nu(-p) \rangle & \sim & 
\frac{i}{C_A \gamma^2} \left[ 1 + \left[ \frac{5}{8} - \frac{3}{8} 
\ln \left[ \frac{C_A\gamma^4}{\mu^4} \right] \right] C_A a \right] 
f^{abc} P_{\mu\nu}(p) \nonumber \\
\langle A^a_\mu(p) \rho^{bc}_\nu(-p) \rangle & = & 0 \nonumber \\ 
\langle \xi^{ab}_\mu(p) \xi^{cd}_\nu(-p) \rangle & \sim & -~ \left[ 
\frac{4\gamma^2}{\pi\sqrt{C_A} (p^2)^2 a} \right. \nonumber \\
&& \left. ~~~~~-~ \frac{1}{42 \pi^2 C_A p^2 a} \left[ 6 \pi^2 C_A^4 a_4 
- 12 \pi^2 C_A^4 b_3 + \pi^2 C_A^5 b_4 - 432 \pi^2 C_A^2 a_3 b_4 
\right. \right. \nonumber \\
&& \left. \left. ~~~~~~~~~~~~~~~~~~~~~~~~~~-~ 72 \pi^2 C_A^3 a_3 
+ 432 \pi^2 C_A^2 a_4 b_3 + 84 C_A a_2 b_4 
\right. \right. \nonumber \\
&& \left. \left. ~~~~~~~~~~~~~~~~~~~~~~~~~~-~ 84 C_A a_4 b_2 
- 1008 a_2 b_3 + 1008 a_3 b_2 \right] \right. \nonumber \\
&& \left. ~~~~~~~~~\times \left[ 12 a_2 b_3 - C_A a_2 b_4 - 12 a_3 b_2 
+ C_A a_4 b_2 \right]^{-1} \frac{}{} \right] \delta^{ac}\delta^{bd} 
P_{\mu\nu}(p) \nonumber \\
&& -~ \left[ \frac{4\gamma^2}{\pi\sqrt{C_A} (p^2)^2 a} \right. \nonumber \\
&& \left. ~~~~~-~ \frac{1}{90 \pi^2 C_A p^2 a} \left[ 6 \pi^2 C_A^4 a_4 
- 12 \pi^2 C_A^4 b_3 + \pi^2 C_A^5 b_4 - 432 \pi^2 C_A^2 a_3 b_4 
\right. \right. \nonumber \\
&& \left. \left. ~~~~~~~~~~~~~~~~~~~~~~~~~~-~ 72 \pi^2 C_A^3 a_3 
+ 432 \pi^2 C_A^2 a_4 b_3 + 84 C_A a_2 b_4 
\right. \right. \nonumber \\
&& \left. \left. ~~~~~~~~~~~~~~~~~~~~~~~~~~-~ 180 C_A a_4 b_2 
- 2160 a_2 b_3 + 2160 a_3 b_2 \right] \right. \nonumber \\
&& \left. ~~~~~~~~~\times \left[ 12 a_2 b_3 - C_A a_2 b_4 - 12 a_3 b_2 
+ C_A a_4 b_2 \right]^{-1} \frac{}{} \right] \delta^{ac}\delta^{bd} 
L_{\mu\nu}(p) \nonumber \\
&& +~ \frac{2 C_A [ 72 a_3 - 6 C_A a_4 + 12 C_A b_3 - C_A^2 b_4 ]}
{7[ 12 a_2 b_3 - C_A a_2 b_4 - 12 a_3 b_2 + C_A a_4 b_2 ] p^2 a} 
f^{ace} f^{bde} P_{\mu\nu}(p) \nonumber \\
&& +~ \frac{2 C_A [ 72 a_3 - 6 C_A a_4 + 12 C_A b_3 - C_A^2 b_4 ]}
{15[ 12 a_2 b_3 - C_A a_2 b_4 - 12 a_3 b_2 + C_A a_4 b_2 ] p^2 a} 
f^{ace} f^{bde} L_{\mu\nu}(p) \nonumber \\
&& +~ \left[ \frac{8\gamma^2}{\pi\sqrt{C_A} (p^2)^2 a} \right. \nonumber \\
&& \left. ~~~~~-~ \frac{1}{7 \pi^2 C_A^2 p^2 a}
\left[ 6 \pi^2 C_A^4 a_4 - 12 \pi^2 C_A^4 b_3 + \pi^2 C_A^5 b_4 
- 72 \pi^2 C_A^3 a_3 \right. \right. \nonumber \\
&& \left. \left. ~~~~~~~~~~~~~~~~~~~~~~~~+~ 336 a_2 b_3 - 28 C_A a_2 b_4 
- 336 a_3 b_2 + 28 C_A a_4 b_2 \right] \right. \nonumber \\
&& \left. ~~~~~~~~~\times
\left[ 12 a_2 b_3 - C_A a_2 b_4 - 12 a_3 b_2 + C_A a_4 b_2 \right]^{-1}
\frac{}{} \right] f^{abe} f^{cde} P_{\mu\nu}(p) \nonumber \\
&& -~ \frac{2 C_A [ 72 a_3 - 6 C_A a_4 + 12 C_A b_3 - C_A^2 b_4 ]}
{15[ 12 a_2 b_3 - C_A a_2 b_4 - 12 a_3 b_2 + C_A a_4 b_2 ] p^2 a} 
f^{abe} f^{cde} L_{\mu\nu}(p) \nonumber \\
&& +~ \frac{12 C_A [ C_A b_4 - 12 C_A b_3 ]}
{7[ 12 a_2 b_3 - C_A a_2 b_4 - 12 a_3 b_2 + C_A a_4 b_2 ] p^2 a} 
d_A^{abcd} P_{\mu\nu}(p) \nonumber \\
&& +~ \frac{4 C_A [ C_A b_4 - 12 C_A b_3 ]}
{5[ 12 a_2 b_3 - C_A a_2 b_4 - 12 a_3 b_2 + C_A a_4 b_2 ] p^2 a} 
d_A^{abcd} L_{\mu\nu}(p) \nonumber \\
&& +~ \frac{12 [ 12 a_1 b_3 - C_A a_1 b_4 - 12 a_3 b_1 + C_A a_4 b_1 ]}
{7 C_A [ 12 a_2 b_3 - C_A a_2 b_4 - 12 a_3 b_2 + C_A a_4 b_2 ] p^2 a} 
\delta^{ab} \delta^{cd} P_{\mu\nu}(p) \nonumber \\
&& +~ \frac{4 [ 12 a_1 b_3 - C_A a_1 b_4 - 12 a_3 b_1 + C_A a_4 b_1 ]}
{5 C_A [ 12 a_2 b_3 - C_A a_2 b_4 - 12 a_3 b_2 + C_A a_4 b_2 ] p^2 a} 
\delta^{ab} \delta^{cd} L_{\mu\nu}(p) \nonumber \\
&& +~ \left[ \frac{4\gamma^2}{\pi\sqrt{C_A} (p^2)^2 a} \right. \nonumber \\
&& \left. ~~~~~+~ \frac{1}{42 \pi^2 C_A p^2 a} \left[ 6 \pi^2 C_A^4 a_4 
- 12 \pi^2 C_A^4 b_3 + \pi^2 C_A^5 b_4 - 432 \pi^2 C_A^2 a_3 b_4 
\right. \right. \nonumber \\
&& \left. \left. ~~~~~~~~~~~~~~~~~~~~~~~~~~-~ 72 \pi^2 C_A^3 a_3 
+ 432 \pi^2 C_A^2 a_4 b_3 - 84 C_A a_2 b_4 
\right. \right. \nonumber \\
&& \left. \left. ~~~~~~~~~~~~~~~~~~~~~~~~~~+~ 84 C_A a_4 b_2 
+ 1008 a_2 b_3 - 1008 a_3 b_2 \right] \right. \nonumber \\
&& \left. ~~~~~~~~~\times \left[ 12 a_2 b_3 - C_A a_2 b_4 - 12 a_3 b_2 
+ C_A a_4 b_2 \right]^{-1} \frac{}{} \right] \delta^{ad}\delta^{bc} 
P_{\mu\nu}(p) \nonumber \\
&& +~ \left[ \frac{4\gamma^2}{\pi\sqrt{C_A} (p^2)^2 a} \right. \nonumber \\
&& \left. ~~~~~+~ \frac{1}{90 \pi^2 C_A p^2 a} \left[ 6 \pi^2 C_A^4 a_4 
- 12 \pi^2 C_A^4 b_3 + \pi^2 C_A^5 b_4 - 432 \pi^2 C_A^2 a_3 b_4 
\right. \right. \nonumber \\
&& \left. \left. ~~~~~~~~~~~~~~~~~~~~~~~~~~-~ 72 \pi^2 C_A^3 a_3 
+ 432 \pi^2 C_A^2 a_4 b_3 - 180 C_A a_2 b_4 
\right. \right. \nonumber \\
&& \left. \left. ~~~~~~~~~~~~~~~~~~~~~~~~~~+~ 180 C_A a_4 b_2 
+ 2168 a_2 b_3 - 2168 a_3 b_2 \right] \right. \nonumber \\
&& \left. ~~~~~~~~~\times \left[ 12 a_2 b_3 - C_A a_2 b_4 - 12 a_3 b_2 
+ C_A a_4 b_2 \right]^{-1} \frac{}{} \right] \delta^{ad}\delta^{bc} 
L_{\mu\nu}(p) \nonumber \\
\langle \xi^{ab}_\mu(p) \rho^{cd}_\nu(-p) \rangle & = & 0 \nonumber \\ 
\langle \rho^{ab}_\mu(p) \rho^{cd}_\nu(-p) \rangle & \sim & -~ 
\left[ \frac{8 \sqrt{C_A} \gamma^2}{\pi C_A (p^2)^2 a} ~+~ 
\frac{4}{\pi^2 C_A p^2 a} \right] \delta^{ac} \delta^{bd} \eta_{\mu\nu} 
\nonumber \\
\langle \omega^{ab}_\mu(p) \bar{\omega}^{cd}_\nu(-p) \rangle & \sim & -~ 
\left[ \frac{8 \sqrt{C_A} \gamma^2}{\pi C_A (p^2)^2 a} ~+~ 
\frac{4}{\pi^2 C_A p^2 a} \right] \delta^{ac} \delta^{bd} \eta_{\mu\nu} 
\end{eqnarray} 
For the colour group $SU(3)$ the Bose ghost propagator at $O(1/p^2)$ becomes 
\begin{eqnarray}
\left. \langle \xi^{ab}_\mu(p) \xi^{cd}_\nu(-p) \rangle \right|_{SU(3)} 
& \sim & -~ \left[ \frac{4\gamma^2}{\sqrt{3}\pi (p^2)^2 a} ~+~ 
\frac{2[ 7 + 78 \pi^2 ]}{21\pi^2 p^2 a} \right] \delta^{ac}\delta^{bd} 
P_{\mu\nu}(p) \nonumber \\
&& +~ \frac{16}{7 p^2 a} f^{ace} f^{bde} P_{\mu\nu}(p) \nonumber \\
&& +~ \left[ \frac{8\gamma^2}{3\sqrt{3} \pi (p^2)^2 a} ~+~ 
\frac{4[ 7 - 18 \pi^2 ]}{63 \pi^2 p^2 a} \right] f^{abe} f^{cde} P_{\mu\nu}(p)
\nonumber \\
&& +~ \frac{32}{7 p^2 a} d_A^{abcd} P_{\mu\nu}(p) ~-~ 
\frac{24}{7 p^2 a} \delta^{ab} \delta^{cd} P_{\mu\nu}(p) \nonumber \\
&& +~ \left[ \frac{4\gamma^2}{\sqrt{3} (p^2)^2 a} ~+~ 
\frac{2[ 7 - 78 \pi^2 ]}{21\pi^2 p^2 a} \right] \delta^{ad}\delta^{bc} 
P_{\mu\nu}(p) \nonumber \\
&& -~ \left[ \frac{4\gamma^2}{\sqrt{3}\pi (p^2)^2 a} ~+~ 
\frac{2[ 5 + 26 \pi^2 ]}{15\pi^2 p^2 a} \right] \delta^{ac}\delta^{bd} 
L_{\mu\nu}(p) \nonumber \\
&& +~ \frac{16}{15 p^2 a} f^{ace} f^{bde} L_{\mu\nu}(p) ~-~ 
\frac{8}{15 p^2 a} f^{abe} f^{cde} L_{\mu\nu}(p) \nonumber \\
&& +~ \frac{32}{15 p^2 a} d_A^{abcd} L_{\mu\nu}(p) ~-~ 
\frac{8}{5 p^2 a} \delta^{ab} \delta^{cd} L_{\mu\nu}(p) \nonumber \\
&& +~ \left[ \frac{4\gamma^2}{\sqrt{3} (p^2)^2 a} ~+~ 
\frac{2[ 5 - 26 \pi^2 ]}{15\pi^2 p^2 a} \right] \delta^{ad}\delta^{bc} 
L_{\mu\nu}(p) 
\end{eqnarray}
as $p^2$~$\rightarrow$~$0$. Repeating the analogous calculation to 
(\ref{adjprojth}), we find a similar structure in four dimensions since 
\begin{eqnarray}
f^{apq} f^{brs} \langle \xi^{pq}_\mu(p) \xi^{rs}_\nu(-p) \rangle & \sim & 
-~ \delta^{ab} \left[ \frac{p^2}{\gamma^4} ~+~ 
\frac{69 \pi C_A^2 a}{128 \sqrt{C_A} \gamma^2} \right] P_{\mu\nu}(p)
\nonumber \\
&& -~ \delta^{ab} \left[ \frac{8 C_A \gamma^2}{\pi \sqrt{C_A} (p^2)^2 a} ~+~ 
\frac{4}{\pi^2 p^2 a} ~+~ \frac{2}{\pi^3 \sqrt{C_A} \gamma^2 a} \right] 
L_{\mu\nu}(p)
\label{props}
\end{eqnarray} 
for the leading order behaviour for each structure. Again the transverse
enhancement disappears and overall the transverse projection also freezes to a 
non-zero value. There is longitudinal enhancement in keeping with \cite{24} and
our observation on the massless poles in the original propagator. 

\sect{Discussion.}

The main motivation of the article was to provide the loop analysis of the
three dimensional Gribov-Zwanziger Lagrangian, (\ref{laggz}), to the same order
in perturbation theory which is currently available in four dimensions. In 
having achieved this we note that many of the key properties are preserved. For
example, the enhancement of the fermionic ghosts and certain colour channels of
the imaginary part of the Bose localizing field is evident. Moreover, we have
provided the complete analysis of the construction of the latter to one loop 
for both the transverse and longitudinal parts. This additional enhancement of 
a {\em Bose} field is in keeping with the Kugo-Ojima ethos underlying 
(\ref{laggz}), \cite{12,13,14,24}, which has been argued to be a necessary 
criterion for confinement. However, the actual mechanism of how this is 
realised in practical terms in (\ref{laggz}) is still not resolved. In the 
pioneering ideas of Mandelstam and others, \cite{65,66,67,68}, for four 
dimensional Yang-Mills it was believed that the rising potential was due to the
single exchange of a colour valued field between heavy quarks whose propagator 
was of a dipole form in the infrared. Then it was the matter of a simple 
Fourier transform to coordinate space to produce the rising potential. With the
emergence of the enhancement of certain colour channels of the $\xi^{ab}_\mu$ 
propagator observed in \cite{24,25} the exchange of this colour quanta was 
explored to see if the dipole behaviour emerged. However, it transpired that 
the vertex colour structure nullified the colour structure of the associated 
dipole term of the infrared part of the propagator. In addition the momentum 
dependence of the vertex would also have led to a diminishing of the power of 
the momentum in the exchange. In repeating the analogous analysis here we have 
confirmed at one loop an underlying feature of Zwanziger's all orders BRST 
reasoning. That is the enhancement of the Bose ghost is independent of the 
spacetime dimension. In other words one retains the dipole behaviour in the 
infrared. If one were to have the exchange of a dipole type term to produce a 
linearly rising potential in three dimensions then it is clear that the single 
Bose ghost exchange would not be the simple explanation. This is simply because
one has to have a $1/(p^2)^{3/2}$ behaviour in the infrared for the zero 
momentum limit in order to have a linear dependence on the radial distance upon
performing the Fourier transform. Of course, this is on the understanding that 
the underlying mechanism is effectively dimension independent. To manufacture 
the necessary extra powers of momentum to alter the $\xi^{ab}_\mu$ enhanced 
form via say vertex correction momentum dependence would appear to be difficult
because to be balanced one would have to have two factors of $(p^2)^{1/4}$. 

Whilst anomalous dimensions of vertices and fields can in principle acquire 
large corrections in the infrared, to explore this further would require a 
summation of a significant set of Feynman diagrams, for instance. Moreover,
this would seem an unlikely avenue since the three dimensional theory is
superrenormalizable being less ultraviolet divergent than the four dimensional
counterpart. So an anomalous dimension could be trivial in three dimensions. 
Instead the point of view might be that the actual dynamics of how the rising 
potential emerges in (\ref{laggz}) rests in the enhancement of the Bose ghosts 
residing within Feynman diagrams themselves. Several ways to perhaps achieve 
this could be worth considering. One might be the summation of ladder type 
diagrams which could be similar to a colour flux tube. An advantage of this is 
that the dimensionality of the exchange required in both dimensions could be 
naturally accommodated by the Feynman integral measure. We have already touched
on another possibility when we rewrote the definition of the horizon condition 
purely in terms of the Bose ghost which enhances. One could conceive of some 
sort of effective infrared theory involving only the fields which enhance. The 
canonical gauge potential which determines the ultraviolet dynamics would 
appear as a bound state of the Bose ghost, $\xi^{ab}_\mu$. Such a scenario of 
the gluon or gauge potential being regarded as a bound state has been 
considered before but in other contexts. For instance, in \cite{69} the gluon 
was interpreted as the bound state of quarks. Whilst the analogous
interpretation here would be a bound state of $\xi^{ab}_\mu$ fields, the work 
of \cite{69} demonstrated that the (perturbative) structure of $d$-dimensional 
QCD was equivalent to the non-abelian Thirring model for calculations. Although
this was primarily true only at a non-trivial fixed point of the 
renormalization group flow in $d$-dimensions, one could in effect use the 
simpler non-abelian Thirring model. In other words it was an effective theory 
but which is non-renormalizable in dimensions greater than two. It is not 
inconceivable therefore, that underlying the Gribov-Zwanziger theory there is a
similar but clearly more complicated non-renormalizable effective theory which 
could have the structure of a nonlinear $\sigma$ model with an infinite set of 
interactions. In essence the quark gluon vertex could be replaced by quarks 
interacting with $n$ $\xi^{ab}_\mu$ fields. However, it is difficult to see how
such speculative ideas could be realised practically in the short-term. In 
passing we note the curiosity that the non-abelian Thirring model has a 
formulation which involves a dimension two operator built from a spin-$1$
auxiliary field carrying an adjoint colour index. 

\vspace{1cm}
\noindent
{\bf Acknowledgement.} The author thanks Prof. Y. Schr\"{o}der and Prof. D. 
Zwanziger for useful discussions. 

\appendix

\sect{Transverse parts.}

In this appendix we record the explicit one loop contributions to the 
transverse parts of the $2$-point functions. We have 
\begin{eqnarray}
X &=& \left[ ~-~ \frac{\pi}{8} T_F \Nf ~+~ \frac{21\pi}{512} C_A ~+~ 
\left[ \frac{1}{32} \eta_1(p^2) - \frac{5}{256} \eta_2(p^2)
+ \frac{69}{512} \eta_3(p^2) \right] C_A \right. \nonumber \\ 
&& \left. ~+~ \left[ \frac{9}{128} \eta_1(p^2) + \frac{5}{256} \eta_2(p^2)
\right] \frac{C_A^2 \gamma^4}{(p^2)^2} ~+~ \left[ \frac{9}{512} \eta_4(p^2) 
+ \frac{25}{256} \eta_5(p^2) \right] \frac{C_A^{3/2} \gamma^2}{p^2} \right. 
\nonumber \\ 
&& \left. ~+~ \left[ -~ \frac{3}{512} \eta_4(p^2) + \frac{19}{1024} \eta_5(p^2)
\right] \frac{\sqrt{C_A} p^2}{\gamma^2} \right. \nonumber \\ 
&& \left. ~+~ \left[ -~ \frac{25}{2048} \eta_1(p^2) + \frac{1}{128} \eta_2(p^2)
+ \frac{9}{4096} \eta_3(p^2) \right] \frac{(p^2)^2}{\gamma^4} ~-~ 
\frac{\pi}{256} \frac{(p^2)^2}{\gamma^4} \right. \nonumber \\
&& \left. ~-~ \frac{3 \sqrt{2} C_A^{5/4}\gamma}{512 \sqrt{p^2}} ~-~ 
\frac{7 \sqrt{2} C_A^{7/4}\gamma^3}{128 \sqrt{(p^2)^3}} ~+~ 
\frac{81 \sqrt{2} C_A^{3/4} \sqrt{p^2}}{1024 \gamma} \right]
\frac{\sqrt{p^2}}{\pi} g^2 ~+~ O(g^4) \\
U &=& -~ i \left[ ~-~ \left[ \frac{1}{1024} \eta_4(p^2) + \frac{13}{512}
\eta_5(p^2) \right] \frac{\sqrt{C_A}}{\gamma^2} ~-~
\frac{5C_A^{3/2}\gamma^2}{2048(p^2)^2} \eta_4(p^2) \right. \nonumber \\
&& \left. ~~~~~~+~ \left[ \frac{13}{128} \eta_1(p^2) + \frac{1}{128} \eta_2(p^2)
\right] \frac{C_A}{p^2} ~+~ \left[ \frac{1}{128} \eta_2(p^2) 
- \frac{13}{512} \eta_1(p^2) + \frac{7}{1024} \eta_3(p^2) \right] 
\frac{p^2}{\gamma^4} \right. \nonumber \\
&& \left. ~~~~~~-~ \frac{\pi p^2}{512\gamma^4} ~+~ \left[ \frac{3}{2048} 
\eta_4(p^2) - \frac{3}{4096} \eta_5(p^2) \right] 
\frac{(p^2)^2}{\sqrt{C_A}\gamma^6} \right. \nonumber \\
&& \left. ~~~~~~-~ \frac{175\sqrt{2} C_A^{3/4}}{3072\gamma\sqrt{p^2}} ~-~ 
\frac{5\sqrt{2} C_A^{5/4}\gamma}{1024 (p^2)^{3/2}} ~-~ 
\frac{\sqrt{2} C_A^{1/4} \sqrt{p^2}}{1024 \gamma^3} \right] 
\frac{\sqrt{p^2}}{\pi} g^2 ~+~ O(g^4) \\
V &=& W_\rho ~=~ R_\rho ~=~ S_\rho ~=~ O(g^4)
\\
Q_\xi &=& Q_\rho ~=~ \left[ \frac{C_A}{16} \eta_2(p^2) ~-~
\frac{C_A^{3/2} \gamma^2}{64 p^2} \eta_4(p^2) ~+~
\frac{\sqrt{C_A} p^2}{64\gamma^2} \eta_4(p^2) \right. \nonumber \\
&& \left. ~~~~~~~~~~~~~-~ \frac{\sqrt{2} C_A^{5/4} \gamma}{32 \sqrt{p^2}} ~+~ 
\frac{\sqrt{2}C_A^{3/4} \sqrt{p^2}}{32 \gamma} \right] \frac{\sqrt{p^2}}{\pi} 
g^2 ~+~ O(g^4) \\
W_\xi &=& \left[ \frac{1}{192} \eta_1(p^2) ~-~ \frac{1}{128} \eta_2(p^2) ~-~
\frac{1}{192} \eta_3(p^2) ~+~ \frac{C_A \gamma^4}{768 (p^2)^2} \eta_2(p^2)
\right. \nonumber \\
&& \left. ~+~ \frac{\sqrt{C_A} \gamma^2}{384 p^2} \eta_4(p^2) ~+~
\left[ \frac{1}{1536} \eta_5(p^2) - \frac{1}{384} \eta_4(p^2) \right]
\frac{p^2}{\sqrt{C_A} \gamma^2} \right. \nonumber \\
&& \left. ~+~ \left[ \frac{1}{768} \eta_2(p^2) 
- \frac{1}{768} \eta_1(p^2) \right] \frac{(p^2)^2}{C_A \gamma^4} ~-~
\frac{\pi (p^2)^2}{1536 C_A \gamma^4} \right. \nonumber \\ 
&& \left. ~+~ \frac{11\sqrt{2} C_A^{1/4}\gamma}{2304\sqrt{p^2}} ~-~ 
\frac{\sqrt{2} C_A^{3/4}\gamma^3}{768 (p^2)^{3/2} } ~-~ 
\frac{\sqrt{2} \sqrt{p^2}}{768 C_A^{1/4} \gamma} \right] 
\frac{\sqrt{p^2}}{\pi} g^2 ~+~ O(g^4) 
\\
R_\xi &=& \left[ \frac{1}{192} \eta_1(p^2) ~-~ \frac{1}{128} \eta_2(p^2) ~-~
\frac{1}{192} \eta_3(p^2) ~+~ \left[ \frac{1}{192} \eta_2(p^2) 
- \frac{1}{32} \eta_1(p^2) \right]  \frac{C_A \gamma^4}{(p^2)^2} \right. 
\nonumber \\
&& \left. ~+~ \frac{5\sqrt{C_A} \gamma^2}{768 p^2} \eta_4(p^2) ~+~
\left[ \frac{1}{768} \eta_4(p^2) - \frac{1}{768} \eta_5(p^2) \right]
\frac{p^2}{\sqrt{C_A} \gamma^2} \right. \nonumber \\
&& \left. ~+~ \left[ \frac{1}{1536} \eta_1(p^2) - \frac{1}{384} \eta_2(p^2)
+ \frac{1}{1024} \eta_3(p^2) \right] \frac{(p^2)^2}{C_A \gamma^4} ~+~
\frac{\pi (p^2)^2}{768 C_A \gamma^4} \right. \nonumber \\
&& \left. ~+~ \frac{5\sqrt{2} C_A^{1/4}\gamma}{576\sqrt{p^2}} ~+~ 
\frac{\sqrt{2} C_A^{3/4}\gamma^3}{96 (p^2)^{3/2} } ~-~ 
\frac{\sqrt{2} \sqrt{p^2}}{768 C_A^{1/4} \gamma} \right] 
\frac{\sqrt{p^2}}{\pi} g^2 ~+~ O(g^4) 
\\
S_\xi &=& \left[ \frac{1}{32} \eta_1(p^2) ~-~ \frac{3}{64} \eta_2(p^2) ~-~
\frac{1}{32} \eta_3(p^2) ~+~ \frac{C_A \gamma^4}{128(p^2)^2} \eta_2(p^2) ~+~ 
\frac{\sqrt{C_A} \gamma^2}{64 p^2} \eta_4(p^2) \right. \nonumber \\
&& \left. ~+~ \left[ \frac{1}{256} \eta_5(p^2) - \frac{1}{64} \eta_4(p^2) 
\right] \frac{p^2}{\sqrt{C_A} \gamma^2} ~+~ \left[ \frac{1}{128} \eta_2(p^2) 
- \frac{1}{128} \eta_1(p^2) \right] \frac{(p^2)^2}{C_A \gamma^4} \right.
\nonumber \\
&& \left. ~-~ \frac{\pi (p^2)^2}{256 C_A \gamma^4} ~+~ \frac{11\sqrt{2} 
C_A^{1/4}\gamma}{384\sqrt{p^2}} ~-~ 
\frac{\sqrt{2} C_A^{3/4}\gamma^3}{128 (p^2)^{3/2} } ~-~ 
\frac{\sqrt{2} \sqrt{p^2}}{128 C_A^{1/4} \gamma} \right] 
\frac{\sqrt{p^2}}{C_A\pi} g^2 ~+~ O(g^4) \,. 
\end{eqnarray}
The zero momentum limit of these quantities is
\begin{eqnarray}
X &=& \left[ ~-~ \frac{\pi}{8} T_F \Nf ~+~ \frac{21\pi}{512} C_A ~-~
\frac{53 \sqrt{2} C_A^{5/4} \gamma}{384\sqrt{p^2}} ~+~ 
\frac{1231 \sqrt{2} C_A^{3/4} \sqrt{p^2}}{7680 \gamma} \right] 
\frac{\sqrt{p^2}}{\pi} g^2 ~+~ O(g^2) \nonumber \\
U &=& -~ i \left[ \frac{5 \sqrt{2} C_A^{1/4} p^2}{256 \gamma^3} ~-~
\frac{\sqrt{2} (p^2)^{3/2}}{512 \gamma^4} \right] \frac{g^2}{\pi} ~+~ 
O(g^4) ~~~,~~~ 
V ~=~ W_\rho ~=~ R_\rho ~=~ S_\rho ~=~ O(g^4) \nonumber 
\\
Q_\xi &=& Q_\rho ~=~ \frac{\sqrt{2} C_A^{3/4} p^2}{12\pi \gamma} g^2 ~+~ 
O(g^4) ~~~,~~~
W_\xi ~=~ -~ \frac{7\sqrt{2} p^2}{2880\pi C_A^{1/4} \gamma} g^2 ~+~ 
O(g^4) \nonumber \\
R_\xi &=& -~ \frac{\sqrt{2} p^2}{720\pi C_A^{1/4} \gamma} g^2 ~+~ 
O(g^4) ~~~,~~~ 
S_\xi ~=~ -~ \frac{7\sqrt{2} p^2}{480\pi C_A^{5/4} \gamma} g^2 ~+~ O(g^4) ~. 
\end{eqnarray}

\sect{Longitudinal parts.}

In this appendix we record the explicit one loop contributions to the 
longitudinal parts of the $2$-point functions. We have 
\begin{eqnarray}
X^L &=& \left[ \frac{47\pi}{512} C_A ~-~ \left[ \frac{9}{256} \eta_1(p^2) 
+ \frac{19}{256} \eta_2(p^2) \right] C_A ~-~ \left[ \frac{9}{64} \eta_1(p^2) 
+ \frac{5}{128} \eta_2(p^2) \right] \frac{C_A^2 \gamma^4}{(p^2)^2} 
\right. \nonumber \\ 
&& \left. ~+~ \frac{7C_A^{3/2} \gamma^2}{512p^2} \eta_4(p^2) ~-~ 
\frac{7 \sqrt{2} C_A^{5/4}}{64 \sqrt{p^2}} ~+~ 
\frac{7 \sqrt{2} C_A^{7/4}\gamma^3}{64 \sqrt{(p^2)^3}} \right]
\frac{\sqrt{p^2}}{\pi} g^2 ~+~ O(g^4) \\
U^L &=& -~ i \left[ ~-~ \frac{13\sqrt{C_A}}{1024\gamma^2} \eta_4(p^2) ~+~
\frac{5C_A^{3/2}\gamma^2}{1024(p^2)^2} \eta_4(p^2) ~+~ 
\left[ \frac{3}{64} \eta_1(p^2) - \frac{1}{32} \eta_2(p^2) \right] 
\frac{C_A}{p^2} \right. \nonumber \\
&& \left. ~~~~~~+~ \left[ \frac{1}{256} \eta_2(p^2) 
- \frac{3}{256} \eta_1(p^2) \right] \frac{p^2}{\gamma^4} ~+~ 
\frac{\pi p^2}{512\gamma^4} 
\right. \nonumber \\
&& \left. ~~~~~~+~ \frac{7\sqrt{2} C_A^{3/4}}{1536\gamma} ~+~ 
\frac{5\sqrt{2} C_A^{5/4}\gamma}{512 (p^2)^{3/2}} ~-~ 
\frac{\sqrt{2} C_A^{1/4} \sqrt{p^2}}{128 \gamma^3} \right] 
\frac{\sqrt{p^2}}{\pi} g^2 ~+~ O(g^4) \\
V^L &=& -~ \left[ \frac{\sqrt{C_A}}{128\gamma^2} \eta_4(p^2) ~+~
\frac{C_A}{128p^2} \eta_2(p^2) ~-~ \frac{p^2}{128\gamma^4} \eta_1(p^2) \right.
\nonumber \\
&& \left. ~~~~~~+~ \frac{\pi p^2}{128\gamma^4} ~-~ 
\frac{\sqrt{2} C_A^{3/4}}{128\gamma} ~-~ 
\frac{\sqrt{2} C_A^{1/4} \sqrt{p^2}}{128 \gamma^3} \right] 
\frac{\sqrt{p^2}}{\pi} g^2 ~+~ O(g^4) 
\\
W_\rho^L &=& R_\rho^L ~=~ S_\rho^L ~=~ O(g^4)
\\
Q_\xi^L &=& Q_\rho^L ~=~ \left[ \frac{C_A}{16} \eta_2(p^2) ~-~
\frac{C_A^{3/2} \gamma^2}{64 p^2} \eta_4(p^2) ~+~
\frac{\sqrt{C_A} p^2}{64\gamma^2} \eta_4(p^2) \right. \nonumber \\
&& \left. ~~~~~~~~~~~~~-~ \frac{\sqrt{2} C_A^{5/4} \gamma}{32 \sqrt{p^2}} ~+~ 
\frac{\sqrt{2}C_A^{3/4} \sqrt{p^2}}{32 \gamma} \right] \frac{\sqrt{p^2}}{\pi} 
g^2 ~+~ O(g^4) \\
W_\xi^L &=& \left[ ~-~ \frac{1}{48} \eta_2(p^2) ~+~ \frac{1}{48} 
\eta_3(p^2) ~-~ \frac{C_A \gamma^4}{384 (p^2)^2} \eta_2(p^2) ~+~ 
\frac{\sqrt{C_A} \gamma^2}{384 p^2} \eta_4(p^2) \right. \nonumber \\
&& \left. ~~+~ \left[ \frac{1}{192} \eta_5(p^2) - \frac{1}{128} \eta_4(p^2) 
\right] \frac{p^2}{\sqrt{C_A} \gamma^2} ~+~ \left[ \frac{1}{384} \eta_2(p^2) 
- \frac{1}{768} \eta_3(p^2) \right] \frac{(p^2)^2}{C_A \gamma^4} 
\right. \nonumber \\
&& \left. ~~-~ \frac{\pi (p^2)^2}{768 C_A \gamma^4} ~+~ 
\frac{7\sqrt{2} C_A^{1/4}\gamma}{1152\sqrt{p^2}} ~+~ 
\frac{\sqrt{2} C_A^{3/4}\gamma^3}{384 (p^2)^{3/2} } ~+~ 
\frac{\sqrt{2} \sqrt{p^2}}{384 C_A^{1/4} \gamma} \right] \! 
\frac{\sqrt{p^2}}{\pi} g^2 \, + \, O(g^4) 
\\
R_\xi^L &=& \left[ ~-~ \frac{1}{32} \eta_1(p^2) ~+~ \frac{5}{192}
\eta_2(p^2) ~-~ \frac{1}{96} \eta_3(p^2) ~+~ \left[ \frac{1}{16} \eta_1(p^2) 
- \frac{1}{96} \eta_2(p^2) \right]  \frac{C_A \gamma^4}{(p^2)^2} \right. 
\nonumber \\
&& \left. ~~-~ \frac{5\sqrt{C_A} \gamma^2}{384 p^2} \eta_4(p^2) ~+~
\left[ \frac{1}{128} \eta_4(p^2) - \frac{1}{384} \eta_5(p^2) \right]
\frac{p^2}{\sqrt{C_A} \gamma^2} \right. \nonumber \\
&& \left. ~~+~ \left[ \frac{1}{256} \eta_1(p^2) - \frac{1}{192} \eta_2(p^2)
+ \frac{1}{1536} \eta_3(p^2) \right] \frac{(p^2)^2}{C_A \gamma^4} ~+~
\frac{\pi (p^2)^2}{384 C_A \gamma^4} \right. \nonumber \\
&& \left. ~~-~ \frac{5\sqrt{2} C_A^{1/4}\gamma}{288\sqrt{p^2}} ~-~ 
\frac{\sqrt{2} C_A^{3/4}\gamma^3}{48 (p^2)^{3/2} } ~+~ 
\frac{\sqrt{2} \sqrt{p^2}}{384 C_A^{1/4} \gamma} \right] 
\frac{\sqrt{p^2}}{\pi} g^2 ~+~ O(g^4) 
\\
S_\xi^L &=& \left[ \frac{1}{8} \eta_3(p^2) ~-~ \frac{1}{8} 
\eta_2(p^2) ~-~ \frac{C_A \gamma^4}{64(p^2)^2} \eta_2(p^2) ~+~ 
\frac{\sqrt{C_A} \gamma^2}{64 p^2} \eta_4(p^2) \right. \nonumber \\
&& \left. ~+~ \left[ \frac{1}{32} \eta_5(p^2) - \frac{3}{64} \eta_4(p^2) 
\right] \frac{p^2}{\sqrt{C_A} \gamma^2} ~+~ \left[ \frac{1}{64} \eta_2(p^2) 
- \frac{1}{128} \eta_3(p^2) \right] \frac{(p^2)^2}{C_A \gamma^4} \right.
\nonumber \\
&& \left. ~-~ \frac{\pi (p^2)^2}{128 C_A \gamma^4} ~+~ 
\frac{7\sqrt{2} C_A^{1/4}\gamma}{192\sqrt{p^2}} ~+~ 
\frac{\sqrt{2} C_A^{3/4}\gamma^3}{64 (p^2)^{3/2} } ~+~ 
\frac{\sqrt{2} \sqrt{p^2}}{64 C_A^{1/4} \gamma} \right] \! 
\frac{\sqrt{p^2}}{C_A\pi} g^2 \, + \, O(g^4) \,. 
\end{eqnarray}
Analogously to the previous appendix, we record the respective zero momentum
limits are
\begin{eqnarray}
X^L &=& \left[ \frac{47\pi}{512} C_A ~-~
\frac{53 \sqrt{2} C_A^{5/4} \gamma}{384\sqrt{p^2}} ~-~ 
\frac{301 \sqrt{2} C_A^{3/4} \sqrt{p^2}}{3840 \gamma} \right] 
\frac{\sqrt{p^2}}{\pi} g^2 ~+~ O(g^2) \nonumber \\
U^L &=& -~ i \left[ \frac{\sqrt{2} C_A^{1/4} p^2}{192 \gamma^3} ~+~
\frac{\sqrt{2} (p^2)^{3/2}}{512 \gamma^4} \right] \frac{g^2}{\pi} ~+~ O(g^4) 
\nonumber 
\\
V^L &=& -~ \left[ ~-~ \frac{\sqrt{2} C_A^{1/4} p^2}{48 \gamma^3} ~+~
\frac{(p^2)^{3/2}}{128 \gamma^4} \right] \frac{g^2}{\pi} ~+~ O(g^4) ~~~,~~~ 
\nonumber \\
W_\rho^L &=& R_\rho^L ~=~ S_\rho^L ~=~ O(g^4) \nonumber 
\\
Q_\xi^L &=& Q_\rho^L ~=~ \frac{\sqrt{2} C_A^{3/4} p^2}{12\pi \gamma} g^2 ~+~ 
O(g^4) ~~~,~~~
W_\xi^L ~=~ -~ \frac{\sqrt{2} p^2}{180\pi C_A^{1/4} \gamma} g^2 ~+~ 
O(g^4) \nonumber \\
R_\xi^L &=& \frac{\sqrt{2} p^2}{360\pi C_A^{1/4} \gamma} g^2 ~+~ 
O(g^4) ~~~,~~~ 
S_\xi^L ~=~ -~ \frac{\sqrt{2} p^2}{30\pi C_A^{5/4} \gamma} g^2 ~+~ O(g^4) ~. 
\end{eqnarray}

\sect{$\rho^{ab}_\mu$ propagator.}

In this appendix we record the formal solution of (\ref{rhoprop}) for the case 
of $SU(N_c)$ as those for an arbitrary colour group are too involved. Unlike 
the solution given earlier for the $A^a_\mu$ and $\xi^{ab}_\mu$ sector we have 
not set $T_\rho$ to zero at the outset. We have, for the transverse sector 
only, 
\begin{eqnarray}
{\cal D}_\rho &=& \frac{1}{2} \left[ N_c^4 ({\cal S}_\rho)^3
-9 N_c^3 ({\cal S}_\rho)^2 {\cal W}_\rho
-72 N_c^2 ({\cal Q}_\rho)^2 {\cal S}_\rho
-114 N_c^2 {\cal Q}_\rho ({\cal S}_\rho)^2
-72 N_c^2 {\cal Q}_\rho {\cal S}_\rho {\cal T}_\rho
\right. \nonumber \\
&& \left. ~~~ 
-~ 36 N_c^2 ({\cal S}_\rho)^3
-18 N_c^2 ({\cal S}_\rho)^2 {\cal T}_\rho
-216 N_c ({\cal Q}_\rho)^2 {\cal W}_\rho
-360 N_c {\cal Q}_\rho {\cal S}_\rho {\cal W}_\rho
\right. \nonumber \\
&& \left. ~~~ 
-~216 N_c {\cal Q}_\rho {\cal T}_\rho {\cal W}_\rho
-108 N_c ({\cal S}_\rho)^2 {\cal W}_\rho
-216 N_c {\cal S}_\rho {\cal T}_\rho {\cal W}_\rho
+108 N_c ({\cal W}_\rho)^3
\right. \nonumber \\
&& \left. ~~~ 
-~ 432 ({\cal Q}_\rho)^3
-864 ({\cal Q}_\rho)^2 {\cal S}_\rho
-864 ({\cal Q}_\rho)^2 {\cal T}_\rho
-432 {\cal Q}_\rho ({\cal S}_\rho)^2
\right. \nonumber \\
&& \left. ~~~ 
-~ 864 {\cal Q}_\rho {\cal S}_\rho {\cal T}_\rho
-432 {\cal Q}_\rho ({\cal T}_\rho)^2
+432 {\cal Q}_\rho ({\cal W}_\rho)^2 \right] \nonumber \\ 
&& ~ \times \left[ 6 {\cal Q}_\rho + N_c^2 {\cal S}_\rho + 3 N_c {\cal W}_\rho 
+ 6 {\cal T}_\rho \right]^{-1} 
\left[ 6 {\cal Q}_\rho + (N_c + 6) {\cal S}_\rho - 6 {\cal W}_\rho 
+ {\cal T}_\rho \right]^{-1} \nonumber \\
&& ~ \times  
\left[ (N_c - 6) {\cal S}_\rho - 6 {\cal Q}_\rho - 6 {\cal T}_\rho 
- 6 {\cal W}_\rho \right]^{-1}
\left[ {\cal Q}_\rho - {\cal T}_\rho \right]^{-1} 
\nonumber \\ 
{\cal J}_\rho &=& 4 \left[ -N_c^3 ({\cal S}_\rho)^2
+3 N_c^2 {\cal S}_\rho {\cal W}_\rho +9 N_c ({\cal S}_\rho)^2
+18 N_c ({\cal W}_\rho)^2 +54 {\cal Q}_\rho {\cal W}_\rho
+54 {\cal T}_\rho {\cal W}_\rho \right] \nonumber \\ 
&& ~ \times \left[ 6 {\cal Q}_\rho + N_c^2 {\cal S}_\rho + 3 N_c {\cal W}_\rho 
+ 6 {\cal T}_\rho \right]^{-1} 
\left[ 6 {\cal Q}_\rho + (N_c + 6) {\cal S}_\rho - 6 {\cal W}_\rho 
+ {\cal T}_\rho \right]^{-1} \nonumber \\
&& ~ \times  
\left[ (N_c - 6) {\cal S}_\rho - 6 {\cal Q}_\rho - 6 {\cal T}_\rho 
- 6 {\cal W}_\rho \right]^{-1}
\nonumber \\ 
{\cal K}_\rho &=& \left[ 4 N_c^4 {\cal Q}_\rho {\cal R}_\rho ({\cal S}_\rho)^2
+2 N_c^4 {\cal Q}_\rho ({\cal S}_\rho)^2 {\cal W}_\rho
-2 N_c^4 {\cal R}_\rho ({\cal S}_\rho)^3
-4 N_c^4 {\cal R}_\rho ({\cal S}_\rho)^2 {\cal T}_\rho
-N_c^4 ({\cal S}_\rho)^3 {\cal W}_\rho
\right. \nonumber \\
&& \left. ~
-~ 2 N_c^4 ({\cal S}_\rho)^2 {\cal T}_\rho {\cal W}_\rho
+4 N_c^3 ({\cal Q}_\rho)^2 ({\cal S}_\rho)^2
-12 N_c^3 {\cal Q}_\rho {\cal R}_\rho {\cal S}_\rho {\cal W}_\rho
-8 N_c^3 {\cal Q}_\rho ({\cal S}_\rho)^2 {\cal T}_\rho
\right. \nonumber \\
&& \left. ~
-~ 6 N_c^3 {\cal Q}_\rho {\cal S}_\rho ({\cal W}_\rho)^2
+18 N_c^3 {\cal R}_\rho ({\cal S}_\rho)^2 {\cal W}_\rho
+12 N_c^3 {\cal R}_\rho {\cal S}_\rho {\cal T}_\rho {\cal W}_\rho
+4 N_c^3 ({\cal S}_\rho)^2 ({\cal T}_\rho)^2
\right. \nonumber \\
&& \left. ~
+~ 9 N_c^3 ({\cal S}_\rho)^2 ({\cal W}_\rho)^2
+6 N_c^3 {\cal S}_\rho {\cal T}_\rho ({\cal W}_\rho)^2
+72 N_c^2 ({\cal Q}_\rho)^2 {\cal R}_\rho {\cal S}_\rho
+24 N_c^2 ({\cal Q}_\rho)^2 {\cal S}_\rho {\cal W}_\rho
\right. \nonumber \\
&& \left. ~
+~ 96 N_c^2 {\cal Q}_\rho {\cal R}_\rho ({\cal S}_\rho)^2
+144 N_c^2 {\cal Q}_\rho {\cal R}_\rho {\cal S}_\rho {\cal T}_\rho
-72 N_c^2 {\cal Q}_\rho {\cal R}_\rho ({\cal W}_\rho)^2
+48 N_c^2 {\cal Q}_\rho ({\cal S}_\rho)^2 {\cal W}_\rho
\right. \nonumber \\
&& \left. ~
+~ 96 N_c^2 {\cal Q}_\rho {\cal S}_\rho {\cal T}_\rho {\cal W}_\rho
-36 N_c^2 {\cal Q}_\rho ({\cal W}_\rho)^3
+72 N_c^2 {\cal R}_\rho ({\cal S}_\rho)^3
+168 N_c^2 {\cal R}_\rho ({\cal S}_\rho)^2 {\cal T}_\rho
\right. \nonumber \\
&& \left. ~
+~ 72 N_c^2 {\cal R}_\rho {\cal S}_\rho ({\cal T}_\rho)^2
+72 N_c^2 {\cal R}_\rho {\cal T}_\rho ({\cal W}_\rho)^2
+36 N_c^2 ({\cal S}_\rho)^3 {\cal W}_\rho
+84 N_c^2 ({\cal S}_\rho)^2 {\cal T}_\rho {\cal W}_\rho
\right. \nonumber \\
&& \left. ~
+~ 24 N_c^2 {\cal S}_\rho ({\cal T}_\rho)^2 {\cal W}_\rho
+36 N_c^2 {\cal T}_\rho ({\cal W}_\rho)^3
-36 N_c ({\cal Q}_\rho)^2 ({\cal S}_\rho)^2
-72 N_c ({\cal Q}_\rho)^2 ({\cal W}_\rho)^2
\right. \nonumber \\
&& \left. ~
+~ 576 N_c {\cal Q}_\rho {\cal R}_\rho {\cal S}_\rho {\cal W}_\rho
+432 N_c {\cal Q}_\rho {\cal R}_\rho {\cal T}_\rho {\cal W}_\rho
+72 N_c {\cal Q}_\rho ({\cal S}_\rho)^2 {\cal T}_\rho
+288 N_c {\cal Q}_\rho {\cal S}_\rho ({\cal W}_\rho)^2
\right. \nonumber \\
&& \left. ~
+~ 360 N_c {\cal Q}_\rho {\cal T}_\rho ({\cal W}_\rho)^2
+216 N_c {\cal R}_\rho ({\cal S}_\rho)^2 {\cal W}_\rho
+576 N_c {\cal R}_\rho {\cal S}_\rho {\cal T}_\rho {\cal W}_\rho
+432 N_c {\cal R}_\rho ({\cal T}_\rho)^2 {\cal W}_\rho
\right. \nonumber \\
&& \left. ~
-~ 216 N_c {\cal R}_\rho ({\cal W}_\rho)^3
-36 N_c ({\cal S}_\rho)^2 ({\cal T}_\rho)^2
+108 N_c ({\cal S}_\rho)^2 ({\cal W}_\rho)^2
+288 N_c {\cal S}_\rho {\cal T}_\rho ({\cal W}_\rho)^2
\right. \nonumber \\
&& \left. ~
+~ 144 N_c ({\cal T}_\rho)^2 ({\cal W}_\rho)^2
-108 N_c ({\cal W}_\rho)^4
+432 ({\cal Q}_\rho)^3 {\cal R}_\rho
+864 ({\cal Q}_\rho)^2 {\cal R}_\rho {\cal S}_\rho
\right. \nonumber \\
&& \left. ~
+~ 1296 ({\cal Q}_\rho)^2 {\cal R}_\rho {\cal T}_\rho
+432 ({\cal Q}_\rho)^2 {\cal S}_\rho {\cal W}_\rho
+864 ({\cal Q}_\rho)^2 {\cal T}_\rho {\cal W}_\rho
+432 {\cal Q}_\rho {\cal R}_\rho ({\cal S}_\rho)^2
\right. \nonumber \\
&& \left. ~
+~ 1728 {\cal Q}_\rho {\cal R}_\rho {\cal S}_\rho {\cal T}_\rho
+1296 {\cal Q}_\rho {\cal R}_\rho ({\cal T}_\rho)^2
-432 {\cal Q}_\rho {\cal R}_\rho ({\cal W}_\rho)^2
+216 {\cal Q}_\rho ({\cal S}_\rho)^2 {\cal W}_\rho
\right. \nonumber \\
&& \left. ~
+~ 864 {\cal Q}_\rho {\cal S}_\rho {\cal T}_\rho {\cal W}_\rho
+864 {\cal Q}_\rho ({\cal T}_\rho)^2 {\cal W}_\rho
-216 {\cal Q}_\rho ({\cal W}_\rho)^3
+432 {\cal R}_\rho ({\cal S}_\rho)^2 {\cal T}_\rho
\right. \nonumber \\
&& \left. ~
+~ 864 {\cal R}_\rho {\cal S}_\rho ({\cal T}_\rho)^2
+432 {\cal R}_\rho ({\cal T}_\rho)^3
-432 {\cal R}_\rho {\cal T}_\rho ({\cal W}_\rho)^2
\right. \nonumber \\
&& \left. ~
+~ 216 ({\cal S}_\rho)^2 {\cal T}_\rho {\cal W}_\rho
+432 {\cal S}_\rho ({\cal T}_\rho)^2 {\cal W}_\rho
-216 {\cal T}_\rho ({\cal W}_\rho)^3 \right] \nonumber \\ 
&& ~ \times \left[ 6 {\cal Q}_\rho + N_c^2 {\cal S}_\rho + 3 N_c {\cal W}_\rho 
+ 6 {\cal T}_\rho \right]^{-1} 
\left[ 2 N_c {\cal R}_\rho + N_c {\cal W}_\rho + 2 {\cal Q}_\rho 
- 2 {\cal T}_\rho \right]^{-1} \nonumber \\
&& ~ \times  
\left[ 6 {\cal Q}_\rho + (N_c + 6) {\cal S}_\rho - 6 {\cal W}_\rho 
+ {\cal T}_\rho \right]^{-1} 
\left[ (N_c - 6) {\cal S}_\rho - 6 {\cal Q}_\rho - 6 {\cal T}_\rho 
- 6 {\cal W}_\rho \right]^{-1} \! 
\left[ {\cal Q}_\rho - {\cal T}_\rho \right]^{-1} \nonumber \\ 
{\cal L}_\rho &=& 12 \left[ N_c^2 ({\cal S}_\rho)^2
-3 N_c {\cal S}_\rho {\cal W}_\rho +18 {\cal Q}_\rho {\cal S}_\rho
+18 ({\cal S}_\rho)^2 +18 {\cal S}_\rho {\cal T}_\rho
-18 ({\cal W}_\rho)^2 \right] \nonumber \\ 
&& ~ \times \left[ 6 {\cal Q}_\rho + N_c^2 {\cal S}_\rho + 3 N_c {\cal W}_\rho 
+ 6 {\cal T}_\rho \right]^{-1} 
\left[ 6 {\cal Q}_\rho + (N_c + 6) {\cal S}_\rho - 6 {\cal W}_\rho 
+ {\cal T}_\rho \right]^{-1} \nonumber \\
&& ~ \times 
\left[ (N_c - 6) {\cal S}_\rho - 6 {\cal Q}_\rho - 6 {\cal T}_\rho 
- 6 {\cal W}_\rho \right]^{-1} \nonumber \\
{\cal M}_\rho &=& 6 \left[ -6 N_c^4 {\cal P}_\rho ({\cal S}_\rho)^2
-5 N_c^4 ({\cal S}_\rho)^3 +18 N_c^3 {\cal P}_\rho {\cal S}_\rho {\cal W}_\rho
+9 N_c^3 ({\cal S}_\rho)^2 {\cal W}_\rho
-144 N_c^2 {\cal P}_\rho {\cal Q}_\rho {\cal S}_\rho
\right. \nonumber \\
&& \left. ~~~ 
-~ 168 N_c^2 {\cal P}_\rho ({\cal S}_\rho)^2
-144 N_c^2 {\cal P}_\rho {\cal S}_\rho {\cal T}_\rho
+108 N_c^2 {\cal P}_\rho ({\cal W}_\rho)^2
-126 N_c^2 {\cal Q}_\rho ({\cal S}_\rho)^2 
\right. \nonumber \\
&& \left. ~~~ 
-~ 144 N_c^2 ({\cal S}_\rho)^3
-126 N_c^2 ({\cal S}_\rho)^2 {\cal T}_\rho
+108 N_c^2 {\cal S}_\rho ({\cal W}_\rho)^2
-108 N_c {\cal P}_\rho {\cal Q}_\rho {\cal W}_\rho
\right. \nonumber \\
&& \left. ~~~ 
+~ 72 N_c {\cal P}_\rho {\cal S}_\rho {\cal W}_\rho
- 108 N_c {\cal P}_\rho {\cal T}_\rho {\cal W}_\rho
-216 N_c {\cal Q}_\rho {\cal S}_\rho {\cal W}_\rho
-108 N_c ({\cal S}_\rho)^2 {\cal W}_\rho
\right. \nonumber \\
&& \left. ~~~ 
-~216 N_c {\cal S}_\rho {\cal T}_\rho {\cal W}_\rho 
+ 108 N_c ({\cal W}_\rho)^3
+ 216 {\cal P}_\rho ({\cal Q}_\rho)^2
+432 {\cal P}_\rho {\cal Q}_\rho {\cal S}_\rho
+432 {\cal P}_\rho {\cal Q}_\rho {\cal T}_\rho
\right. \nonumber \\
&& \left. ~~~ 
+~ 216 {\cal P}_\rho ({\cal S}_\rho)^2
+432 {\cal P}_\rho {\cal S}_\rho {\cal T}_\rho
+ 216 {\cal P}_\rho ({\cal T}_\rho)^2
-216 {\cal P}_\rho ({\cal W}_\rho)^2 \right] \nonumber \\
&& ~ \times \left[ 6 {\cal Q}_\rho + N_c^2 {\cal S}_\rho + 3 N_c {\cal W}_\rho 
+ 6 {\cal T}_\rho \right]^{-1} 
\nonumber \\
&& ~ \times 
\left[ 6 (N_c^2 - 6) {\cal P}_\rho + 5 N_c^2 {\cal S}_\rho 
+ 6 N_c {\cal W}_\rho + 6 {\cal Q}_\rho + 6 {\cal T}_\rho \right]^{-1} 
\nonumber \\
&& ~ \times 
\left[ 6 {\cal Q}_\rho + (N_c + 6) {\cal S}_\rho - 6 {\cal W}_\rho 
+ {\cal T}_\rho \right]^{-1} 
\left[ (N_c - 6) {\cal S}_\rho - 6 {\cal Q}_\rho - 6 {\cal T}_\rho 
- 6 {\cal W}_\rho \right]^{-1}
\nonumber \\ 
{\cal N}_\rho &=& \frac{1}{2} \left[ -N_c^4 ({\cal S}_\rho)^3
+9 N_c^3 ({\cal S}_\rho)^2 {\cal W}_\rho
+18 N_c^2 {\cal Q}_\rho ({\cal S}_\rho)^2
+72 N_c^2 {\cal Q}_\rho {\cal S}_\rho {\cal T}_\rho
+36 N_c^2 ({\cal S}_\rho)^3 
\right. \nonumber \\
&& \left. ~~~ 
+~ 114 N_c^2 ({\cal S}_\rho)^2 {\cal T}_\rho
+72 N_c^2 {\cal S}_\rho ({\cal T}_\rho)^2
+216 N_c {\cal Q}_\rho {\cal S}_\rho {\cal W}_\rho
+216 N_c {\cal Q}_\rho {\cal T}_\rho {\cal W}_\rho
\right. \nonumber \\
&& \left. ~~~ 
+~ 108 N_c ({\cal S}_\rho)^2 {\cal W}_\rho
+360 N_c {\cal S}_\rho {\cal T}_\rho {\cal W}_\rho
+216 N_c ({\cal T}_\rho)^2 {\cal W}_\rho -108 N_c ({\cal W}_\rho)^3
\right. \nonumber \\
&& \left. ~~~ 
+~ 432 ({\cal Q}_\rho)^2 {\cal T}_\rho
+864 {\cal Q}_\rho {\cal S}_\rho {\cal T}_\rho
+864 {\cal Q}_\rho ({\cal T}_\rho)^2 +432 ({\cal S}_\rho)^2 {\cal T}_\rho
\right. \nonumber \\
&& \left. ~~~ 
+~ 864 {\cal S}_\rho ({\cal T}_\rho)^2 +432 ({\cal T}_\rho)^3
-432 {\cal T}_\rho ({\cal W}_\rho)^2 \right] \nonumber \\ 
&& ~ \times \left[ 6 {\cal Q}_\rho + N_c^2 {\cal S}_\rho + 3 N_c {\cal W}_\rho 
+ 6 {\cal T}_\rho \right]^{-1} 
\left[ 6 {\cal Q}_\rho + (N_c + 6) {\cal S}_\rho - 6 {\cal W}_\rho 
+ {\cal T}_\rho \right]^{-1} \nonumber \\
&& ~ \times 
\left[ (N_c - 6) {\cal S}_\rho - 6 {\cal Q}_\rho - 6 {\cal T}_\rho 
- 6 {\cal W}_\rho \right]^{-1} 
\left[ {\cal Q}_\rho - {\cal T}_\rho \right]^{-1} ~.
\end{eqnarray} 
One of the reasons for recording these expressions is that they illustrate how
cumbersome the full propagator structure for the two colour index field is
even in the case where there is no matrix of $2$-point functions. However, a
more important feature is to illustrate the role of the implementation of the
gap equation to extract the enhanced propagators. In these final expressions
the enhancement will derive from the factors $\left[ {\cal Q}_\rho - 
{\cal T}_\rho \right]^{-1}$ in the ${\cal D}_\rho$, ${\cal K}_\rho$ and 
${\cal N}_\rho$ channels. If it transpired from explicit calculations at any
loop order that ${\cal T}_\rho$ was non-zero or not proportional to the gap
equation itself then there would be no enhancement of $\rho^{ab}_\mu$. The
same situation would occur for $\xi^{ab}_\mu$ but in respect of ${\cal T}_\xi$
in that case independent of the complication of having to analyse the
$2$~$\times$~$2$ matrix for the gluon and $\xi^{ab}$ sector. Similar comments
apply to the longitudinal situation as the equations for the longitudinal
piece are formally similar.

\end{document}